\documentclass[prd, aps, twocolumn, showpacs,superscriptaddress,nofootinbib]{revtex4-1}

\usepackage{amsbsy,amsmath,amssymb,amsthm,wasysym,amsfonts}
\usepackage{graphicx}
\usepackage{psfrag}
\usepackage{epstopdf}
\usepackage{color}
\usepackage{mathrsfs}
\usepackage{comment}
\usepackage{bm}
\usepackage[normalem]{ulem}
\usepackage[utf8x]{inputenc}

\DeclareSymbolFontAlphabet{\mathrsfs}{rsfs}
\DeclareMathAlphabet\mathbfcal{OMS}{cmsy}{b}{n}

\newcommand{\be}{\begin{equation}}  
\newcommand{\ee}{\end{equation}}
\newcommand{\bea}{\begin{eqnarray}}           
\newcommand{\eea}{\end{eqnarray}} 
\newcommand{\beqn}{\begin{eqnarray*}}
\newcommand{\eeqn}{\end{eqnarray*}}
\newcommand{\ba}{\begin{align}}
\newcommand{\ea}{\end{align}}

\def\lm{{\ell m}}   

\def\f{{\hat{f}}}
\def\ha{{\hat{a}}}
\def\ta{{\tilde{a}}}
\def\to{{\tilde{a}_{1}}}
\def\tt{{\tilde{a}_{2}}}

\def\p4{{\psi_4}} 

\usepackage{float}
\definecolor{cyan}{rgb}{0,0.9,0.9}
\definecolor{orange}{rgb}{0.9,0.5,0}
\definecolor{magenta}{rgb}{1,0,1}
\definecolor{purple}{rgb}{0.8,0.4,0.8}

\begin{document}

\title{Factorization and resummation: A new paradigm to improve\\ 
gravitational wave amplitudes. II: the higher multipolar modes.}

\author{Francesco \surname{Messina}}
\affiliation{Dipartimento di Fisica, Universit\'a degli studi di Milano Bicocca, Piazza della Scienza 3, 20126 Milano, Italy}
\affiliation{INFN, Sezione di Milano Bicocca, Piazza della Scienza 3, 20126 Milano, Italy}
\author{Alberto \surname{Maldarella}}
\affiliation{Dipartimento di Fisica, Universit\`a di Torino, via P. Giuria 1, I-10125 Torino, Italy}
\author{Alessandro \surname{Nagar}}
\affiliation{Centro Fermi - Museo Storico della Fisica e Centro Studi e Ricerche ``Enrico Fermi'', 00184 Roma, Italy}
 \affiliation{INFN Sezione di Torino, Via P.~Giuria 1, 10125 Torino, Italy}
\affiliation{Institut des Hautes Etudes Scientifiques, 91440 Bures-sur-Yvette, France}
\begin{abstract}
  The factorization and resummation approach of Nagar and Shah~[Phys.~Rev.~D~94 (2016), 104017],
  designed to improve the strong-field behavior of the post-Newtonian (PN)
  residual waveform amplitudes $f_\lm$'s entering the effective-one-body, circularized,
  gravitational waveform for spinning coalescing binaries, is here improved and generalized to all
  multipoles up to $\ell=6$. For a test-particle orbiting a Kerr black hole, each multipolar
  amplitude is truncated at relative 6~post Newtonian (PN) order, both for the orbital
  (nonspinning) and spin factors. By taking a certain Pad\'e approximant
  (typically the $P^4_2$ one) of the orbital factor in conjuction with the inverse Taylor
  (iResum) representation of the spin factor, it is possible to push the analytical/numerical
  agreement of the energy fluxe at the level of $5\%$ at the last-stable-orbit
  for a quasi-maximally spinning black hole with dimensionless spin parameter $+0.99$.
  When the procedure is generalized to comparable-mass binaries, each orbital factor is
  kept at relative $3^{+3}$PN order, i.e. the 3PN comparable-mass terms are hybridized
  with higher PN test-particle terms up to 6PN relative order. The same Pad\'e resummation is used
  for continuity. By contrast, the spin factor is only kept at the highest comparable-mass
  PN-order currently available. We illustrate that the consistency between different truncations
  in the spin content of the waveform amplitudes is stronger in the resummed case than
  when using the standard Taylor-expanded form of Pan et al.~[Phys.~Rev.~D~83 (2011) 064003].
  We finally introduce a method to consistently hybridize comparable-mass and test-particle
  information {\it also} in the presence of spin (including the spin of the particle),
  discussing it explicitly for the $\ell=m=2$ spin-orbit and spin-square terms.
  The improved, factorized and resummed, multipolar waveform amplitudes presented
  here are expected to set a new standard for effective-one-body-based
  gravitational waveform models. 
\end{abstract}
\date{\today}


\maketitle

\section{Introduction}
The parameter estimation of gravitational wave
events~\cite{Abbott:2016blz,Abbott:2016nmj,Abbott:2016nmj,Abbott:2017oio,Abbott:2017vtc,Abbott:2017gyy,TheLIGOScientific:2017qsa}
relies on analytical waveforms models, possibly calibrated
(or informed) by Numerical Relativity simulations~\cite{Taracchini:2013rva,Hannam:2013oca,Schmidt:2014iyl,Khan:2015jqa,Nagar:2015xqa,Babak:2016tgq,Bohe:2016gbl,Nagar:2017jdw}.
The effective-one-body (EOB) model is currently the only analytical model available
that can be consistently used for analyzing both black hole binaries and neutron star
binaries~\cite{Damour:2009wj,Bernuzzi:2014owa,Lackey:2016krb,Hinderer:2016eia,Steinhoff:2016rfi,Dietrich:2017feu}.
One of the central building blocks of the model is the factorized and
resummed (circularized) multipolar post-Newtonian (PN) waveform
introduced in~\cite{Damour:2008gu} for nonspinning binaries.
This approach was then straightforwardly generalized in~\cite{Pan:2010hz}
to spinning binaries. Already Ref.~\cite{Pan:2010hz} pointed out that,
in the test-particle limit, the amplitude of such resummed waveform gets
inaccurate in the strong-field, fast velocity regime, when the spin of
the central black hole is $\gtrsim 0.7$. In the same study, an alternative
factorization to improve the test-mass waveform behavior also for larger
values of the spin was discussed. More pragmatically, Ref.~\cite{Taracchini:2013wfa}
finally suggested to improve the analytical multipolar waveform amplitude
(and fluxes) of~\cite{Pan:2010hz} by fitting a few parameters, describing
effective high-PN orders, to the highly-accurate fluxes obtained solving
numerically the Teukolsky equation~\cite{Hughes:2005qb}.
Although this approach is certainly useful to reliably improve the
radiation reaction force that drives the transition from quasi-circular
inspiral to plunge~\cite{Taracchini:2014zpa,Harms:2014dqa,Nagar:2014kha}
for a large mass-ratio binary, the question remains whether
the domain of validity of purely analytical results can be
enlarged in some way. This question makes special sense
nowadays, since PN calculations of the fluxes are available
at high order~\cite{Fujita:2014eta,Shah:2014tka} and one would
like to use then at best. In addition, following for example the seminal
attitude of Refs.~\cite{Damour:2007xr,Damour:2008gu}, one has to keep
in mind that the test-particle limit should always be seen as a
useful {\it theoretical laboratory} to implement new methods and
test new ideas that could be transferred, after suitable modifications,
to the case of comparable-mass binaries.

Reference~\cite{Nagar:2016ayt} gave a fresh cut to this problem by exploring
a new way of treating the residual, PN-expanded, amplitude corrections
to the waveforms (i.e., the outcome of the factorization of Refs.~\cite{Damour:2008gu,Pan:2010hz})
that consists of: (i) factorizing it in a purely orbital and a purely
spin-dependent part; (ii) separately resumming each factor in various ways,
notably using the inverse Taylor (``iResum'') approximant for the spin-dependent
factor. Using the test-particle limit to probe the approach,
Ref.~\cite{Nagar:2016ayt} showed that such factorization--and--resummation
paradigm yields a rather good agreement between the $\ell=2$ numerical
and analytical waveform amplitudes up to (and often beyond)
the last stable orbit (LSO). The contextual preliminary analysis
of the comparable-mass case of~\cite{Nagar:2016ayt} also suggests
that such improved waveform amplitudes are more robust
than the standard ones and may eventually need less important
NR-calibration via the next-to-quasi-circular correction
factor~\cite{Damour:2007xr}.

The purpose of this paper is to deepen and refine the investigation
of Ref.~\cite{Nagar:2016ayt} as well as to generalize it to higher
multipoles up to $\ell=6$.
The paper is organized as follows. In Sec.~\ref{sec:iRlm} we review
and improve the test-particle results of~\cite{Nagar:2016ayt} and
generalize the procedure up to $\ell=6$ modes. Section~\ref{sec:PNhlm}
brings together all the PN-expanded results currently available for
the spin-dependent waveform amplitudes~\cite{Bohe:2013cla,Marsat:2014xea,Bohe:2015ana},
notably written in multipolar form, while Sec.~\ref{sec:factorization}
explicitly shows the spin-dependent part of the factorized residual amplitudes,
both in the standard form of~\cite{Damour:2008gu,Pan:2010hz}, and with
the factorization of the orbital terms. The approach to the resummation
is undertaken in Sec.~\ref{sec:resum}, in particular by discussing the
{\it hybridization} (notably of the orbital terms) with the test-particle
information. After the conclusions, Sec.~\ref{sec:conclusions} the paper
is completed by an Appendix that lists all the currently known, PN-expanded,
$\nu$-dependent, energy fluxes up to next-to-next-to-leading order in the
spin orbit interaction (which include also next-to-leading-order for the
spin-spin-terms and leading-order for spin-cube terms). We use units with
$c=G=1$.

\section{Test-particle limit: improving the residual multipolar amplitudes}
\label{sec:iRlm}
The purpose of this Section is to review and improve the test-particle results of 
Ref.~\cite{Nagar:2016ayt} for the $\ell=2$ multipole and then generalize them
to all multipoles up to $\ell=6$. Let us recall our notation for the multipolar waveform
for a circularized, nonprecessing, binary with total mass $M$ and (dimensionful)
spins $S_1$ and $S_2$. To start with, following Ref.~\cite{Damour:2008gu}
(see e.g. Eq.~(75)-(78) there), each waveform multipole is written as
\be
\label{eq:hlm0}
h_{\ell m}(x)=h_{\ell m}^{(N,\epsilon)}\hat{h}_{\ell m}^{(\epsilon)},
\ee
where $x = (GM\Omega/c^3)^{2/3}={\cal O}(c^{-2})$ is the PN-ordering
frequency parameter ($\Omega$ is the orbital frequency) [we recall that
$n$-PN order means ${\cal O}(c^{-2n})$]; $h_{\ell m}^{(N,\epsilon)}$ is 
the Newtonian (leading-order) contribution to the given $(\ell,m)$ 
multipole, where $\epsilon=0,1$ is the parity of $\ell+m$ (see Eq.~(78) 
of~\cite{Damour:2008gu} and Eq.~\eqref{eq:hlmNexpl} in Appendix),
while  $\hat{h}_{\ell m}^{(\epsilon)}$ is the PN correction. 
Such PN correction is then written in factorized
form~\cite{Damour:2008gu} as
\be
\hat{h}_{\ell m}(x) =\hat{S}_{\rm eff}^{(\epsilon)}\hat{h}_{\ell m}^{\rm tail}f_{\ell m}(x,S_1,S_2).
\ee
Here, the first factor, $\hat{S}^{(\epsilon)}_{\rm eff}$ , is the parity-dependent effective source 
term~\cite{Damour:2008gu}, define as the EOB effective energy along circular orbits,
for $\epsilon=0$, or the Newton-normalized orbital angular momentum,
for $\epsilon=1$; the second factor, $\hat{h}_\lm^{\rm tail}\equiv T_\lm e^{\rm i \delta_\lm}$
is a complex factor that accounts for the effect of the tails and other
phase-related effects~\cite{Damour:2008gu,Mano:1996vt,Faye:2014fra}; 
the third factor, $f_\lm$ is the residual amplitude correction. This latter
factor can be further resummed in various ways, that notably depend,
when  $\nu\neq 0$ and  $S_{1,2}\neq 0$, on the parity of $m$. 
For example, the original proposal of~\cite{Damour:2008gu}, 
implemented when the objects are nonspinning, was to first
compute from the $f_{\ell m}$ the (Taylor-expanded) functions
\be
\rho_{\ell m} \equiv T_n\left[(f_\lm)^{1/\ell}\right],
\ee
where $T_n[\dots]$ indicates the Taylor expansion up to $x^n$ and then define the
resummed $f_\lm$ by replacing their Taylor expansions with $(\rho_{\lm})^\ell$.
When spins are present, the $\rho_\lm$ functions are naturally written as the
sum of an orbital (spin-independent) and a spin-dependent contribution as
\be
\rho_\lm = \rho_\lm^{\rm orb}+\rho_\lm^{\rm S}.
\ee
\begin{figure*}[t]
  \center
 \includegraphics[width=0.21\textwidth]{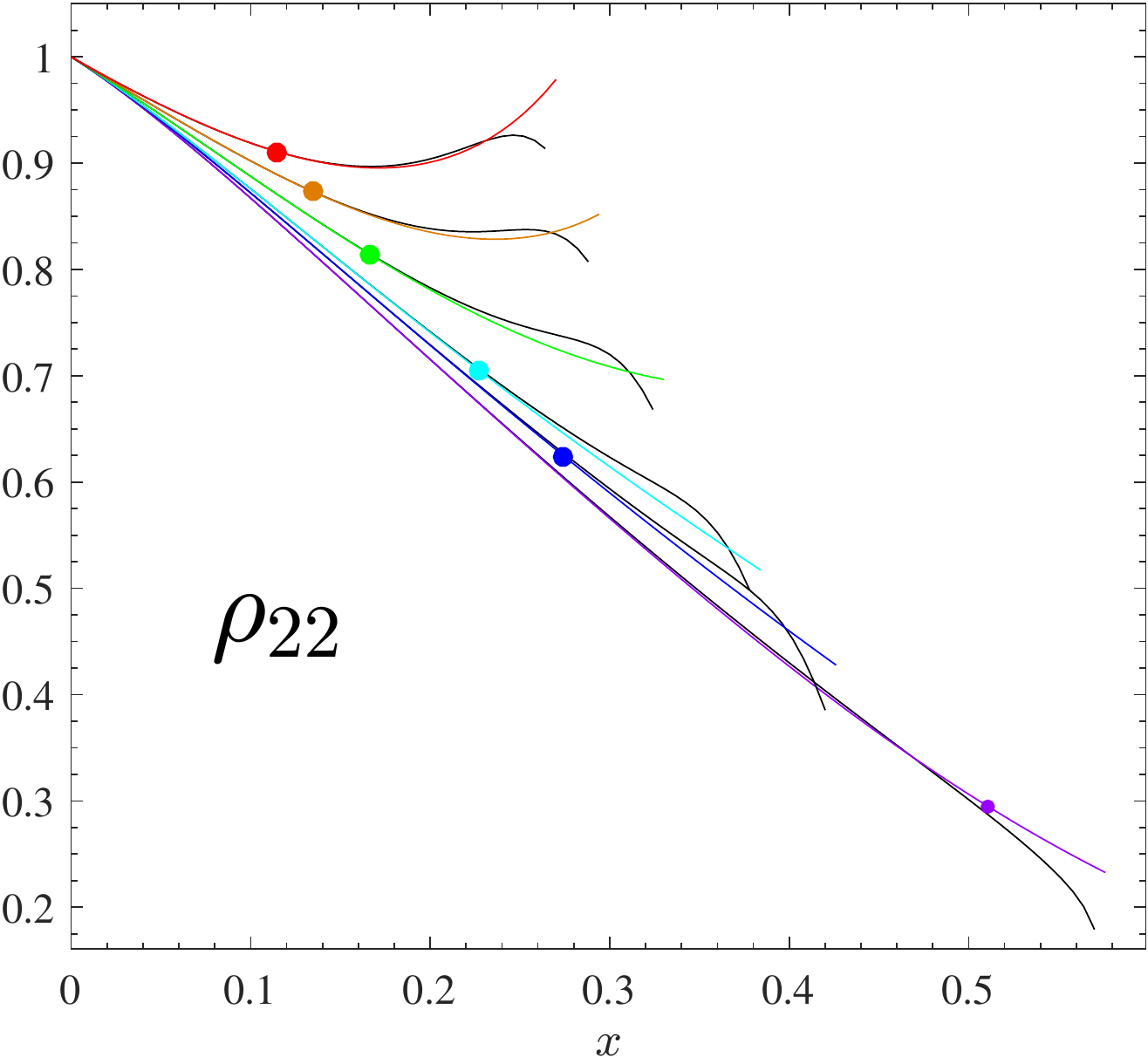}
  \includegraphics[width=0.21\textwidth]{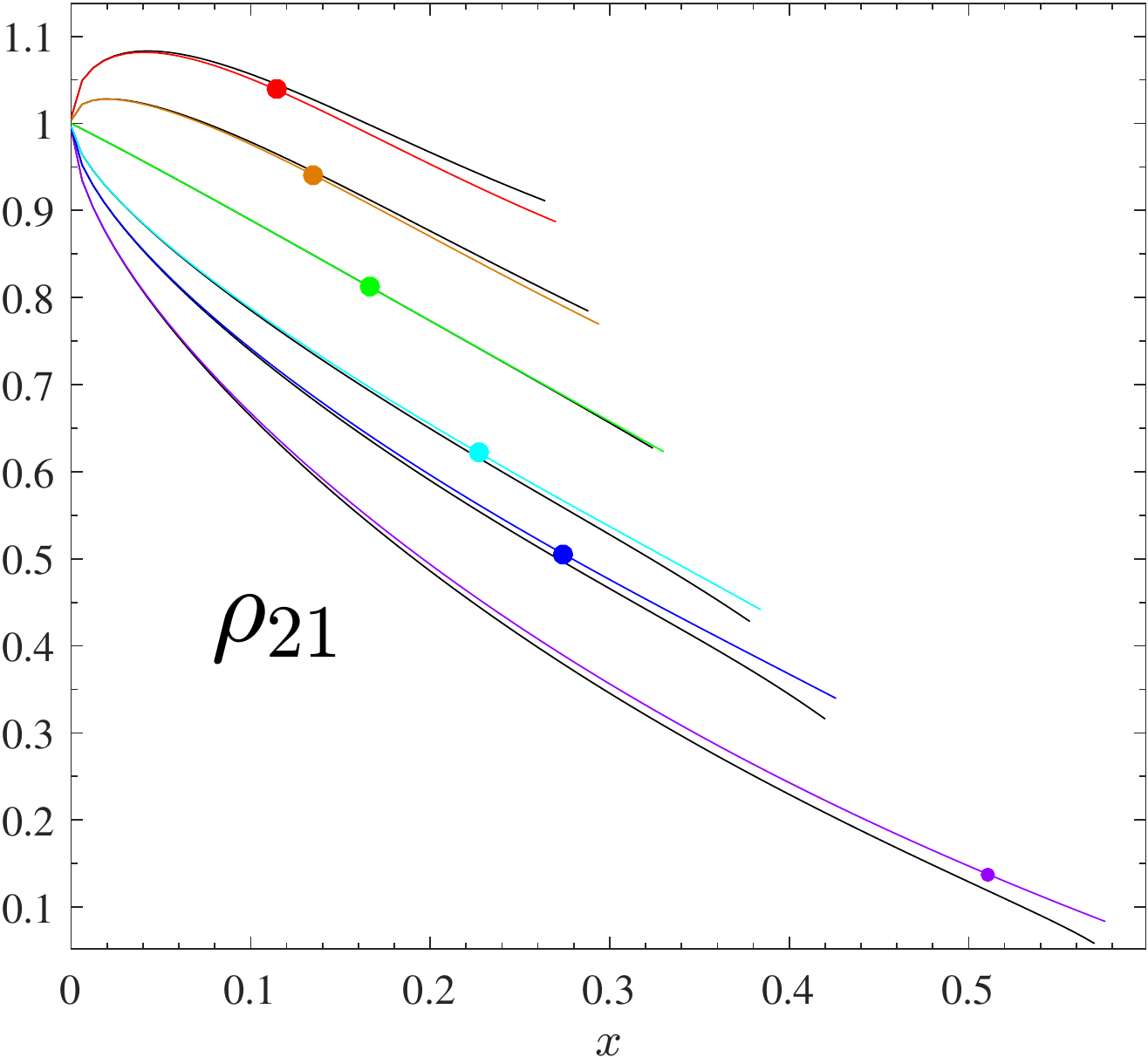}
  \includegraphics[width=0.21\textwidth]{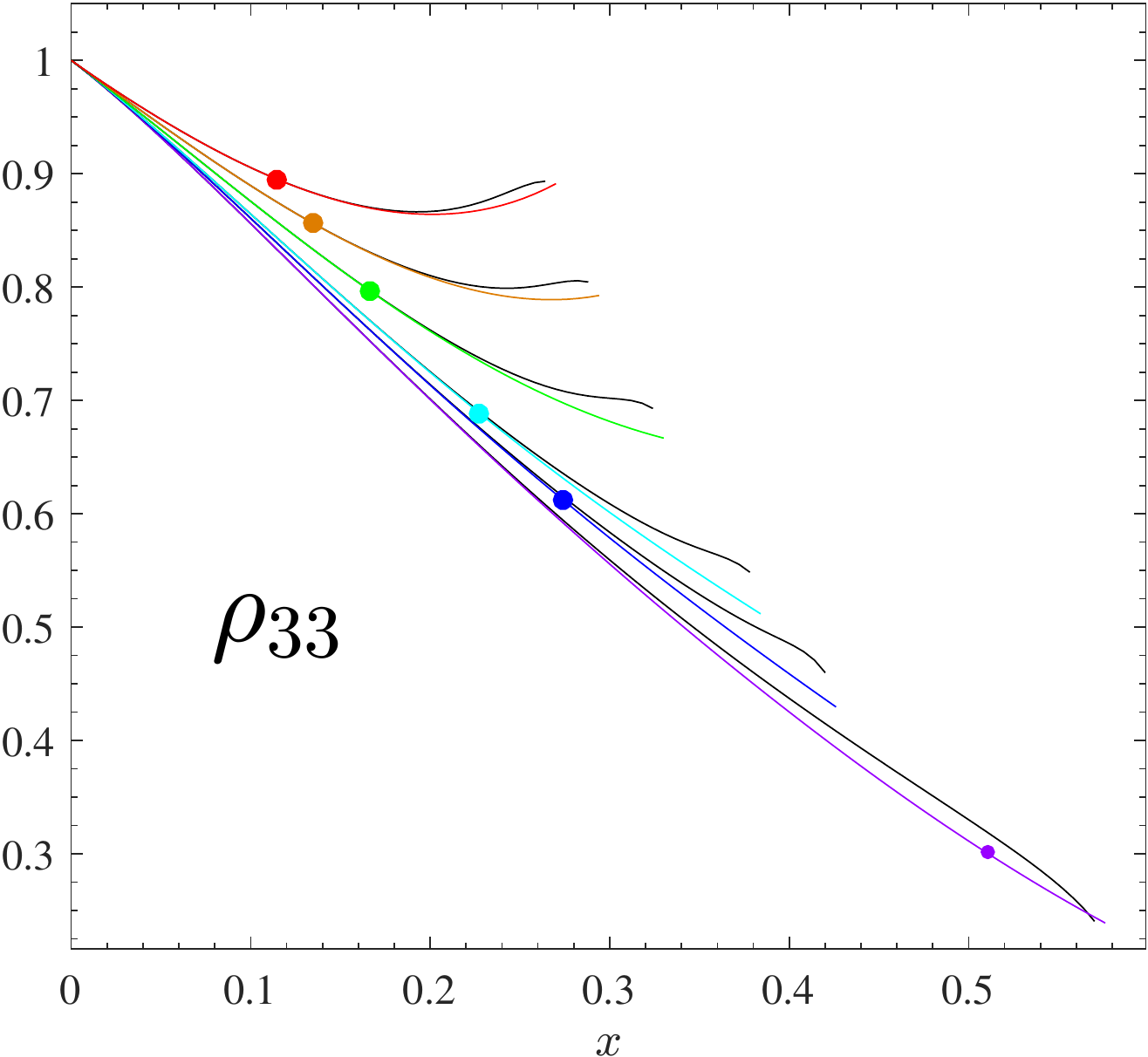}
  \includegraphics[width=0.21\textwidth]{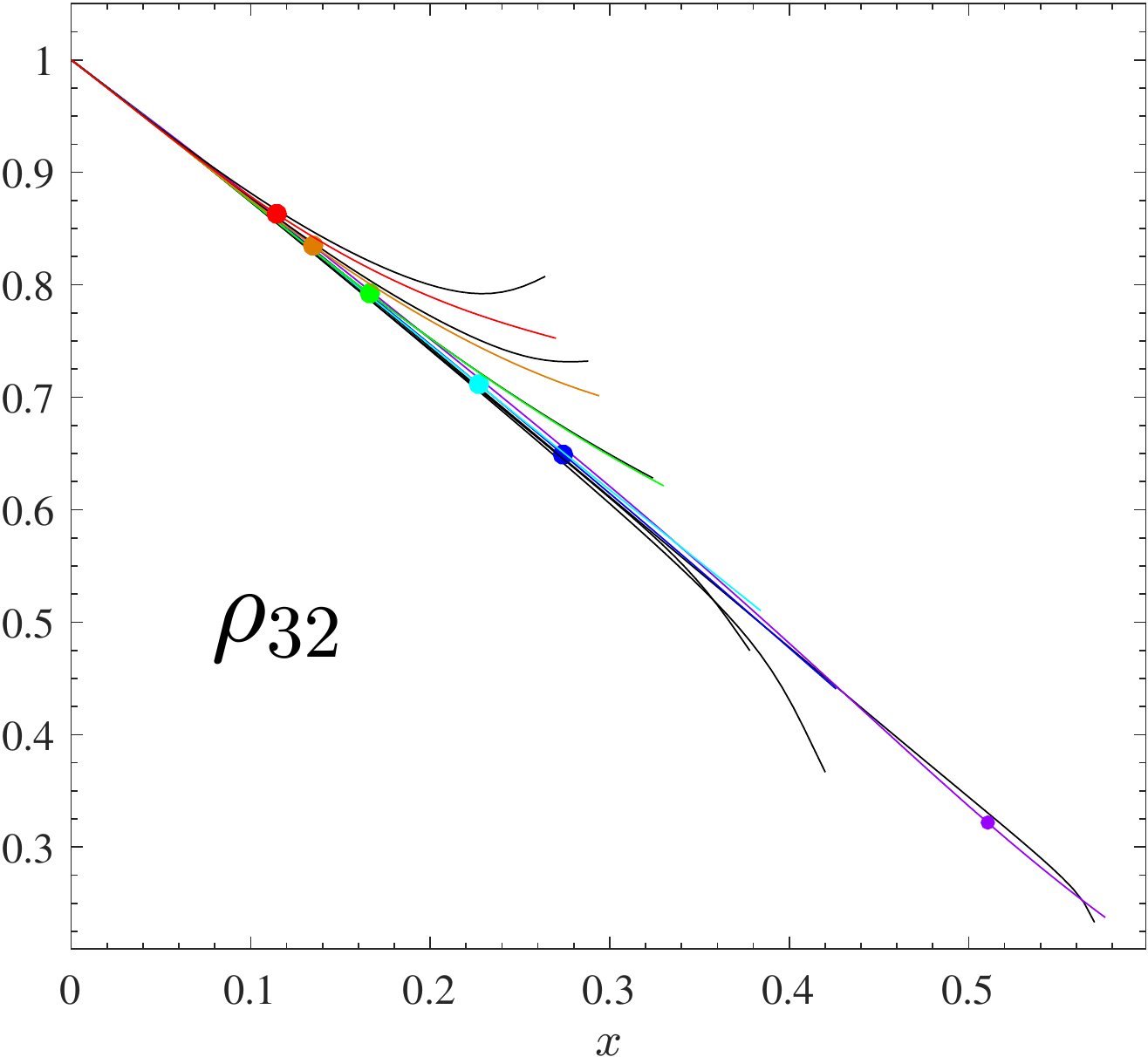}\\
  \includegraphics[width=0.21\textwidth]{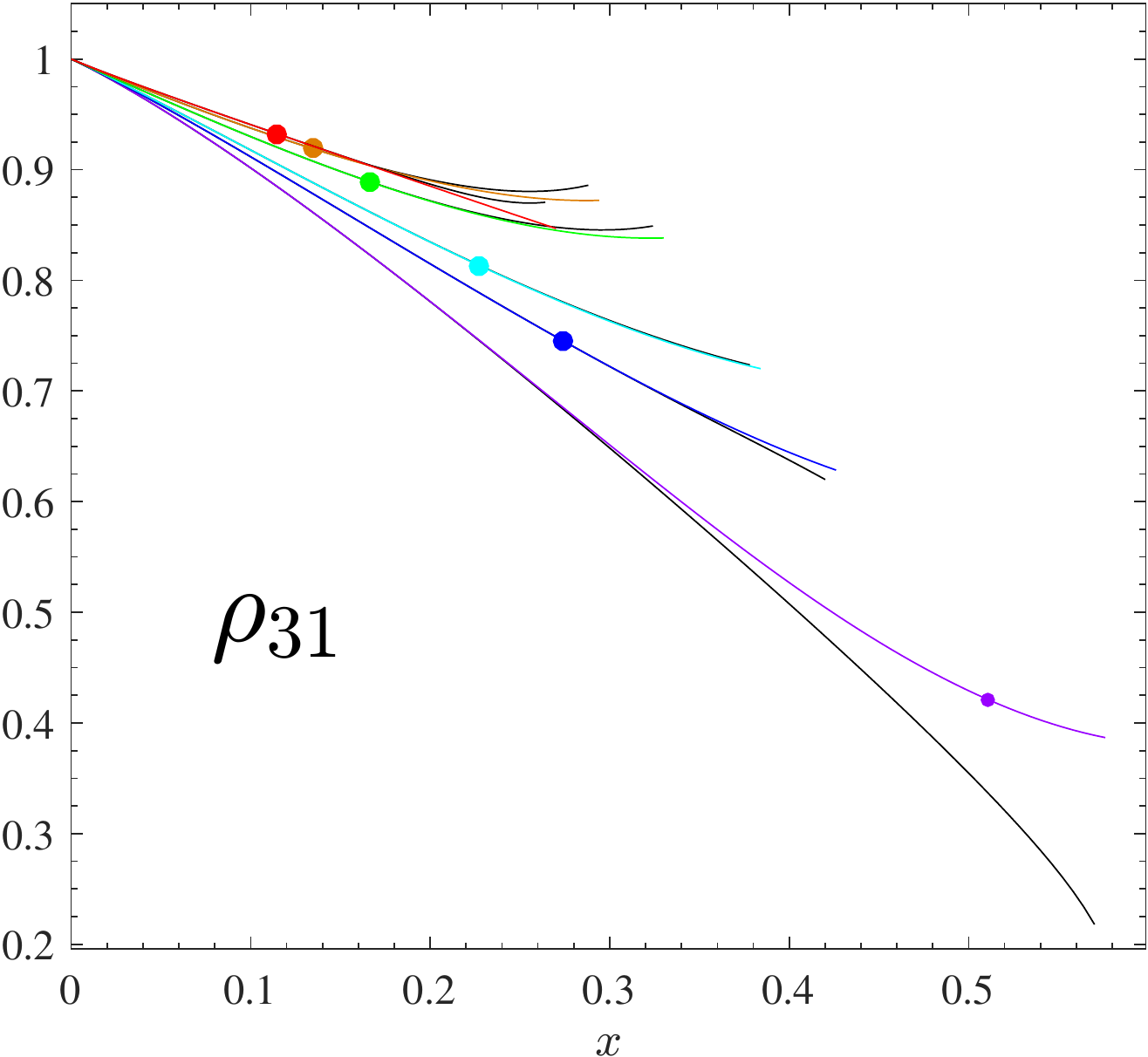}
  \includegraphics[width=0.21\textwidth]{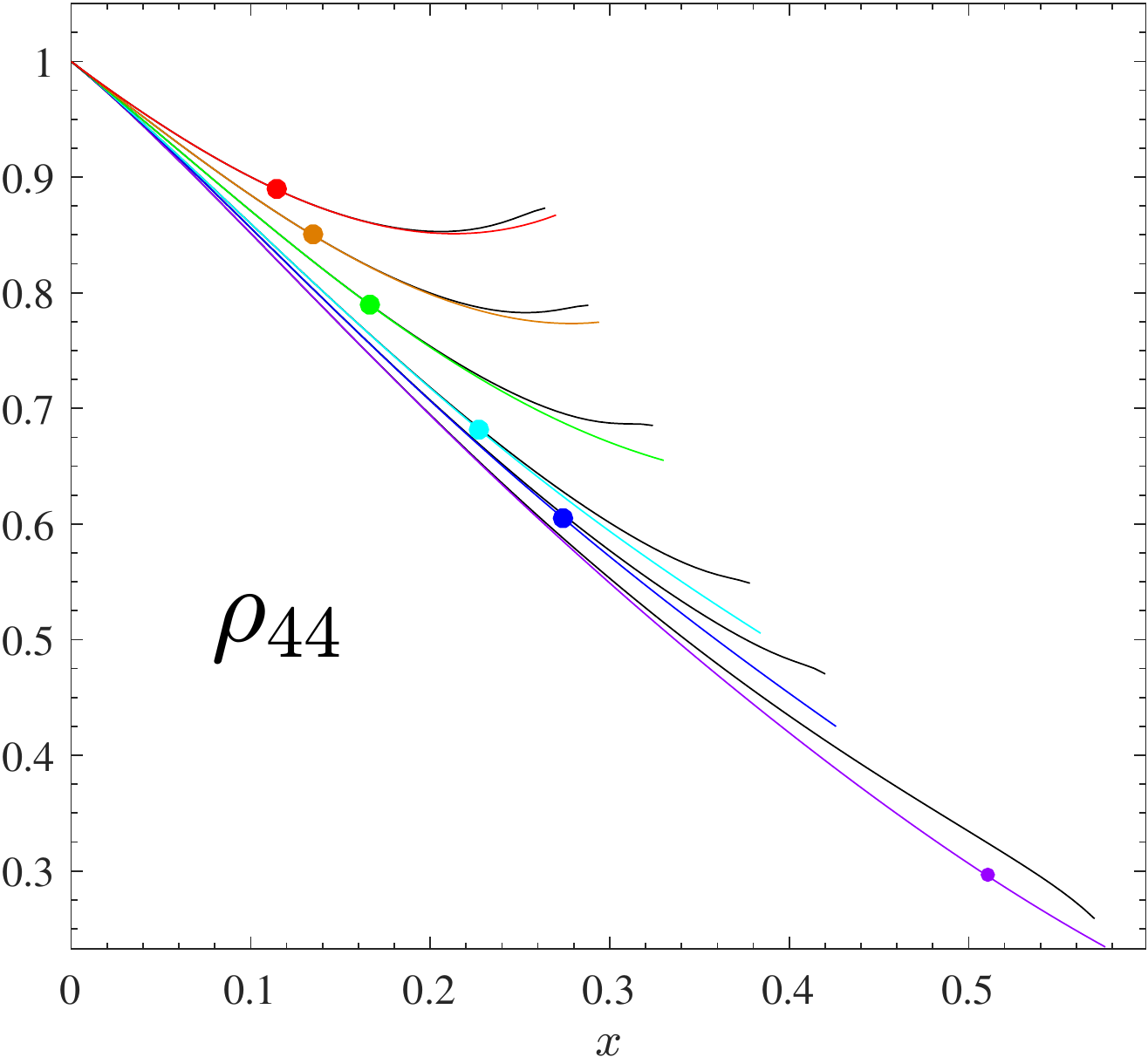}
  \includegraphics[width=0.21\textwidth]{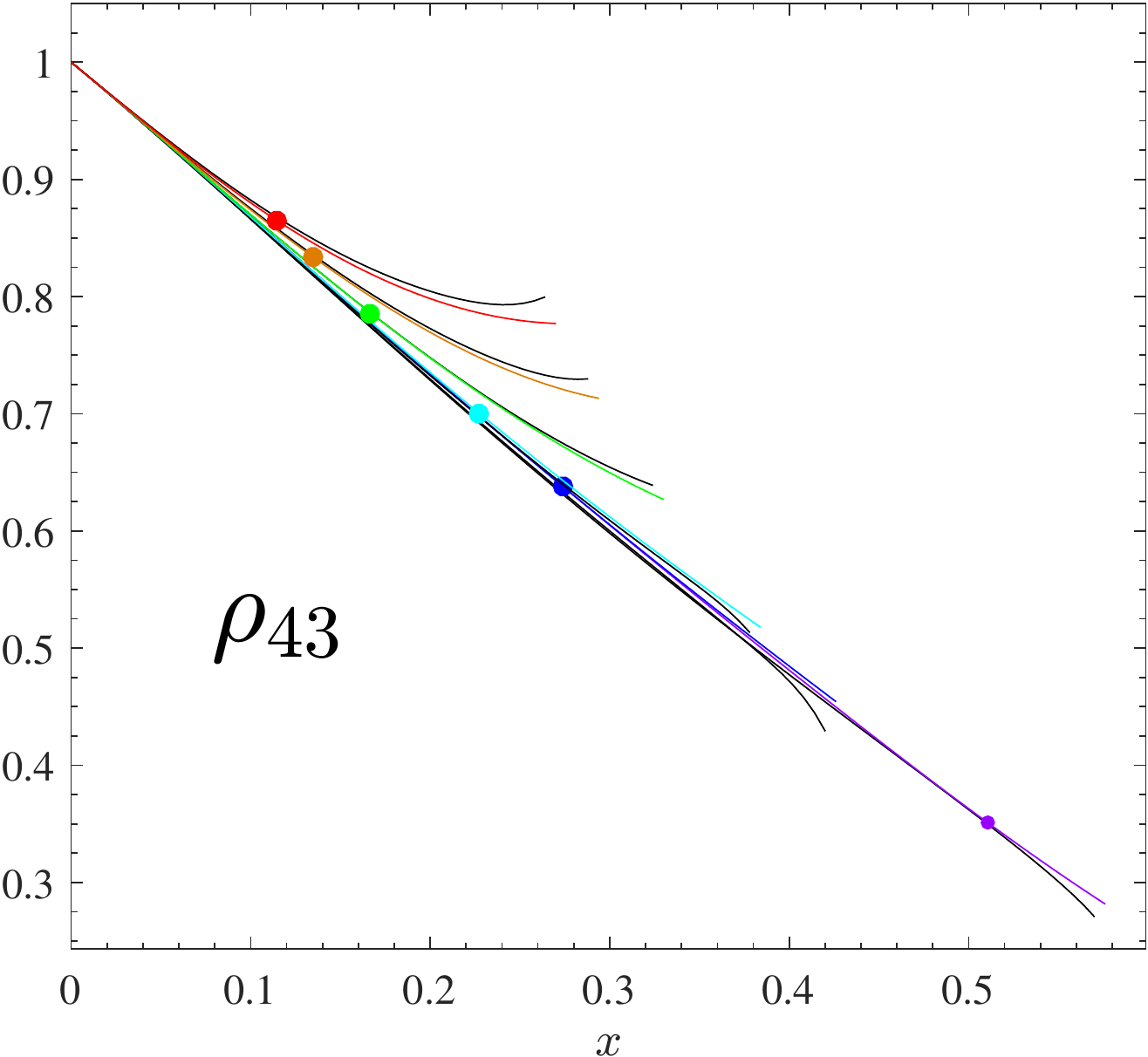}
  \includegraphics[width=0.21\textwidth]{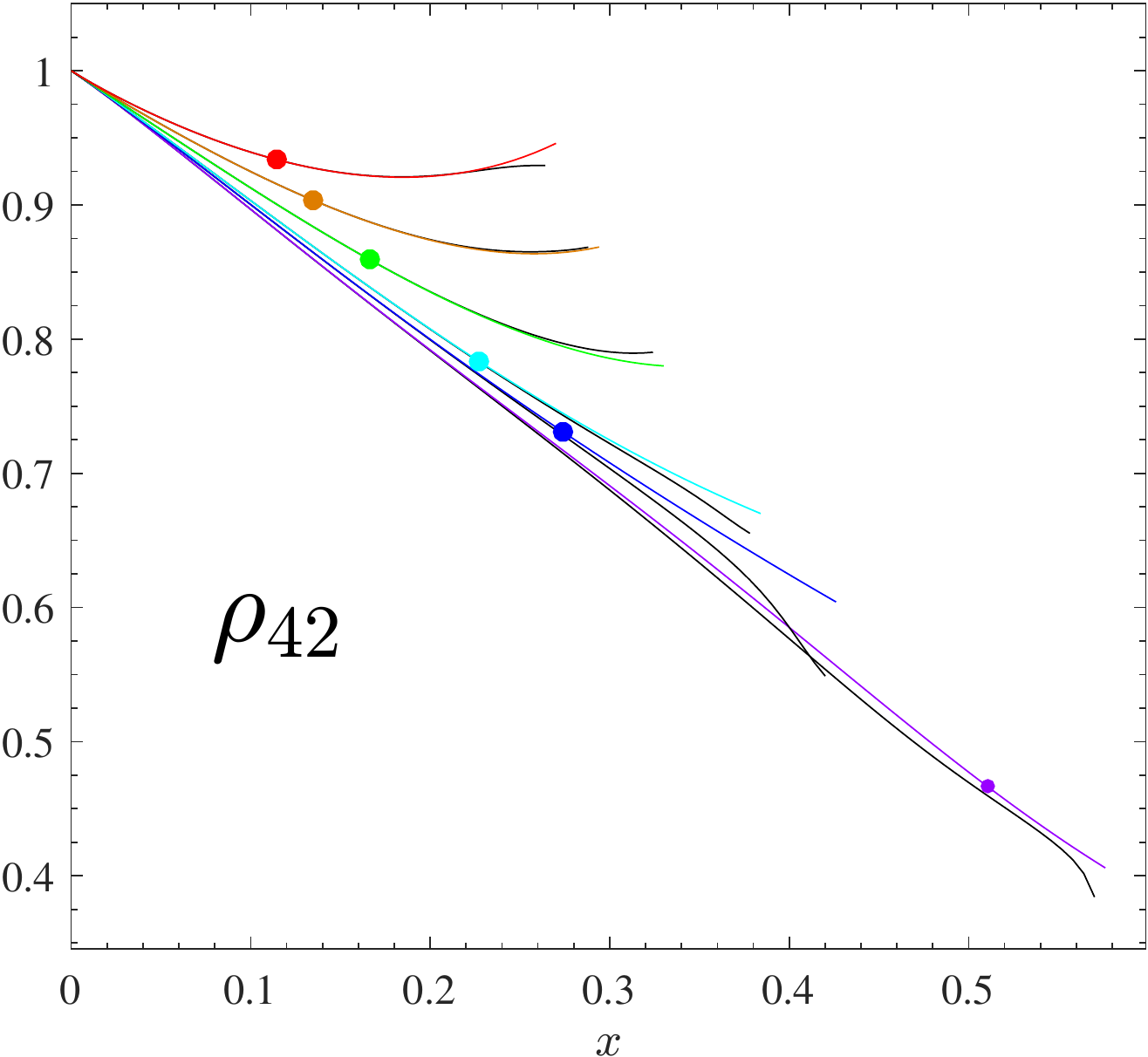}\\
  \includegraphics[width=0.21\textwidth]{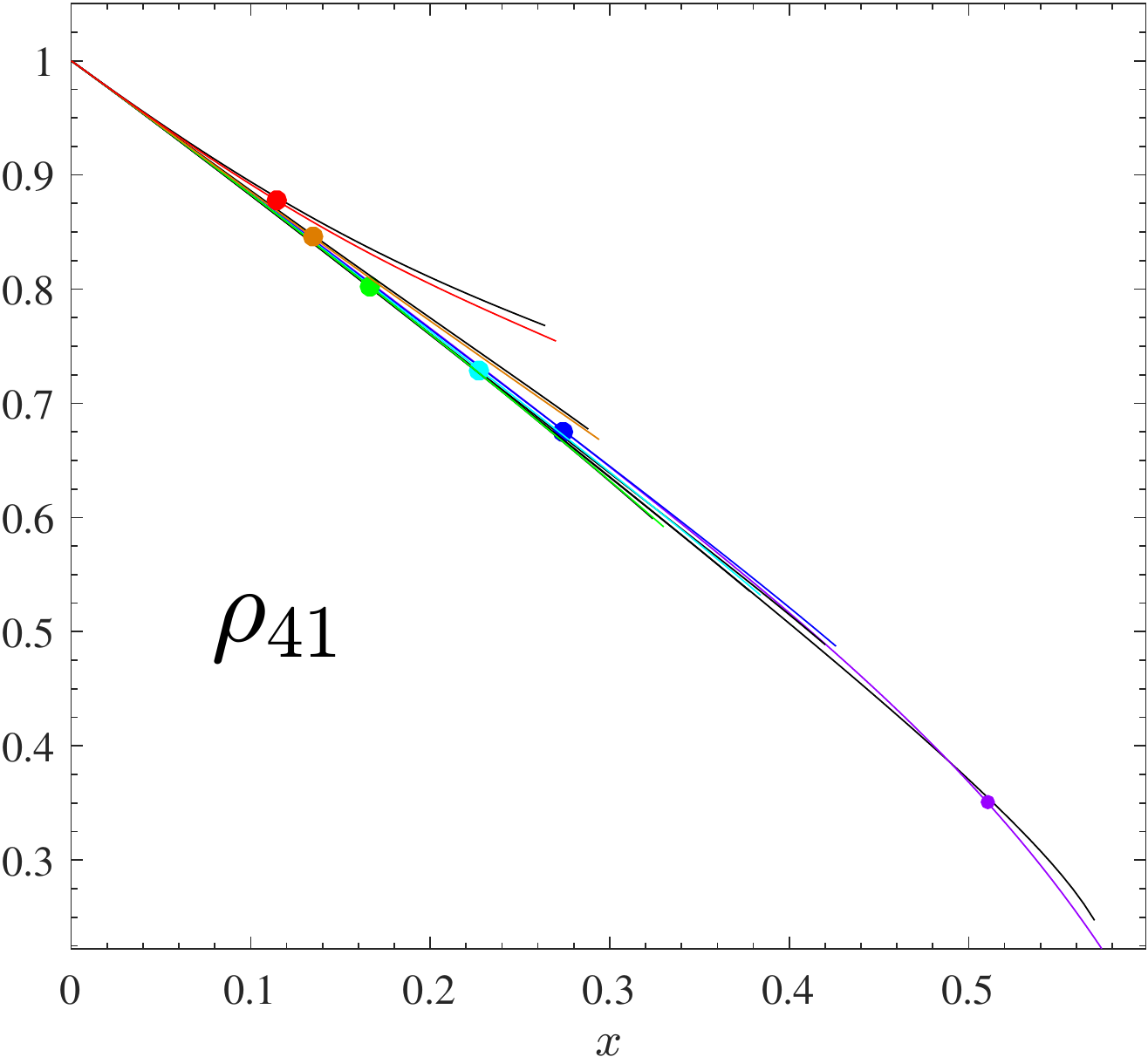}
  \includegraphics[width=0.21\textwidth]{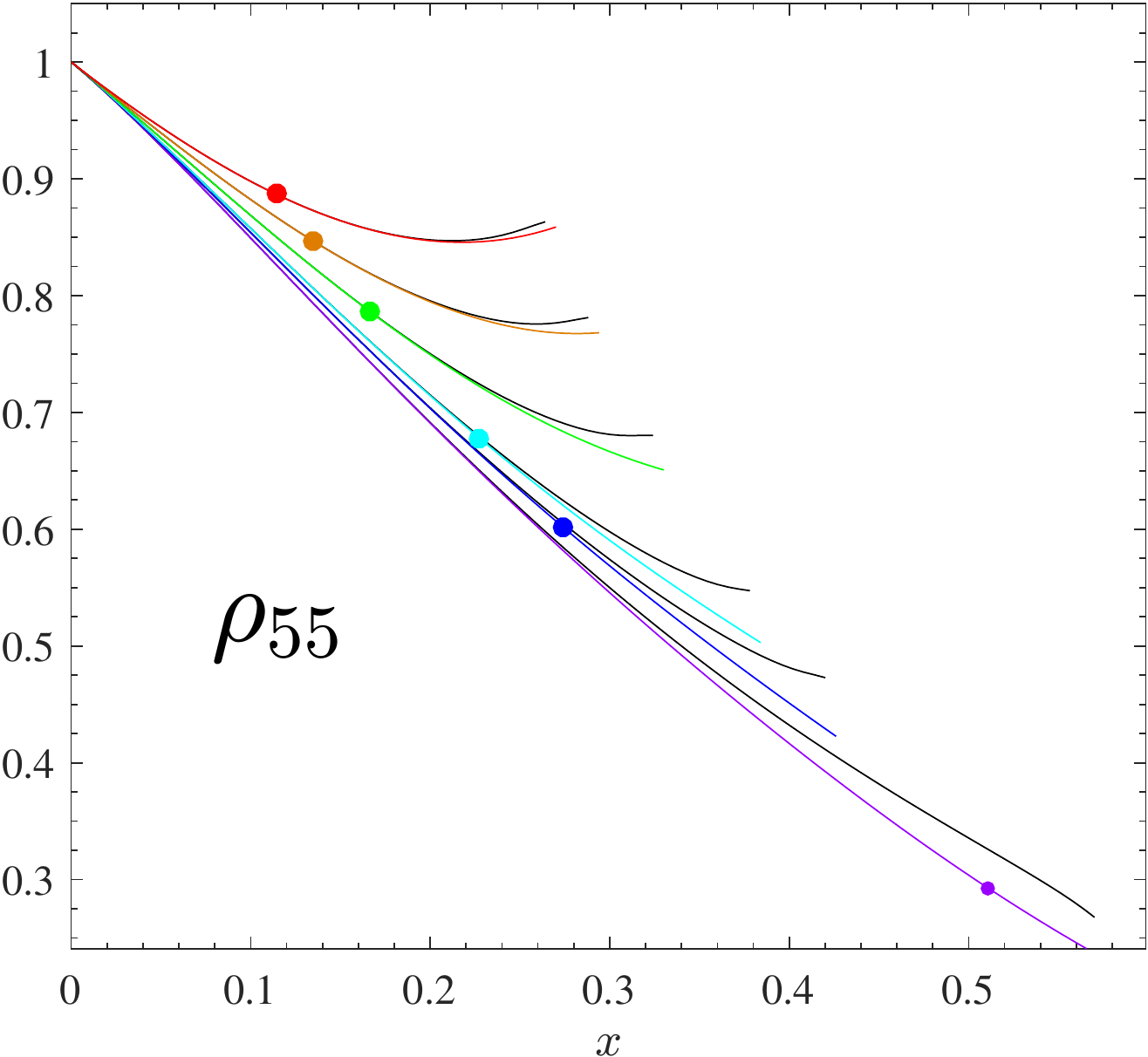}
  \includegraphics[width=0.21\textwidth]{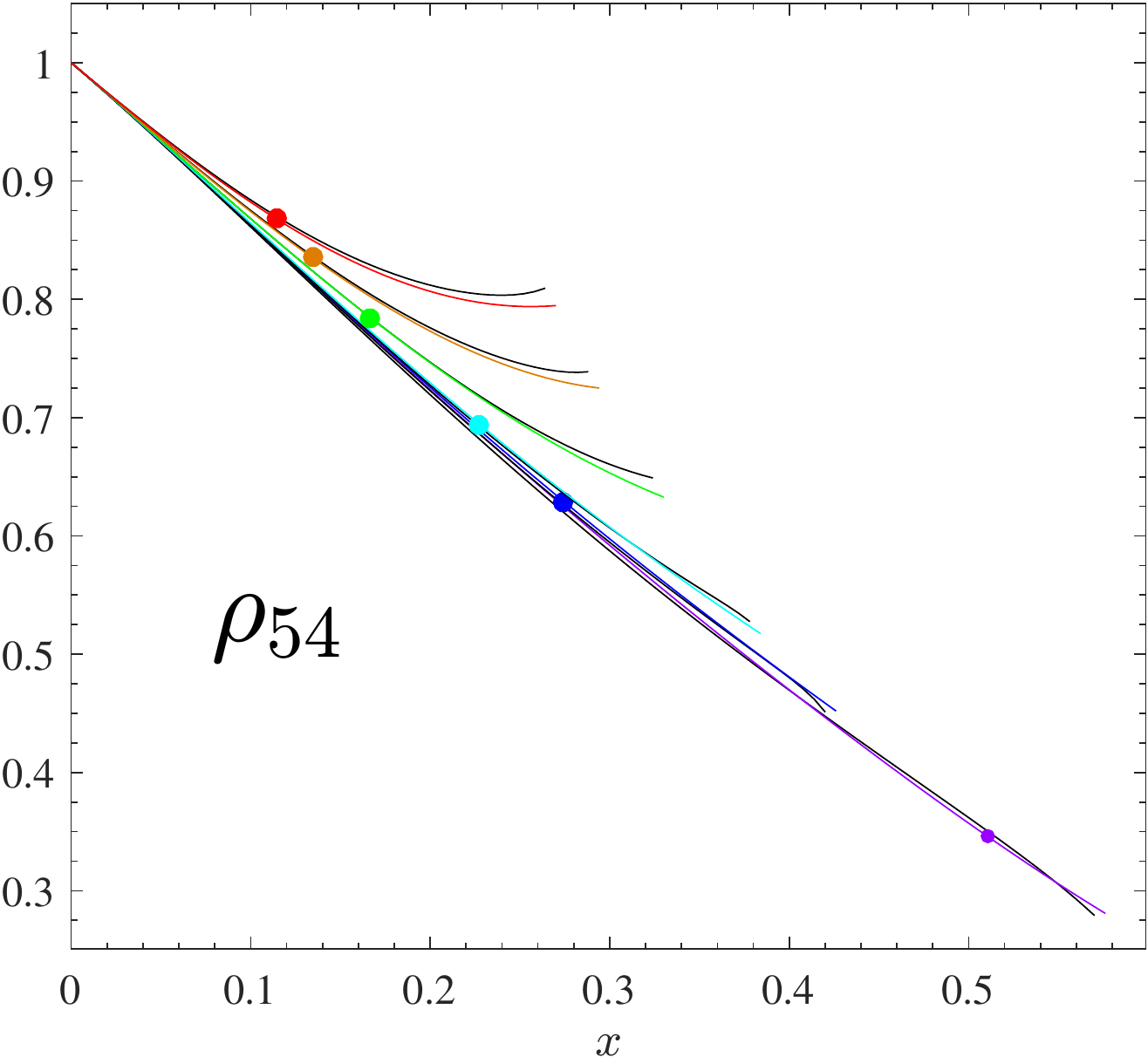}
  \includegraphics[width=0.21\textwidth]{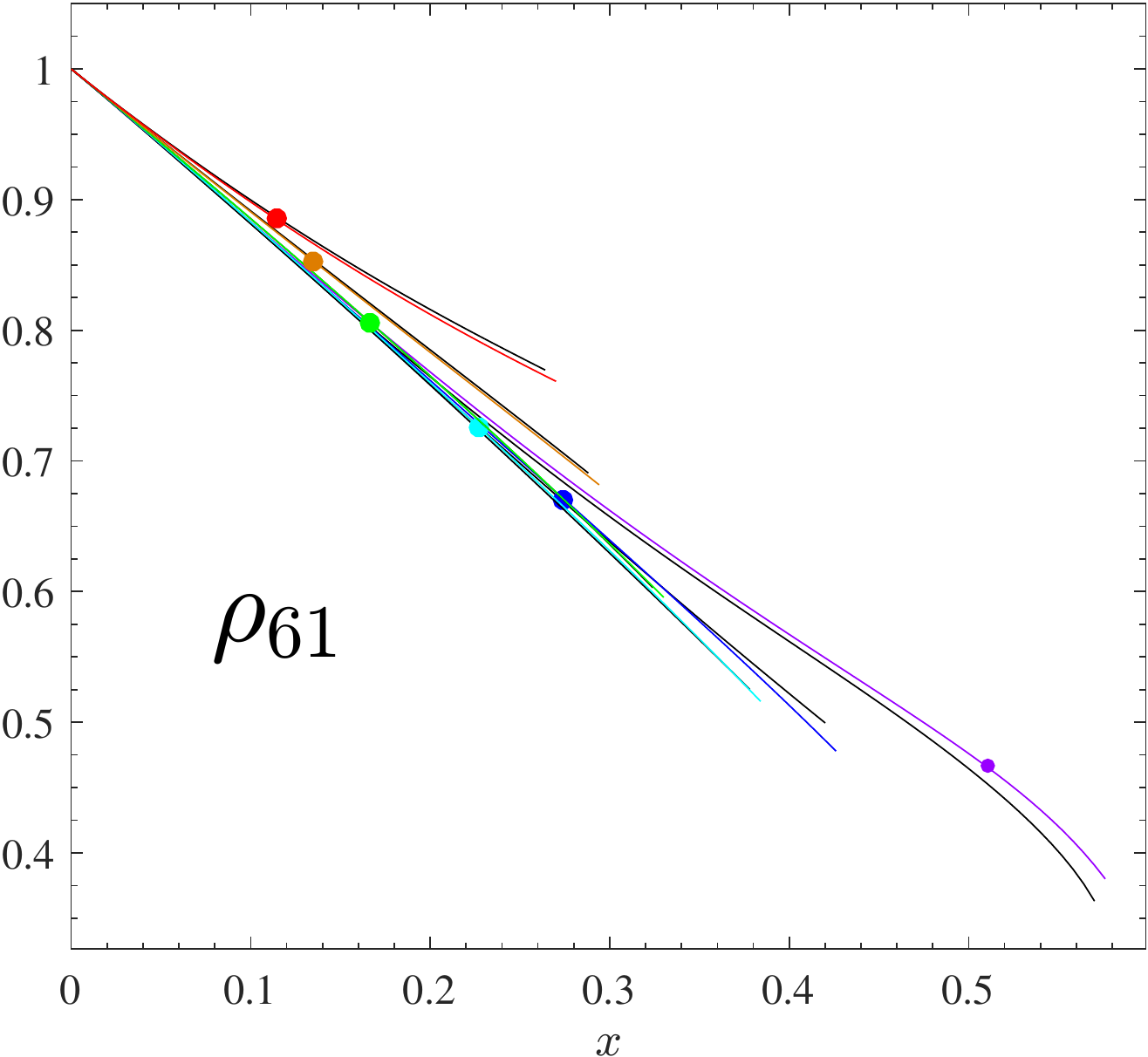} \\
  \includegraphics[width=0.21\textwidth]{fig01.pdf}
  \includegraphics[width=0.21\textwidth]{fig01.pdf}
  \includegraphics[width=0.21\textwidth]{fig01.pdf}
   \includegraphics[width=0.21\textwidth]{fig01.pdf}\\
  \includegraphics[width=0.21\textwidth]{fig01.pdf}
  \includegraphics[width=0.21\textwidth]{fig01.pdf}
  \includegraphics[width=0.21\textwidth]{fig01.pdf}
  \includegraphics[width=0.21\textwidth]{fig01.pdf}
  \caption{\label{fig:rholms}Comparison between the factorized and resummed analytical
    $\rho_{\ell m}$'s (colored online) and the corresponding numerical (exact) functions
    (black online) up to $\ell=6$ for values of the BH dimensionless spin
    $\ha{a}=(-0.99, -0.5, 0, +0.5,+0.99)$ (red, orange, green, cyan, blue and purple
    respectively). The filled circles mark the LSO location. This plot is obtained using relative
    6PN information for all modes except $\rho_{32}$ that employs 5PN relative accuracy
    for $\rho_{32}^{\rm orb}$. The Pad\'e approximants used on $\rho_{\ell m}^{\rm orb}$
    are listed in second column of Table~\ref{tab:rhoLSO}. The same table also lists the
    numerical/analytical relative difference at the LSO. The agreement remains good
    (except for few exceptions, see text for details) also for $\ha=+0.99$.}
\end{figure*}
Reference~\cite{Nagar:2016ayt} proposed then to improve the strong-field behavior of the
$\rho_\lm$'s functions by  (i) writing them as the product of a purely orbital and
purely spin-dependent factors as
\be
\label{eq:rho_factorized}
\rho_{\lm}=\rho_{\lm}^{\rm orb}\hat{\rho}^{\rm S}_\lm,
\ee
where $\hat{\rho}^{\rm S}_{\lm}\equiv T_n[1+\rho_\lm^{\rm S}/\rho^{\rm orb}_\lm]$,
and then resumming each separate factor in a certain way that we detail below
\footnote{To simplify the notation, note that we are using here the same
  symbol $\rho_\lm$, for both the orbital-additive and orbital-factorized amplitudes.
  By contrast, Ref.~\cite{Nagar:2016ayt} was addressing with $\tilde{\rho}_\lm$ the
  orbital-factorized amplitudes.}
Although there is no first-principle reason for treating the orbital and spin
contributions as separate multiplicative factors, such representation proved useful for
interpreting the global behavior of the $\rho_{\ell m}$'s as well as for improving
it near (or even below) the LSO. For instance, it was argued that a sort of
compensation between the spin and orbital factors should occurr in order to
guarantee a good agreement between the numerical and analytical functions
close to the LSO, especially for large and positive values of the black hole spin.
To accomplish such effect, it is necessary to {\it resum} each factor (or at least 
the spin-dependent one), that is given by a truncated Taylor series, in a specific way.
In particular, it was suggested~\cite{Nagar:2016ayt} that a simple and efficient 
method to temperate the divergent behavior of $\hat{\rho}^{\rm S}_\lm$ towards the LSO
is to take its inverse Taylor series (or inverse resummed representation, ``iResum'')
defined as
\be
\label{eq:iRrhoS} 
\bar{\rho}_{\ell m}^{\rm S}=\left(T_n\left[\left(\hat{\rho}^{\rm S}_{\ell m}(x)\right)^{-1}\right]\right)^{-1}.
\ee
Reference~\cite{Nagar:2016ayt} illustrated that, due to the large amount of PN 
information available, it is possible to achieve satisfactory numerical/analytical 
agreement using different truncated PN series as a starting point, though lower-PN 
orders (e.g. 6PN) are preferable with respect to high-PN orders 
(e.g. 10PN or 20PN)\footnote{It has to be stressed that the impact
  of high-PN information, i.e. larger than $10$~PN, has not been assessed throughly yet,
  except for preliminary investigations reported in Ref.~\cite{Nagar:2016ayt}. We are
  not going to do this in the current work, but we postpone it to future studies.}.
The analysis~\cite{Nagar:2016ayt} also showed that, once that the factorization
and resummation paradigm is assumed, one is free to choose at what PN order to work, 
provided the resummed amplitude shows a good agreement with the numerical curves.
For consistency with previous, EOB-related, works~\cite{Damour:2014sva},
in~\cite{Nagar:2016ayt} it was chosen to keep the orbital part at 5PN order,
and in Taylor-expanded form, together with the spin-dependent factor truncated 
at 3.5PN. This choice was made so to be consistent with the spin-dependent
information used in the comparable-mass case. For the $\ell=m=2$ multipole, 
this yielded rather acceptable analytical/numerical agreement ($\simeq 1\%$)
up to the LSO for all spin values between $-0.99$
and $+0.99$ (see Fig.~4 of~\cite{Nagar:2016ayt}).
\begin{figure}[t]
  \center
 \includegraphics[width=0.45\textwidth]{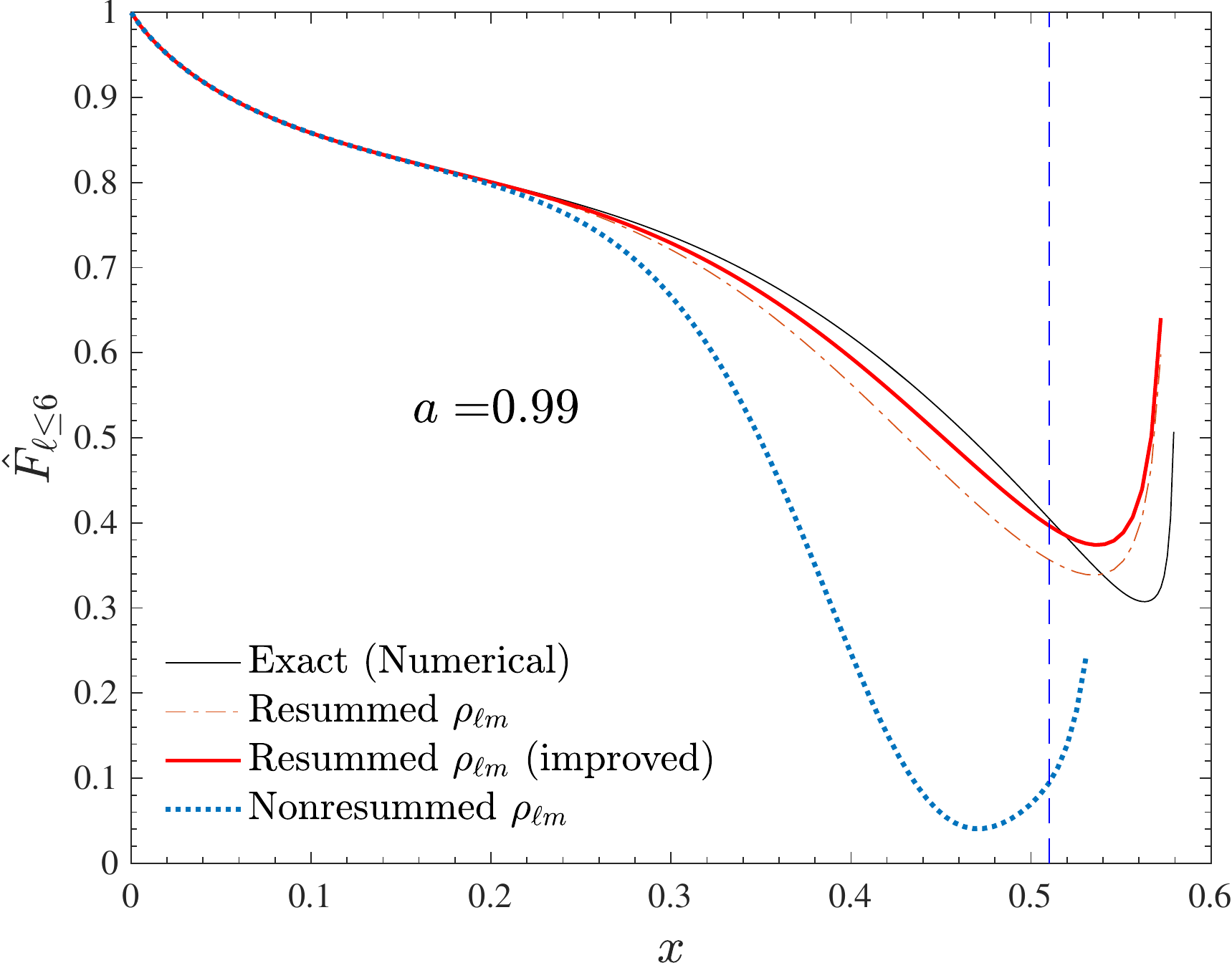}
 \caption{\label{fig:flux}Test-particle limit: comparison between analytical and exact
   (numerical) fluxes for dimensionless black-hole spin $\hat{a}=+0.99$. The functions
   $\rho_{33}^{\rm orb}$, $\rho_{44}^{\rm orb}$ and $\rho_{55}^{\rm orb}$ are either resummed
   using (4,2) Pad\'e approximants (orange, dot-dashed line) or kept in PN-expanded form
   up to (relative) 6PN order. This second choice improves the agreement with the numerical
   curve. The dotted line represents the analytical flux where the total
   6PN-accurate $\rho_{\ell m}$'s are kept in the standard, nonfactorized,
   Taylor-expanded form.}
\end{figure}
\begin{table*}[t]
  \caption{\label{tab:rhoLSO}Fractional differences between the resummed and the numerical $\rho_\ell$'s at the LSO.
    From left to right the columns report: the multipolar order; the Pad\'e approximant chosen for the orbital factor;
    the PN order chosen for the spin-dependent factors; the fractional difference
    $(\rho^{\rm num}-\rho^{\rm anlyt})/\rho^{\rm num}$ at $x_{\rm LSO}$.}
\begin{center}
\begin{ruledtabular}  
\begin{tabular}{lcccccccc}
\rule[0,6cm]{0cm}{0cm}
$(\ell,m)$ & $P^i_j(\rho^{orb}_{\ell m})$ & $iR$(PN) & \multicolumn{6}{c}{$\Delta(x)\rho(x)/\rho\vert_{x=x_{\rm LSO}}$ for $-0.99 \le \hat{a} \le +0.99$}\\
\hfill & \hfill &\hfill & $-0.99$ & $-0.5$ & $ 0 $ & $+0.5$ & $+0.7$ & $+0.99$\\
\hline
\hline
$(2,2)$ & $P^4_2$ & $6$ & $-1 \times 10^{-4}$ & $-3 \times 10^{-4}$ & $-6 \times 10^{-4}$ & $-0.002$ & $-0.004$ & $0.026$\\
$(2,1)$ & $P^5_1$ & $6$ & $-0.006$ & $-0.004$ & $2 \times 10^{-5}$ & $0.010$ & $0.019$ & $0.159$\\
$(3,3)$ & $P^4_2$ & $6$ & $-4 \times 10^{-5}$ & $-1 \times 10^{-4}$ & $-4 \times 10^{-4}$ & $-0.002$ & $-0.005$ & $-0.058$\\
{\bf (3,3)} & $P^6_0$ & $6$ & $-2 \times 10^{-5}$ & $-6 \times 10^{-5}$ & $-2 \times 10^{-4}$ & $-0.001$ & $-0.002$ & $-0.023$\\
$(3,2)$ & $P^4_2$ & $6$ & $-0.004$ & $-0.003$ & $-5 \times 10^{-5}$ & $0.006$ & $0.012$ & $-0.026$\\
$(3,1)$ & $P^3_2$ & $6$ & $-3 \times 10^{-5}$ & $-6 \times 10^{-5}$ & $-1 \times 10^{-4}$ & $-3 \times 10^{-4}$ & $-8 \times 10^{-5}$ & $0.248$\\
{\bf (3,1)} & $P^3_1$ & $8$ & $-2 \times 10^{-5}$ & $-4 \times 10^{-5}$ & $-1 \times 10^{-4}$ & $-8 \times 10^{-4}$ & $-0.002$ & $-0.0017$\\
$(4,4)$ & $P^4_2$ & $6$ & $-2 \times 10^{-5}$ & $-8 \times 10^{-5}$ & $-3 \times 10^{-4}$ & $-0.002$ & $-0.005$ & $-0.088$\\
{\bf (4,4)} & $P^6_0$ & $6$ & $-4 \times 10^{-6}$ & $-3 \times 10^{-5}$ & $-2 \times 10^{-4}$ & $-0.001$ & $-0.002$ & $-0.046$\\
$(4,3)$ & $P^4_2$ & $6$ & $-0.003$ & $-0.002$ & $-1 \times 10^{-4}$ & $0.004$ & $0.008$ & $0.004$\\
$(4,2)$ & $P^6_0$ & $6$ & $-1 \times 10^{-5}$ & $-2 \times 10^{-5}$ & $-5 \times 10^{-5}$ & $6 \times 10^{-4}$ & $0.003$ & $0.015$\\
$(4,1)$ & $P^4_2$ & $6$ & $-0.003$ & $-0.002$ & $8 \times 10^{-6}$ & $0.005$ & $0.008$ & $-0.013$\\
(5,5) & $P^4_2$ & $6$ & $-2 \times 10^{-5}$ & $-7 \times 10^{-5}$ & $-3 \times 10^{-4}$ & $-0.002$ & $-0.037$ & $-0.101$\\ 
{\bf (5,5)} & $P^6_0$ & $6$ & $2 \times 10^{-6}$ & $-2 \times 10^{-5}$ & $-1 \times 10^{-4}$ & $-0.001$ & $-0.034$ & $-0.054$\\ 
$(5,4)$ & $P^4_2$ & $6$ & $-0.003$ & $-0.002$ & $-2 \times 10^{-4}$ & $0.003$ & $0.005$ & $-0.013$\\ 
$(5,3)$ & $P^4_2$ & $6$ & $-2 \times 10^{-5}$ & $-6 \times 10^{-5}$ & $-2 \times 10^{-4}$ & $-6 \times 10^{-4}$ & $-7 \times 10^{-4}$ & $-0.043$\\
$(5,2)$ & $P^4_2$ & $6$ & $-0.002$ & $-0.002$ & $8 \times 10^{-6}$ & $0.004$ & $0.009$ & $0.077$\\
$(5,1)$ & $P^6_0$ & $6$ & $1 \times 10^{-5}$ & $1 \times 10^{-5}$ & $3 \times 10^{-5}$ & $1 \times 10^{-4}$ & $5 \times 10^{-4}$ & $0.101$\\
$(6,6)$ & $P^4_2$ & $5$ & $-4 \times 10^{-5}$ & $-1 \times 10^{-4}$ & $-3 \times 10^{-4}$ & $-5 \times 10^{-5}$ & $0.002$ & $0.064$\\
$(6,5)$ & $P^4_2$ & $6$ & $-0.002$ & $-0.001$ & $-2 \times 10^{-4}$ & $0.002$ & $0.003$ & $-0.029$\\
$(6,4)$ & $P^4_2$ & $6$ & $-2 \times 10^{-5}$ & $-6 \times 10^{-5}$ & $-2 \times 10^{-4}$ & $-8 \times 10^{-4}$ & $-0.001$ & $-0.035$\\
$(6,3)$ & $P^4_2$ & $6$ & $-0.002$ & $-0.001$ & $-1 \times 10^{-5}$ & $0.003$ & $0.007$ & $0.053$\\
$(6,2)$ & $P^4_2$ & $6$ & $-4 \times 10^{-6}$ & $-2 \times 10^{-5}$ & $-5 \times 10^{-5}$ & $2 \times 10^{-5}$ & $0.001$ & $0.213$\\
{\bf (6,2)} & $P^6_2$ & $8$ & $-6 \times 10^{-7}$ & $-8 \times 10^{-7}$ & $-7 \times 10^{-7}$ & $8 \times 10^{-5}$ & $9 \times 10^{-4}$ & $0.015$\\
$(6,1)$ & $P^6_0$ & $6$ & $-0.002$ & $-0.001$ & $7 \times 10^{-6}$ & $0.003$ & $0.003$ & $0.028$\\ 
\end{tabular}
\end{ruledtabular}
\end{center}
\label{tab:fitting}
\end{table*}
Here we relax the constraint of being consistent with previous EOB-related
works and present, instead, a new recipe to further improve $\ell=m=2$ results of 
Ref.~\cite{Nagar:2016ayt} and extend them to higher multipolar modes. 
To do so, we: (i) generally increase the PN order, possibly requiring
it to be the same for both the spin and orbital factors; 
(ii) resum the orbital factor using some Pad\'e approximant, to be chosen
according to the PN order and the multipole; (iii) resum the spin factor 
taking its inverse Taylor approximant (iResum) as proposed 
in~\cite{Nagar:2016ayt},  see Eq.~\eqref{eq:iRrhoS} above.
We find that, modulo a few exceptions to be detailed below, 
a good compromise is reached by working at {\it relative}
6PN order for each mode\footnote{This means that the functions
  $\hat{h}^{(\epsilon)}_\lm$ in Eq.~\eqref{eq:hlm0} are taken at 6PN,
  i.e. as 6th-order polynomials in $x$. This implies that the
  {\it global} PN-accuracy we retain is actually higher than $6PN$,
  because of the presence of the Newtonian prefactors $h_{\ell m}^{(N,\epsilon)}$.}
and taking a Pade $(4,2)$ approximant for the
orbital factor\footnote{A priori one would like to use diagonal Pad\'e approximants,
  since they are known to be the most reliable ones. However, we found that spurious
  poles are always present in this case. This fact prevents us from making this choice
  to preserve the simplicity of the approach.}.
There are exceptions to this choice (see 2nd column of Table~\ref{tab:rhoLSO}).
For example, the (2,1) mode is better represented using a $P^5_1$
approximant, the (3,1) using a $P^3_2$ (i.e. keeping $\rho_{32}^{\rm orb}$ at 5PN accuracy),
while for $(4,2)$, $(5,1)$ and $(6,1)$ the orbital factor in Taylor-expanded 
form is preferable. These choices are made so that the analytical $\rho_{\lm}$'s
remain as close as possible to the numerical one up to the LSO (and possibly beyond).
This is  illustrated in Fig.~\ref{fig:rholms}, which displays all $\rho_{\ell m}$'s
functions up $\ell=6$. The figure collects five values of the dimensionless
black-hole spin, $\ha=(-0.99,-0.5,0,+0.5,+0.99)$. The analytical functions are
depicted as colored curves, while the numerical data are black. Both curves extend
up to the light-ring, while the filled circle mark the LSO location.
The $\ha=+0.99$ curves extend up to the highest-frequency (purple, lowest curve)
while the $\ha=-0.99$ is at the top of each panel and is depicted red.
The information encoded in the figure is complemented 
by Table~\ref{tab:rhoLSO}, that lists, for each multipole, the Pad\'e approximant
adopted together with the numerical/analytical fractional difference computed
at the LSO. The figure also highlights that the numerical/analytical 
agreement looks improvable (for large values of $\ha$) for some subdominant
modes, especially $\rho_{31}$ and $\rho_{62}$, where the analytical functions
are systematically above the 
numerical ones towards the boundary of the $x$-domain considered. 
The reason behind this behavior is that both $P^3_2(\rho^{\rm orb}_{32})$ 
and $P^{4}_2(\rho^{\rm orb}_{62})$ develop a spurious pole on the 
real $x$-axis, at $x\approx 0.86$ for the former and at $x\approx 0.82$
for the latter. In this respect, we stress that our choices about the PN
truncation order and the consequent 
resummation strategies should be seen as a compromise between simplicity 
(i.e. using relatively low-order PN-information)  and achievable accuracy 
(i.e. good global agreement with the numerical functions).
For example, one finds that the numerical/analytical disagreement for $\rho_{31}$
at the LSO for $\hat{a}=+0.99$ can be reduced to just $-0.17\%$ by: (i) taking 
$\rho_{31}^{\rm orb}$ at only 4PN order and resumming it with a $P^3_1$ approximant, 
while (ii) $\hat{\rho}^{\rm S}_{31}$ is taken up to 8PN order and then resummed as usual
with its inverse Taylor series. Similarly, one also easily finds that the global behavior
of the $(3,3)$, $(4,4)$ and $(5,5)$ modes can be improved by just keeping the orbital factor 
in Taylor-expanded form instead of replacing it with its $(4,2)$ Pad\'e approximant.
To figure out the relevance of any of these improvements, it is convenient to
inspect the total energy flux reconstructed using the resummed $\rho_{\ell m}$'s.
At a practical level, some analytical/numerical differences that look large on the 
$\rho_{\ell m}$'s are subdominant within the flux and can be practically ignored.
Figure~\ref{fig:flux} compares Newton-normalized energy fluxes, with all multipoles
summed together up to $\ell=6$ included, as follows: (i) the exact (numerical) flux;
(ii) the analytical flux  that is obtained from the $\rho_{\ell m}$'s shown in
Fig.~\ref{fig:rholms}, where the choices for the Pad\'e of the orbital part are
listed as non-bold face in Table~\ref{tab:rhoLSO} (dot-dashed, orange line);
(iii) the analytical flux obtained by taking the $\rho_{33}^{\rm orb}$,
$\rho_{44}^{\rm orb}$ and $\rho_{55}^{\rm orb}$ as plain 6PN-accurate Taylor expansions;
(iv) the analytical flux where the 6PN-accurate $\rho_{\ell m}$ are neither further
factorized nor resummed, following the original paradigm of Refs.~\cite{Damour:2008gu,Pan:2010hz}.
The vertical line marks the LSO location.
The figure illustrates how changing the treatment of the orbital part of
the subdominant  modes mentioned above allows one to reduce the fractional 
difference around the LSO from $10\%$ to approximately $5\%$. It is also to
be noticed the good qualitative behavior of the flux also below the LSO, close
to the light ring where the flux diverges.
By contrast, the flux obtained using the standard, nonresummed, $\rho_{\ell m}$
amplitudes in the form of~\cite{Damour:2008gu,Pan:2010hz}, though pushed to higher
PN order as discussed above, is reliable only up to $x\approx 0.2$.
We also mention that, even though the choice of $P^3_1(\rho^{\rm orb}_{31})\bar{\rho}_{31}^{\rm S}$
with $\hat{\rho}^{\rm S}_{31}$ at 8PN order can strongly reduce the numerical/analytical
differences displayed in Fig.~\ref{fig:rholms}, in practice this does not have
any notable consequence on the total flux. The same statement also holds 
for the $(6,2)$ mode: the near-LSO behavior of the analytical $\rho_{62}$ 
can be improved by working at 8PN, both in the spin and orbital factors 
(with a $P^6_2$  approximant for this latter), without however producing 
any important impact on the total flux computation.  

Let us finally mention in passing that another way to improve the strong-field
behavior of the $\rho_\lm$'s (and thus of the flux) is by including some effective
high-PN order parameter that can be {\it informed} (i.e., calibrated or even fitted) 
to the numerical data. This approach might be necessary, for example, 
when dealing with precision  calculations that require an accurate representation 
of the radiation reaction in the near-LSO regime, e.g., estimate of the final recoil 
velocity when the central black hole is quasi-extremal with the spin aligned with 
the orbital angular momentum~\cite{Nagar:2014kha}.  
As an exploratory investigation dealing with just $\rho_{22}$, we found  that it is 
sufficient to introduce a 6.5PN (effective) parameter at the denominator of 
$\bar{\rho}^{\rm S}_{22}$ and tune it to reduce by more than an order of 
magnitude the fractional difference between the analytical and numerical 
functions up to the LSO. More precisely, we have that $\bar{\rho}^{\rm S}_{22}$ 
has the structural form $1/(1+\hat{a} x^{3/2} + ......+ \hat{a} c_{13/2}x^{13/2})$, 
where $\hat{a}c_{13/2}x^{13/2}$ is formally the first spin-orbit 
term beyond what we are using in this work. 
One easily checks that the value $c_{13/2}=5.1 $ is sufficient to obtain a 
fractional disagreement of the order $0.15\%$ at the LSO for $\hat{a}=+0.99$. 
This illustrative example suggests that there is a simple, though effective, 
way to incorporate the information encoded in the numerical data within 
the analytical description of the waveform amplitudes. More work will be needed
to put this approach in a more systematic form. In particular one may hope that 
a suitable modification of this method, probably with a few more parameters,
could be used to obtain an accurate, semi-analytic, representation of the 
circularized fluxes also up to the light-ring.

\section{Comparable masses: Post-Newtonian expanded results}
\label{sec:PNhlm}
\subsection{Waveform amplitudes: spin-orbit and quadratic-in-spin terms}
We start by summarizing here new results for the PN-expanded, nonprecessing,
multipolar waveform amplitudes up to: (i) next-to-next-to-leading-order (NNLO)
for the spin-orbit terms; (ii) next-to-leading-order (NLO) for the spin-spin
terms and (iii) for the leading-order (LO) spin-cube terms. These waveform
amplitudes were computed by A.~Boh\'e and S.~Marsat~\cite{marsat:2017}
as part of a project that aims at obtaining the complete waveform at this
PN order (we recall that the corresponding calculation of the PN-expanded
energy flux is complete~\cite{Bohe:2012mr,Marsat:2013caa,Marsat:2014xea,Bohe:2015ana}), and kindly shared with us
before publication. Here we only list the PN-expanded multipolar waveform
amplitudes with their complete, currently known, spin dependence.
For completeness, we also include the known, $\nu$-dependent, orbital 
terms~\cite{Kidder:2007rt}.
To start with, let us set the notations and define our choice of spin variables. 
We denote with $\nu=m_1 m_2/M^2$ the symmetric mass ratio, 
with $M=m_1+m_2$ and we adopt the convention that $m_1\geq m_2$. 
From the conserved norm, dimensionful, spin vectors $({\bf S}_1,{\bf S}_2)$,
PN results are usually expressed in terms of the spin combinations
${\bf S}\equiv {\bf S}_1 + {\bf S}_2$ and ${\bf \Sigma}\equiv M\left({\bf S}_2/m_2-{\bf S}_1/m_1\right)$.
For spin-aligned binaries, where $\bm{\ell}$ indicates the unit vector normal to the 
orbital plane (i.e., the direction of the orbital angular momentum), 
one deals with the projections of the spin-vectors along $\bm{\ell}$, i.e., 
$S_\ell={\bf S}\cdot \bm{\ell}$ and $\Sigma_\ell={\bf \Sigma}\cdot \bm{\ell}$. 
Then, it is common practice to work with dimensionless spin variables 
$\chi_{1,2}\equiv S_{1,2}/(m_{1,2})^{2}$ and in the PN-expansions the 
spin vectors always appear divided by the square of the total mass, 
so that one has
\begin{align}
  \label{eq:Sl}
  \hat{S}_\ell\equiv \dfrac{S_\ell}{M^2}      &= X_1^2 \chi_1 + X_2^2 \chi_2, \\
  \label{eq:Sigmal}
  \hat{\Sigma}_\ell\equiv\dfrac{\Sigma_\ell}{M^2} &= X_2\chi_2 - X_1\chi_1,
\end{align}
where we introduced the usual convenient notation  $X_{i}\equiv m_{i}/M$,
which yields $X_1+X_2=1$, $X_1X_2=\nu$ and, since $X_{1}\geq X_{2}$,
we have $X_{1}=\left(1+\sqrt{1-4\nu}\right)/2$. From the
dimensionless spin variables, the waveform spin-dependence is
sometimes also written via their symmetric and antisymmetric combinations
(see e.g.~\cite{Pan:2010hz,Taracchini:2012ig,Damour:2014sva,Bohe:2016gbl}),
$\chi_S\equiv \left(\chi_1+\chi_2\right)/2$ and $\chi_A\equiv \left(\chi_1-\chi_2\right)/2$.

Here, we express the waveform spin dependence using the Kerr parameters
of the two black holes divided by the total mass of the system, namely
via the variables
\begin{equation}
\label{eq:ai}
\tilde{a}_{i}\equiv \dfrac{a_i}{M}=\dfrac{S_i}{M m_i} = X_{i}\chi_{i}\qquad i=1,2.
\end{equation}
This choice is convenient for two reasons: (i) the analytical expression get
more compact as several factors $\sqrt{1-4\nu}$ are absorbed in the definitions,
and one can more clearly distinguish the sequence of terms that are ``even'',
in the sense that are symmetric under exchange of body 1 with body 2 and
are proportional to the ``total Kerr dimensionless spin''
$\hat{a}_{0}\equiv \tilde{a}_{1}+\tilde{a}_{2}$ from those that are ``odd'',
i.e. change sign under the exchange of body 1 with body 2 and are proportional
to the factor $\sqrt{1-4\nu}(\tilde{a}_{1}-\tilde{a}_{2})$;
(ii) in addition, one can infer the (spinning) test-particle limit
from the general $\nu$-dependent, expressions just by inspecting them visually.
In fact, in this limit, $m_{2}\ll m_{1}$, $\tilde{a}_{12}\rightarrow 0$ and
$\tilde{a}_1$ becomes the dimensionaless spin of the massive black hole
of mass $m_{1}\approx M$, $\tilde{a}_1\rightarrow S_1/(m_1)^2$.
Similarly, the {\it spinning} particle limit around Kerr is simply obtained
by putting $\nu=0$, since $\tilde{a}_{2}$ just reduces to the usual spin-variable
used in PN or numerical
calculations~\cite{Tanaka:1996ht,Harms:2015ixa,Harms:2016ctx,Lukes-Gerakopoulos:2017vkj},
$\sigma \equiv S_{2}/({m_{1} m_{2}})$. To keep the
expressions compact, we also define the following
combinations of the $\tilde{a}_i$ of Eq.~\eqref{eq:ai}
\begin{align}
\ha_{0}&\equiv X_{1}\chi_1+X_{2}\chi_2=\ta_{1}+\ta_{2},\\
\ta_{12}&\equiv \ta_{1}-\ta_{2},\\
X_{12}&\equiv X_1 - X_2=\sqrt{1-4\nu}.
\end{align}
Equations~\eqref{eq:Sl}-\eqref{eq:Sigmal} above then simply read
\begin{align}
  \hat{S}_\ell     &=\dfrac{1}{2}\left(\hat{a}_0+\tilde{a}_{12}X_{12}\right),\\
  \hat{\Sigma}_\ell&=-\tilde{a}_{12}.
\end{align}
We report below the complete modulus of $\hat{h}_{\ell m}$ up to NNLO
in the spin-orbit coupling and up to NLO in the spin-spin coupling.
Note however that for the $m={\rm odd}$ multipoles we defactorized the
factor $X_{12}$ (that is usually seen as part of the Newtonian
prefactor $h_{\ell m}^{(N,\epsilon)}$, see Eq.~\eqref{eq:hlm0})
to avoid the appearence of a fictitious singularity when $\nu=1/4$
in the spin-dependent terms proportional to $\tilde{a}_{12}$
(see also~\cite{Taracchini:2012ig}). To have a consistent notation,
when $m={\rm odd}$ we focus on the quantities
\be
\tilde{h}_{\ell m}^{(\epsilon)}=X_{12}\hat{h}_{\ell m}^{(\epsilon)}.
\ee
In conclusion, the modulus of the Newton-normalized PN-expanded, 
multipolar, waveform we use as starting point reads:
\begin{widetext}
  \begin{align}
    \label{eq:h22}
|\hat{h}_{22}^{(0)}|&=1+\left(-\frac{107}{42}+\frac{55}{42}\nu\right)x-\biggl[-2\pi + \hat{a}_0+\frac{1}{3}\tilde{a}_{12}X_{12}\biggr]x^{3/2}
+\biggl[\hat{a}_0^2-\frac{2173}{1512}-\frac{1069}{216}\nu +\frac{2047}{1512}\nu ^2\biggr]x^2\nonumber\\
&+\biggl[-\pi\left(\dfrac{107}{21} -\dfrac{34}{21}\nu\right)-\hat{a}_0\left(\frac{163}{126}+\frac{46}{63}\nu\right)
-\tilde{a}_{12}X_{12}\left(\frac{157}{126}+\frac{22}{21}\nu\right)\biggr]x^{5/2}\nonumber\\
&+\Biggl\{\frac{27027409}{646800}-\frac{113}{63}\left(\tilde{a}_1^2+\frac{271}{113}\tilde{a}_1\tilde{a}_2+\tilde{a}_2^2\right)-2\pi\left(\hat{a}_0+\frac{1}{3}\tilde{a}_{12}X_{12}\right)+\frac{2}{3}\pi ^2
+\frac{121}{63}\hat{a}_0\tilde{a}_{12}X_{12}-\frac{856}{105}{\rm eulerlog}_2(x)\nonumber\\
&+\biggl[-\frac{278185}{33264}+\frac{20}{21}\left(\tilde{a}_1^2+\frac{24}{5}\tilde{a}_1\tilde{a}_2+\tilde{a}_2^2\right)+\frac{41}{96}\pi^2\biggr]\nu
-\frac{20261}{2772}\nu^2 +\frac{114635}{99792}\nu ^3\Biggr\}x^3\nonumber\\
&+\biggl[\hat{a}_0\left(\frac{1061}{168}+\frac{4043}{168}\nu +\frac{499}{168}\nu ^2\right)+\tilde{a}_{12}X_{12}\left(\frac{241}{216}+\frac{5135}{1512}\nu -\frac{79}{72}\nu ^2\right)\biggr]x^{7/2},
\end{align}
\begin{align}
|\tilde{h}_{21}^{(1)}|&=X_{12}-\frac{3}{2}\tilde{a}_{12}x^{1/2}
+X_{12}\left(-\frac{17}{28}+\frac{5}{7}\nu\right)x
+\biggl[\tilde{a}_{12}\left(\frac{18}{7}+\frac{33}{14}\nu\right)
+X_{12}\left(-\frac{43}{14}\hat{a}_0+\pi\right)\biggr]x^{3/2}\nonumber\\
&+\biggl[\tilde{a}_{12}\left(\hat{a}_0-\frac{3}{2}\pi\right)
+X_{12}\biggl(-\frac{43}{126}+2\hat{a}_0^2-2\tilde{a}_1\tilde{a}_2
-\frac{509}{126}\nu+\frac{79}{168}\nu^2\biggr)\biggr]x^2\nonumber\\
&+\biggl[\tilde{a}_{12}\left(-\frac{131}{72}+\frac{5483}{504}\nu +\frac{179}{126}\nu^2\right)
  +\hat{a}_0X_{12}\left(-\frac{331}{504}+\frac{193}{63}\nu\right)\biggr]x^{5/2},\\
\nonumber\\
|\tilde{h}_{33}^{(0)}|&=X_{12}+X_{12}(-4+2\nu)x
+\biggl[\tilde{a}_{12}\left(-\frac{1}{4}+\frac{5}{2}\nu\right)
+X_{12}\left(-\frac{7}{4}\hat{a}_0+3\pi\right)\biggr]x^{3/2}\nonumber\\
&+X_{12}\biggl(\frac{3}{2}\hat{a}_0^2+\frac{123}{110}-\frac{1838}{165}\nu+\frac{887}{330}\nu^2\biggr)x^2\nonumber\\
&+\biggl[\tilde{a}_{12}\left(-\frac{119}{60}+\frac{27}{20}\nu+\frac{241}{30}\nu^2\right)
+\hat{a}_0X_{12}\left(\frac{139}{60}-\frac{83}{60}\nu\right)\biggr]x^{5/2},\\
\nonumber\\
|\hat{h}_{32}^{(1)}|&=1+\frac{1}{1-3\nu}\biggl\{\left(\hat{a}_0-\tilde{a}_{12}X_{12}\right)x^{1/2}
+\left(-\frac{193}{90}+\frac{145}{18}\nu -\frac{73}{18}\nu^2\right)x\nonumber\\
&+\frac{1}{6}\biggl[\hat{a}_0(-39+73\nu)
+\tilde{a}_{12}X_{12}(23+13\nu)+12\pi(1-3\nu)\biggr]x^{3/2}\biggr\},\\
\nonumber\\
|\tilde{h}_{31}^{(0)}|&=X_{12}-\frac{2}{3}X_{12}(4+\nu)x
+\biggl[\tilde{a}_{12}\left(-\frac{9}{4}+\frac{13}{2}\nu\right)
+X_{12}\left(\frac{1}{4}\hat{a}_0+\pi\right)\biggr]x^{3/2}\nonumber\\
&+\biggl[-4(\tilde{a}_1^2+\tilde{a}_2^2)+X_{12}\left(\frac{607}{198}-\frac{136}{99}\nu-\frac{247}{198}\nu^2+\frac{3}{2}\hat{a}_0^2\right)\biggr]x^2\nonumber\\
&+\biggl[\tilde{a}_{12}\left(\frac{73}{12}-\frac{641}{36}\nu-\frac{5}{2}\nu^2\right)+X_{12}\hat{a}_0\left(-\frac{79}{36}+\frac{443}{36}\nu\right)\biggr]x^{5/2},
\end{align}
\begin{align}
|\hat{h}_{44}^{(0)}|&=1+\frac{1}{1-3\nu}\biggl\{\frac{1}{330}(-1779+6365\nu- 2625\nu^2)x\nonumber\\
&+\frac{1}{15}\biggl[\hat{a}_0(-38+114\nu)+60\pi(1-3\nu)-2\tilde{a}_{12}X_{12}(1-21\nu)\biggr]x^{3/2}\nonumber\\
&+\biggl(\frac{1068671}{200200}-\frac{1088119}{28600}\nu+\frac{146879}{2340}\nu^2-\frac{226097}{17160}\nu^3\biggr)x^2\biggr\},\\
\nonumber\\
|\tilde{h}_{43}^{(1)}|&=X_{12}+\frac{1}{-1+2\nu}\biggl\{
\frac{5}{4}\biggl[\tilde{a}_{12}(1-2\nu)-\hat{a}_0X_{12}\biggr]x^{1/2}
+X_{12}\left(\frac{39}{11}-\frac{1267}{132}\nu +\frac{131}{33}\nu^2\right)x\biggr\},\\
\nonumber\\
|\hat{h}_{42}^{(0)}|&=1+\frac{1}{1-3\nu}\biggl\{\frac{1}{330}(-1311+4025\nu-285\nu^2)x\nonumber\\
&-\frac{1}{15}\biggl[2\hat{a}_0(1-3\nu)+\tilde{a}_{12}X_{12}(38-78\nu)
-30\pi(1-3\nu)\biggr]x^{3/2}\nonumber\\
&+\biggl(\frac{1038039}{200200}-\frac{606751}{28600}\nu
+\frac{400453}{25740}\nu^2+\frac{25783}{17160}\nu^3\biggr)x^2\biggr\},\\
\nonumber\\
\label{eq:h41}
|\tilde{h}_{41}^{(1)}|&=X_{12}+\frac{1}{-1+2\nu}\biggl\{
\frac{5}{4}\biggl[\tilde{a}_{12}(1-2\nu)-\hat{a}_0X_{12}\biggr]x^{1/2}
+X_{12}\left(\frac{101}{33}-\frac{337}{44}\nu +\frac{83}{33}\nu^2\right)x\biggr\}.
\end{align}

\end{widetext}

\subsection{Cubic-order spin effects}
\label{sec:cubic}
We are also going to incorporate leading-order spin-cube effects
in the waveform aplitudes. To do so, we start from the corresponding
energy fluxes, that were recently obtained in Ref.~\cite{Marsat:2014xea}.
The analytically fully known spin-dependence of the energy flux
has the following structure
\begin{align}
{\cal F}^S = \dfrac{32}{5}\nu^2
   x^5\biggl[x^{3/2}f_{\rm SO}^{\rm LO}+x^{2}f_{\rm SS}^{\rm LO}+x^{5/2}f_{\rm SO}^{\rm NLO}\nonumber\\
              + x^3 f_{\rm SS}^{\rm NLO} + x^{7/2}\left(f_{\rm SO}^{\rm NNLO}+f_{\rm SSS}^{\rm LO}\right)\biggr].
\end{align}
All terms, except the cubic ones, can be obtained by multiplying each multipolar amplitude
of the previous section by its corresponding ``Newtonian'' term , taking the square and
finally summing them together. The spin-cube information we shall need in the next section
is included in the $f_{\rm SSS}^{\rm LO}$ term above, though one has to remember that $f_{\rm SSS}^{\rm LO}$
is actually given by two independent multipolar contributions, one coming from the
cubic-in-spin mass quadrupole and another from the cubic-in-spin current quadrupole.
The full term is given in Eq.~(6.19) of~\cite{Marsat:2014xea}, but, for the purpose of this
paper, S.~Marsat kindly separated for us the two partial multipolar contributions,
that read
\begin{align}
\label{eq:f22sss}
f^{\rm sss}_{22}&=-\dfrac{2}{3}\left(\hat{a}_0^3+3\hat{a}_0^2\tilde{a}_{12}X_{12}\right)x^{7/2},\\
\label{eq:f21sss}
f^{\rm sss}_{21}&=-\Bigg[\dfrac{1}{12}\hat{a}_0\tilde{a}_{12}^2\nonumber\\
              &\qquad+\left(\dfrac{5}{24}\tilde{a}_1^2+\dfrac{1}{4}\tilde{a}_1\tilde{a}_2+\dfrac{5}{24}\tilde{a}_2^2\right)\tilde{a}_{12}X_{12}\Bigg]x^{7/2}.
\end{align}
It is easy to verify that by taking the sum $f^{\rm SSS}_{22}+f^{\rm SSS}_{21}$ one obtains
Eq.~(6.19) of~\cite{Marsat:2014xea} once specified to the black-hole case, i.e. with
$\kappa_+=2=\lambda_+$, $\kappa_-=0=\lambda_-$ and using
Eqs.~\eqref{eq:Sl}-\eqref{eq:Sigmal} above.

\subsection{PN-expanded energy and angular momentum along circular orbits}
\label{sec:energy}
To implement the factorization of the waveform amplitudes (and fluxes) in order
to extract the $f_\lm$ and $\rho_\lm$ residual amplitude corrections, one needs
the PN-expanded effective source $\hat{S}_{\rm eff}^{(\epsilon)}$, namely the 
effective energy and angular momentum of the system along circular orbits.
In addition, also the total, real, energy is needed, since it enters the tail factor.
Defined as $\mu\equiv m_1 m_2/M$ the reduced-mass of the system,  
the $\mu$-normalized PN-expanded energy along circular orbits reads
\begin{equation}
\hat{E}^{\rm tot}(x)\equiv \dfrac{E^{\rm tot}}{\mu}=\hat{E}^{\rm orb}(x)+\hat{E}^{\rm SO}(x)+\hat{E}^{\rm SS}(x),
\end{equation}
and is written as the sum of an orbital term, a spin-orbit term (SO) and a quadratic-in-spin term (SS). 
The 3PN-accurate orbital term reads
\begin{align}
&\hat{E}^{\rm orb}(x)=1-\frac{1}{2} \nu  x \bigg\{ 1-\left(\frac{3}{4}+\frac{\nu}{12}\right)x\nonumber\\
&+\left(-\frac{27}{8}+\frac{19}{8}\nu-\frac{\nu^2}{24}\right)x^2\nonumber\\
&+\biggl[-\frac{675}{64}+\left(\frac{34445}{576}-\frac{205}{96}\pi^2\right)\nu-\frac{155}{96}\nu^2-\frac{35}{5184}\nu^3\biggr] x^3\bigg\},
\end{align}
while the spin-orbit term is
\begin{align}
&\hat{E}^{\rm SO}(x)=-\frac{1}{6}(7\hat{a}_0+\tilde{a}_{12}X_{12})\nu x^{5/2}\nonumber\\
&+\frac{1}{4}\biggl[-(11\hat{a}_0+5\tilde{a}_{12}X_{12})\nu +\frac{1}{9}(61\hat{a}_0+\tilde{a}_{12}X_{12})\nu^2\biggr]x^{7/2}\nonumber\\
&+\frac{1}{16}\bigg[-(135\hat{a}_0+81\tilde{a}_{12}X_{12})\nu \nonumber\\
    &+(367\hat{a}_0+55\tilde{a}_{12}X_{12})\nu^2+\frac{1}{3}(-29\hat{a}_0+\tilde{a}_{12}X_{12})\nu^3\bigg]x^{9/2},
\end{align}
and finally the quadratic-in-spin contribution
\begin{align}
\hat{E}^{\rm SS}(x)&=\frac{1}{2}\hat{a}_0^2\nu x^3\nonumber\\
&+\frac{1}{36}\biggl\{\biggl[10(\hat{a}_0^2+\tilde{a}_1\tilde{a}_2)+55(\tilde{a}_1^2-\tilde{a}_2^2)X_{12}\biggr]\nu\nonumber\\
&- 35\left(\tilde{a}_1^2-\frac{2}{7}\tilde{a}_1\tilde{a}_2+\tilde{a}_2^2\right)\nu^2\biggr\} x^4.
\end{align}
The Newton-normalized angular momentum incorporating up to NLO spin-orbit terms reads
\begin{align}
&\hat{j}^{\rm tot}(x)=1+\frac{1}{2}\left(3+\frac{\nu}{3}\right)x-\frac{5}{12}(7\hat{a}_0+\tilde{a}_{12}X_{12})x^{3/2}\nonumber\\
&+\biggl[\frac{1}{8}\left(27-19\nu+\frac{\nu^2}{3}\right)+\hat{a}_0^2\biggr]x^2\nonumber\\
&+\frac{1}{16}\biggl[-(77\hat{a}_0+35\tilde{a}_{12}X_{12})+\frac{1}{9}(427\hat{a}_0+7\tilde{a}_{12}X_{12})\nu\biggr]x^{5/2}.
\end{align}
Finally, the PN-expanded effective energy along circular orbits is obtained
by PN-expanding the usual relation between the real and effective, $\mu$-normalized, 
energy along circular orbits~\cite{Buonanno:1998gg},
\begin{equation}
\hat{E}^{\rm eff}=\dfrac{E_{\rm eff}}{\mu}=T_n\left[1+\frac{1}{2\nu}(\hat{E}_{\rm tot}^2-1)\right].
\end{equation}
\section{Factorized waveform amplitudes}
\label{sec:factorization}

\subsection{Factorizing the source and tail factor: the residual amplitudes}
Now that all the necessary analytical elements are introduced,
we can finally compute the residual amplitude corrections
when $\nu\neq 0$ by factorizing tail and source from
Eqs.~\eqref{eq:h22}-\eqref{eq:h41}, \eqref{eq:f22sss} and \eqref{eq:f21sss}.
Focusing first on the even-$m$ case, the PN-expanded $\rho_\lm$'s functions
are obtained as
\be
\rho_{\ell m}(x;\nu,\tilde{a}_1,\tilde{a}_2)=T_n\left[\left(\dfrac{|\hat{h}_{\ell m}(x)|}{|\hat{h}_{\ell m}^{\rm tail}|\hat{S}^{(\epsilon)}_{\rm eff}}\right)^{1/\ell}\right],
\ee
where $\hat{S}^{(\epsilon)}_{\rm eff}(x)$ is either $\hat{E}_{\rm eff}$ 
when $\ell+m$ is even, or $\hat{j}^{\rm tot}$ when $\ell+m$ is odd,
while $|\hat{h}^{\rm tail}|$ is the modulus of the tail factor introduced in Eq.~\eqref{eq:hlm0}
whose explicit expression is given in Eq.~\eqref{eq:modTail} below. The Taylor expansion
$T_n[\dots]$ is truncated at the same $n$-PN order of the $|h_\lm^{(\epsilon)}|$.
The functions $\rho_{\ell m}$ have the form $1+c_1^{\ell m} x + \dots$ and, 
like in the test-particle case, are given as the sum of orbital and spin
terms as
\be
\rho_\lm(x;\nu,\tilde{a}_1,\tilde{a}_2) = \rho_\lm^{\rm orb}(x;\nu) + \rho_\lm^{\rm S}(x;\nu,\tilde{a}_1,\tilde{a}_2)\ .
\ee
For the odd-$m$ case, the same factorization yields the function
\be
\delta m f_{\ell m}=X_{12}f_\lm^{\rm orb} + \tilde{f}_{\ell m}^{\rm S},
\ee
that is obtained as the following Taylor expansion
\begin{equation}
\delta m f_{\ell m}=T_n\left[\frac{|\tilde{h}_{\ell m}|}{|\hat{h}_{\ell m}^{\rm tail}|\hat{S}_{\rm eff}^{(\epsilon)}}\right],
\end{equation} 
where, for consistency with notation used in Ref.~\cite{Nagar:2016ayt},
we also used $\delta m\equiv X_{12}$. Finally, to perform this calculation,
we also need the Taylor expansion of the modulus of
the tail factor, that is given by~\cite{Damour:2014sva}
\begin{equation}
\label{eq:modTail}
|\hat{h}^{\rm tail}_{\ell m}(x)|^{2}=\dfrac{4\pi E^{\rm tot} m x^{3/2}\prod_{s=1}^{\ell}\left(s^{2}+2 E^{\rm tot} m x^{3/2}\right)^{2}}{(\ell !)^{2}
  \left(1-e^{-4\pi m E^{\rm tot}}\right)}.
\end{equation}
When factorizing out that tail and effective source factors from the waveform amplitudes of 
Eqs.~\eqref{eq:h22}-\eqref{eq:h41}, as well as from the spin-cube flux terms, 
Eqs.~\eqref{eq:f22sss} and~\eqref{eq:f21sss}, one  finally finds the following 
spin-dependent terms:
\begin{widetext}
  \begin{align}
    \label{eq:rho22S}
\rho_{22}^{\rm S}&=-\left(\dfrac{\hat{a}_0}{2}+\frac{1}{6}\tilde{a}_{12}X_{12}\right)x^{3/2}
+\dfrac{\hat{a}_0^2}{2}x^2 -\biggl[\hat{a}_0\left(\frac{52}{63}+\frac{19}{504}\nu\right)+\left(\frac{50}{63}+\frac{209}{504}\nu\right)\tilde{a}_{12}X_{12}\biggr]x^{5/2}\\\nonumber
&+\biggl[\left(-\dfrac{11}{21}+\dfrac{103}{504}\nu\right)\hat{a}_0^2+\left(-\dfrac{19}{63}+\dfrac{10}{9}\nu\right)\tilde{a}_1\tilde{a}_2
  +\frac{221}{252}\hat{a}_0\tilde{a}_{12}X_{12}\biggr]x^3\\\nonumber
&+\biggl[\hat{a}_0\left(\frac{32873}{21168}+\frac{477563}{42336}\nu +\frac{147421}{84672}\nu ^2\right)
  -\tilde{a}_{12}X_{12}\biggl(\frac{23687}{63504}-\frac{171791}{127008}\nu+\frac{50803}{254016}\nu ^2\biggr)
  +\left(\dfrac{7}{12}\hat{a}_0^3-\dfrac{1}{4}\hat{a}_0^2\tilde{a}_{12}X_{12}\right)\biggr]x^{7/2},\\
\nonumber\\
\rho_{32}^{\rm S}&=\frac{1}{3(1-3\nu)}\left(\hat{a}_0-\tilde{a}_{12}X_{12}\right)x^{1/2}\nonumber\\
&+\frac{1}{162(1-3\nu)^2}\biggl[\left(-\frac{1433}{10}+553\nu -\frac{797}{2}\nu^2\right)\hat{a}_0
  -\left(-\frac{1793}{10}+427\nu +\frac{607}{2}\nu ^2\right)\tilde{a}_{12}X_{12}\biggr]x^{3/2},\\
\nonumber\\
\rho_{42}^{\rm S}&=\frac{1}{30}\left(\frac{1}{-1+3\nu}\right)[\hat{a}_0(1-3\nu)+\tilde{a}_{12}X_{12}(19-39\nu)]x^{3/2},\\
\nonumber\\
\rho_{44}^{\rm S}&=\frac{1}{30}\left(\frac{1}{-1+3\nu}\right)[\hat{a}_0(19-57\nu)-\tilde{a}_{12}X_{12}(-1+21\nu)]x^{3/2},
\end{align}
and similarly
\begin{align}
\tilde{f}_{21}^{\rm S}&=-\frac{3}{2}\tilde{a}_{12}x^{1/2}+\biggl[\tilde{a}_{12}\left(\frac{110}{21}+\frac{79}{84}\nu\right)
-\frac{13}{84}\hat{a}_0X_{12}\biggr]x^{3/2}
+\biggl[-\frac{27}{8}(\tilde{a}_1^2-\tilde{a}_2^2)
+\frac{3}{8}X_{12}\left(\tilde{a}_1^2+\frac{10}{3}\tilde{a}_1\tilde{a}_2+\tilde{a}_2^2\right)\biggr]x^2\nonumber\\
&+\biggl[\tilde{a}_{12}\left(-\frac{3331}{1008}-\frac{13}{504}\nu +\frac{613}{1008}\nu ^2\right)
+\hat{a}_0X_{12}\left(-\frac{443}{252}+\frac{1735}{1008}\nu\right)+\frac{3}{4}\hat{a}_0^2\tilde{a}_{12}\biggr]x^{5/2},\\
\nonumber\\
\tilde{f}_{31}^{\rm S}&=\biggl[\tilde{a}_{12}\left(-\frac{9}{4}+\frac{13}{2}\nu\right)+\frac{1}{4}\hat{a}_0X_{12}\biggr]x^{3/2}
+\biggl[-4(\tilde{a}_1^2-\tilde{a}_2^2)+\frac{3}{2}\hat{a}_0^2X_{12}\biggr]x^2\nonumber\\
&+\biggl[\tilde{a}_{12}\left(\frac{41}{8}-\frac{137}{9}\nu-\frac{5}{2}\nu ^2\right)
+\hat{a}_0X_{12}\left(-\frac{65}{72}+\frac{443}{36}\nu\right)\biggr]x^{5/2},\\
\nonumber\\
\tilde{f}_{33}^{\rm S}&=\biggl[\tilde{a}_{12}\left(-\frac{1}{4}+\frac{5}{2}\nu\right)
-\frac{7}{4}\hat{a}_0X_{12}\biggr]x^{3/2}+\frac{3}{2}\hat{a}_0^2X_{12}x^2\nonumber\\
&+\biggl[\tilde{a}_{12}\left(-\frac{233}{120}+\frac{29}{15}\nu +\frac{241}{30}\nu ^2\right)
  +\hat{a}_0X_{12}\left(\frac{313}{120}-\frac{83}{60}\nu\right)\biggr]x^{5/2},\\
\nonumber\\
\tilde{f}_{41}^{\rm S}&=\tilde{f}_{43}^S=\frac{5}{4}\left(\frac{1}{-1+2\nu}\right)[\tilde{a}_{12}(1-2\nu)-\hat{a}_0X_{12}]x^{1/2}.
\end{align}
\end{widetext}
After applying the proper change of variables, one easily checks that the NLO
contributions we computed here do coincide with Eqs.~(85)-(95) of Ref.~\cite{Damour:2014sva}.
Similarly, the NNLO spin-orbit contribution to $\rho_{22}$, that was also
computed in Ref.~\cite{Nagar:2016ayt} is checked with the same term computed
in Ref.~\cite{Bohe:2016gbl}.

\subsection{Factorization of the orbital part}
\label{sec:orbital-factorization}
Likewise the test-particle case above, we now apply
the prescription of Ref.~\cite{Nagar:2016ayt} of
factorizing the orbital parts of $\rho_{\ell m}^{\rm S}$
and $\tilde{f}_{\ell m}^{\rm S}$. After this operation,
the factorized residual amplitudes are written as
\begin{align}
\rho_{\ell m}      & =\rho^{\rm orb}_{\ell m}\hat{\rho}^{\rm S}_{\ell m}\qquad\quad\;\, m={\rm even}\\
\delta m f_{\ell m}& =\left(\rho_{\ell m}^{\rm orb}\right)^\ell \hat{f}^{\rm S}_{\ell m}\qquad m={\rm odd}
\end{align}
where, as in Ref.~\cite{Nagar:2016ayt}, the $m={\rm odd}$ spin factors are written as
the sum of two separate terms
\begin{align}
  \label{flmsplit:21}
  \hat{f}_{21}^{\rm S}&= X_{12}\f^{\rm S_{(0)}}_{21}-\dfrac{3}{2}\tilde{a}_{12}x^{1/2}\f_{21}^{\rm S_{(1)}},\\
  \label{flmsplit:33}
  \hat{f}_{33}^{\rm S} &= X_{12}\f^{\rm S_{(0)}}_{33}+\left(-\dfrac{1}{4}+\dfrac{5}{2}\nu\right)\tilde{a}_{12}x^{3/2}\f_{33}^{\rm S_{(1)}},\\
  \label{flmsplit:31}
  \hat{f}_{31}^{\rm S} &= X_{12}\f^{\rm S_{(0)}}_{31}+\left(-\dfrac{9}{4}+\dfrac{13}{2}\nu\right)\tilde{a}_{12}x^{3/2}\f_{31}^{\rm S_{(1)}},\\
  \label{flmsplit:43}
  \hat{f}_{43}^{\rm S} &= X_{12}\f^{\rm S_{(0)}}_{43}-\dfrac{5}{4}\tilde{a}_{12}x^{1/2}\f_{43}^{\rm S_{(1)}},\\
  \label{flmsplit:41}
  \hat{f}_{43}^{\rm S} &= X_{12}\f^{\rm S_{(0)}}_{41}-\dfrac{5}{4}\tilde{a}_{12}x^{1/2}\f_{41}^{\rm S_{(1)}}.
\end{align}
As shown in Ref.~\cite{Nagar:2016ayt}, we recall that the need of separating the $\hat{f}_{\ell m}$'s function
into two separate terms, one proportional to $X_{12}$ and another to $\tilde{a}_{12}$ times $x$ is
necessary to identify the two functions $\f^{\rm S_{(0)}}_{\ell m}$ and $\f^{\rm S_{(1)}}_{\ell m}$
that can be separately resummed using their inverse Taylor representation.
The $\f_{\ell m}^{\rm S_{(0),(1)}}$ functions read
\begin{align}
\label{eq:hatf210}
\f_{21}^{\rm S_{(0)}}&=1-\frac{13}{84}\hat{a}_0x^{3/2}+\frac{3}{8}\left(\hat{a}_0^2+\frac{4}{3}\tilde{a}_1\tilde{a}_2\right)x^2\nonumber\\
&+\hat{a}_0\left(-\frac{14705}{7056}+\frac{12743}{7056}\nu\right)x^{5/2},\\
\label{eq:hatf211}
\f_{21}^{\rm S_{(1)}}&=1-\left(\frac{349}{252}+\frac{74}{63}\nu\right)x+\frac{9}{4}\hat{a}_0x^{3/2}\nonumber\\
&-\left(\frac{3379}{21168}-\frac{4609}{10584}\nu+\frac{39}{392}\nu^2 +\frac{\hat{a}_0^2}{2}\right)x^2,\\
\nonumber\\
\f_{31}^{\rm S_{(0)}}&=1+\frac{1}{4}\hat{a}_0x^{3/2}
+\frac{3}{2}\hat{a}_0^2x^2+\hat{a}_0\left(-\dfrac{13}{36}+\dfrac{449}{36}\nu\right)x^{5/2},\\
\f_{31}^{\rm S_{(1)}}&=1-\frac{16}{26\nu- 9}\hat{a}_0x^{1/2}\nonumber\\
&+\frac{1}{26\nu-9}\left(1-\frac{95}{9}\nu +\frac{22}{3}\nu ^2\right)x,
\end{align}
\begin{align}
\f_{33}^{\rm S_{(0)}}&=1-\frac{7}{4}\hat{a}_0x^{3/2}+\frac{3}{2}\hat{a}_0^2x^2\nonumber\\
            & +\hat{a}_0\left(-\dfrac{211}{60}+\dfrac{127}{60}\nu\right)x^{5/2},\\
\f_{33}^{\rm S_{(1)}}&=1+\frac{1}{15}\left(\frac{-169+671\nu+182\nu ^2}{10\nu -1}\right)x,\\
\label{eq:4143}
\f_{41}^{\rm S_{(0)}}&=f_{43}^{S_{(0)}}=1-\frac{5}{4}\left(\frac{1}{-1+2\nu}\right)\hat{a}_0x^{1/2},\\
\f_{41}^{\rm S_{(1)}}&=f_{43}^{S_{(1)}}=1.
\end{align}
Equations~\eqref{flmsplit:21},~\eqref{eq:hatf210} correspond to Eqs.~(9)-(10) of~\cite{Nagar:2016ayt},
while Eq.~\eqref{eq:hatf211} presents an additional term, $\hat{a}_0^2/2$, that is the leading-order
spin-cube that was omitted in~\cite{Nagar:2016ayt}.
Finally the $m={\rm even}$ spin factors read
  \begin{align}
  \label{eq:rho22}
  \hat{\rho}_{22}^{\rm S} &= 1 - \left(\dfrac{\hat{a}_0}{2}+\dfrac{1}{6}\tilde{a}_{12}X_{12}\right)x^{3/2} + \dfrac{\hat{a}_0^2}{2}x^2\nonumber\\
  &+ \left[\left(-\dfrac{337}{252}+\dfrac{73}{252}\nu\right)\hat{a}_0-\left(\dfrac{27}{28}+\dfrac{11}{36}\nu\right)\tilde{a}_{12}X_{12}\right]x^{5/2}\nonumber\\
  & +\Bigg[\dfrac{221}{252}\hat{a}_0\tilde{a}_{12}X_{12}-\left(\frac{1}{84}+\dfrac{31}{252}\nu\right)\hat{a}_0^2\nonumber\\
  &+\left(-\dfrac{19}{63}+\dfrac{10}{9}\nu\right)\tilde{a}_1\tilde{a}_2\Bigg]x^3\nonumber\\
  &+\Bigg[\left(-\dfrac{2083}{2646}+\dfrac{123541}{10584}\nu+\dfrac{4717}{2646}\nu^2\right)\hat{a}_0\nonumber\\
    &+\left(-\dfrac{13367}{7938}+\dfrac{22403}{15876}\nu+\dfrac{25}{324}\nu^2\right)\tilde{a}_{12}X_{12}\nonumber\\
    &+\dfrac{7}{12}\hat{a}_0^3-\dfrac{1}{4}\hat{a}_0^2\tilde{a}_{12}X_{12}\Bigg]x^{7/2},\\
\nonumber\\
\label{eq:rho32}
\hat{\rho}_{32}^{\rm S}&=1+\left(\frac{1}{3(1-3\nu)}\right)\left(\hat{a}_0-\tilde{a}_{12}X_{12}\right)x^{1/2}\nonumber\\
&+\frac{1}{27(1-3\nu)^2}\bigg[\hat{a}_0\left(-\frac{259}{20}+55\nu -\frac{223}{4}\nu^2\right)\nonumber\\
  &-\tilde{a}_{12}X_{12}\left(-\frac{379}{20}+34\nu +\frac{245}{4}\nu ^2\right)\bigg]x^{3/2},\\
  \hat{\rho}_{42}^{\rm S}&= 1 + \rho_{42}^{\rm S},\\
  \hat{\rho}_{44}^{\rm S} &= 1 + \rho_{44}^{\rm S}.
  \end{align}
Note that our Eq.~\eqref{eq:rho22} above corrects an error in the published $\tilde{a}_1\tilde{a}_2$
NLO term of Eq.~(8) of Ref.~\cite{Nagar:2016ayt}.

\section{Resummation}
\label{sec:resum}
\begin{figure*}[t]
  \center
 \includegraphics[width=0.29\textwidth]{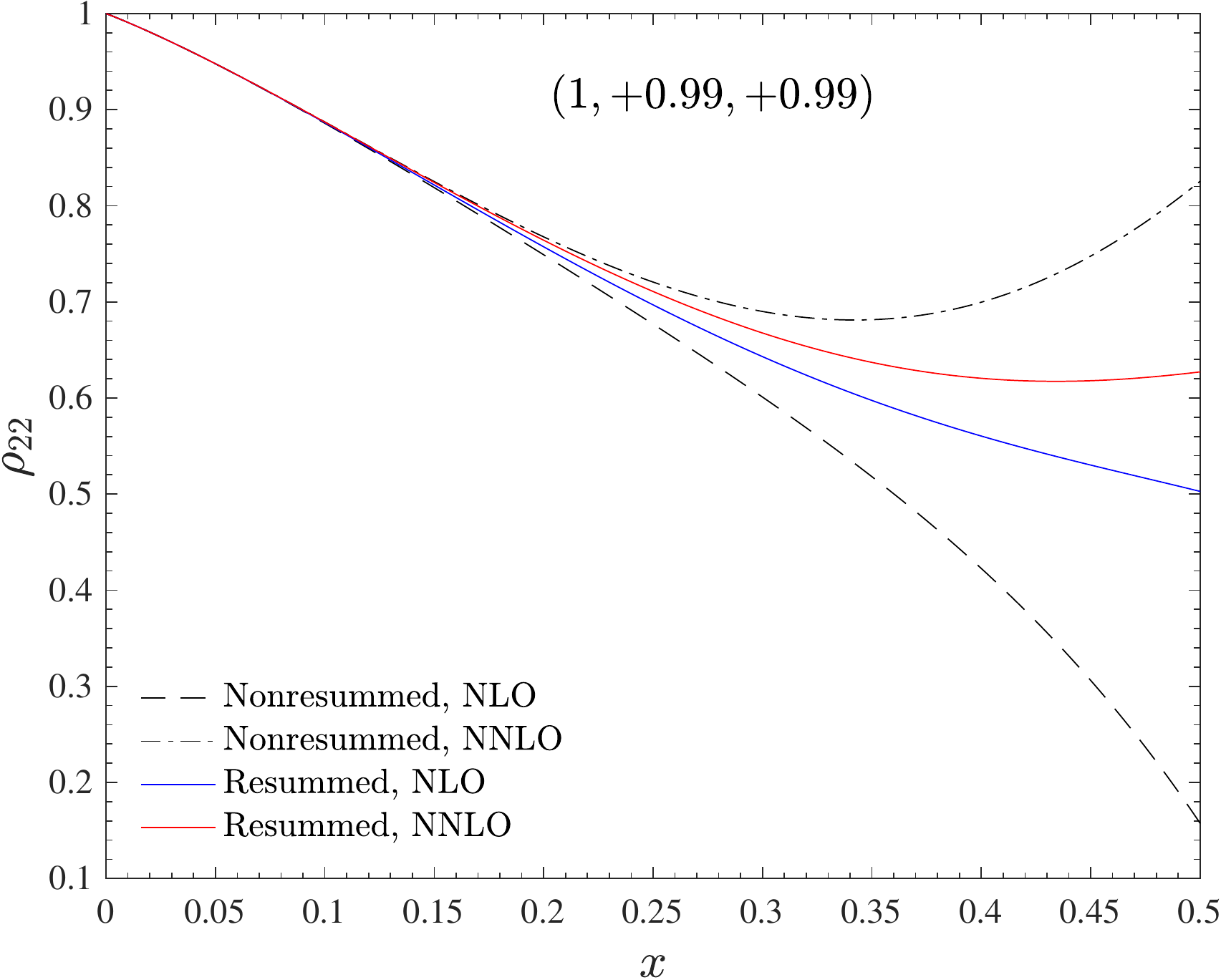}
 \hspace{1 mm}
  \includegraphics[width=0.29\textwidth]{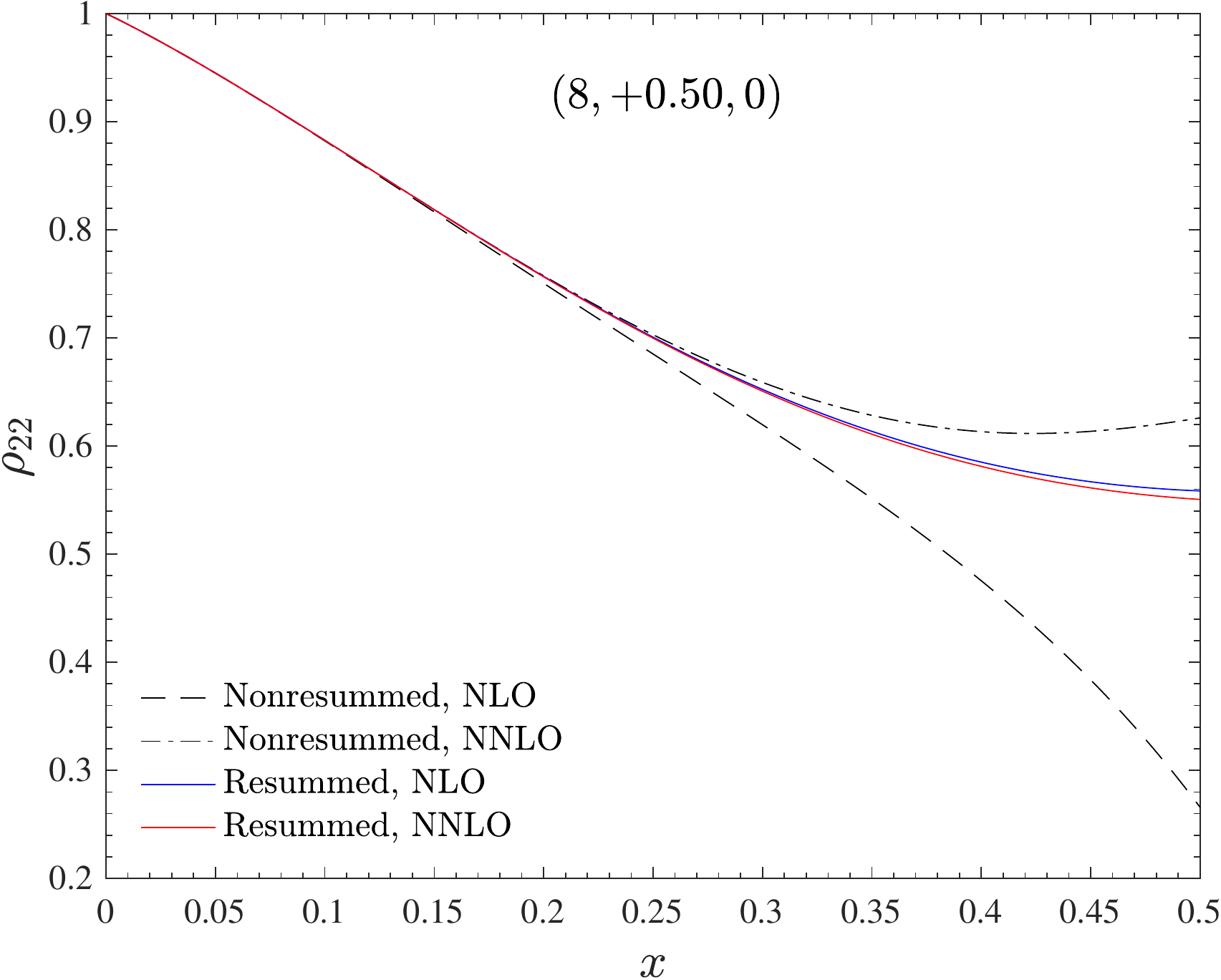}
   \hspace{1 mm}
   \includegraphics[width=0.29\textwidth]{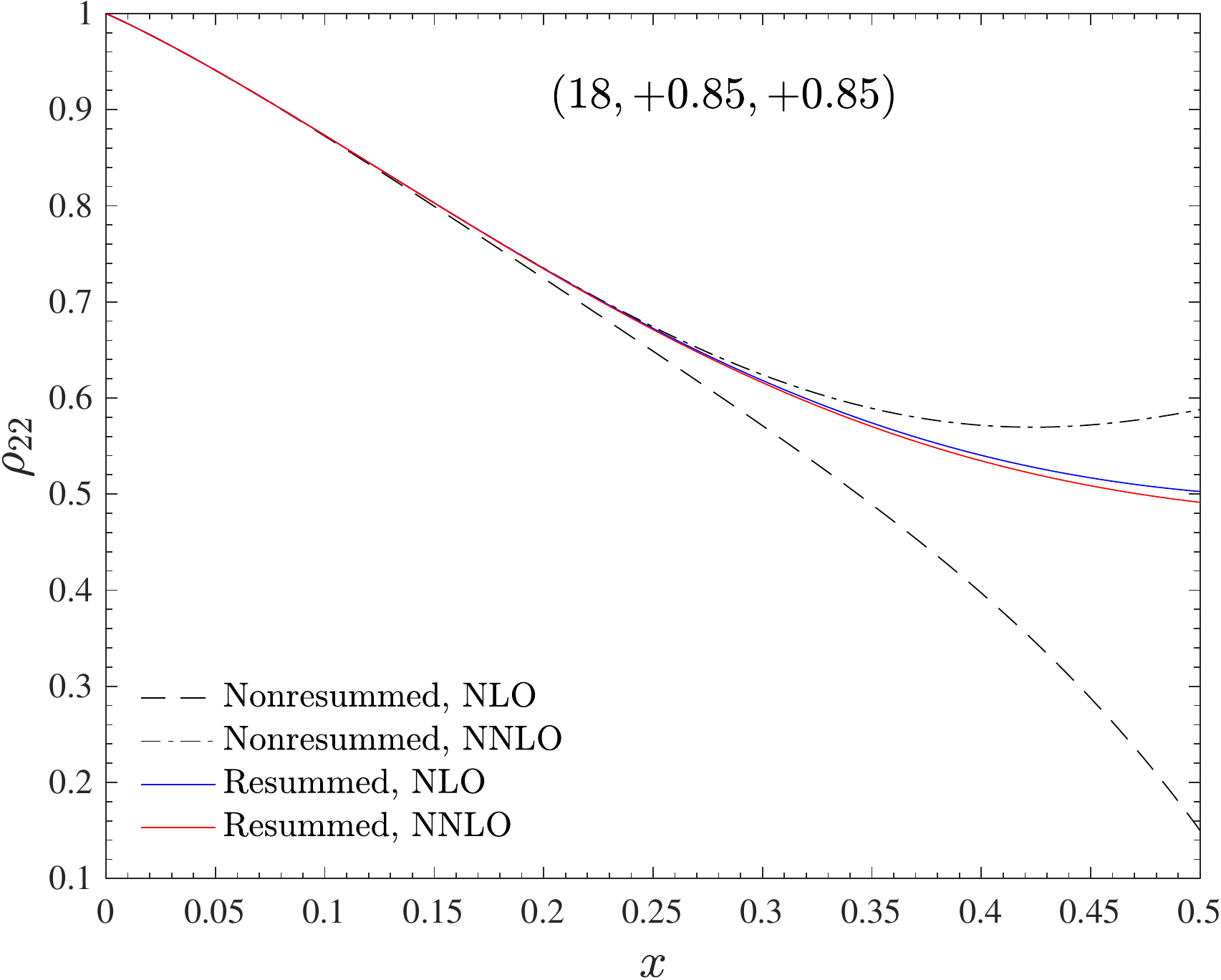}\\
   \vspace{2 mm}
  \includegraphics[width=0.29\textwidth]{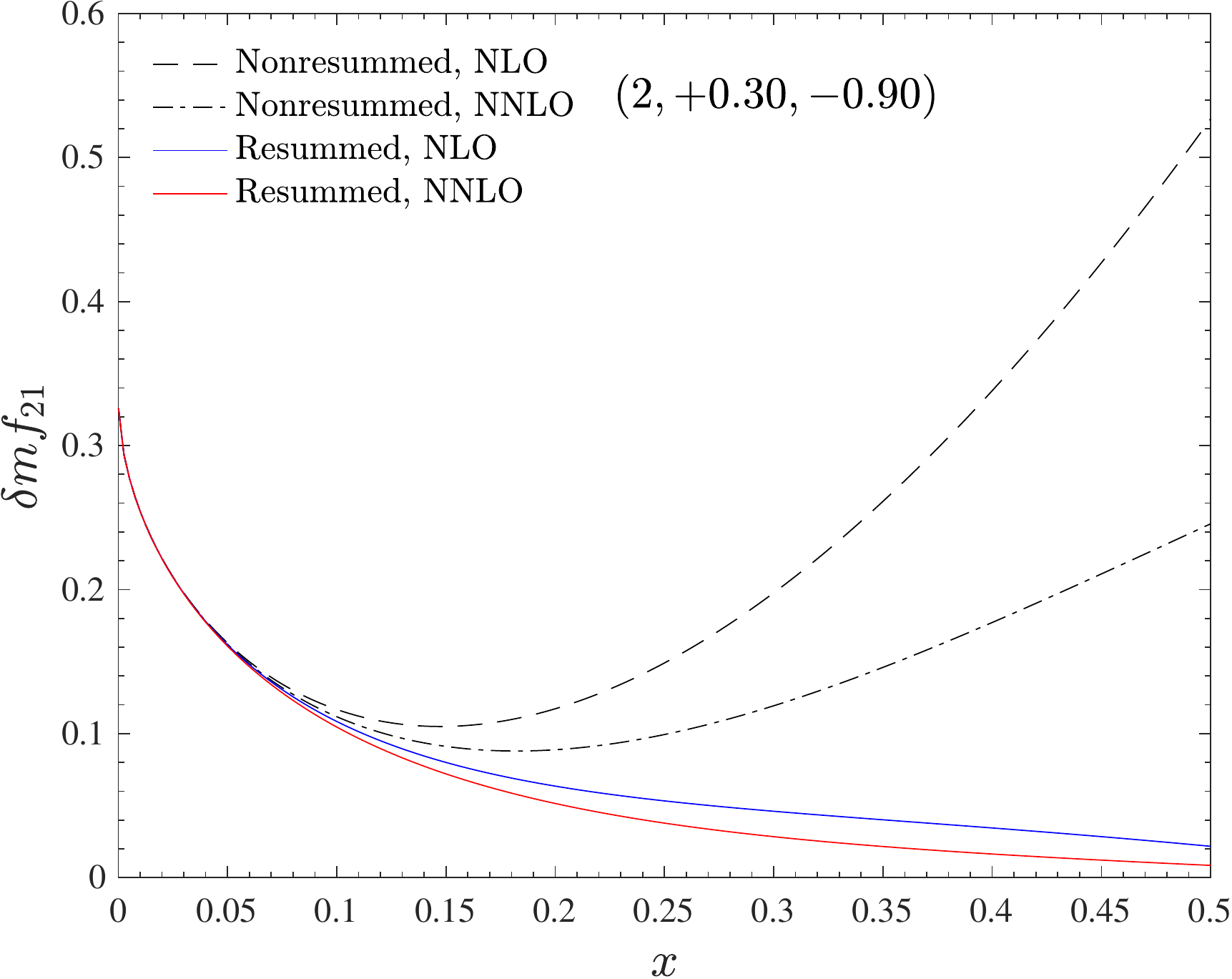}
   \hspace{1 mm}
  \includegraphics[width=0.29\textwidth]{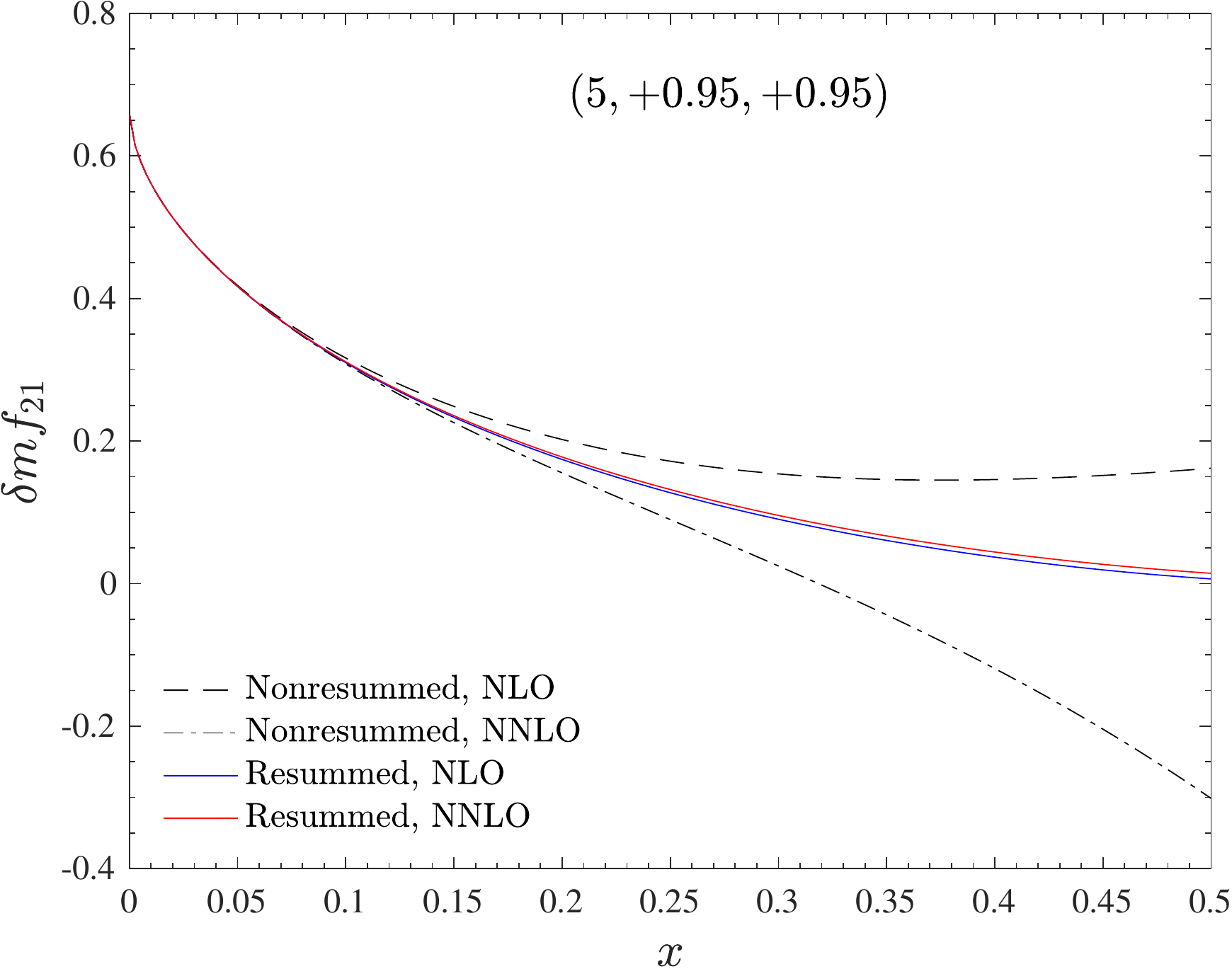}
   \hspace{1 mm}
  \includegraphics[width=0.29\textwidth]{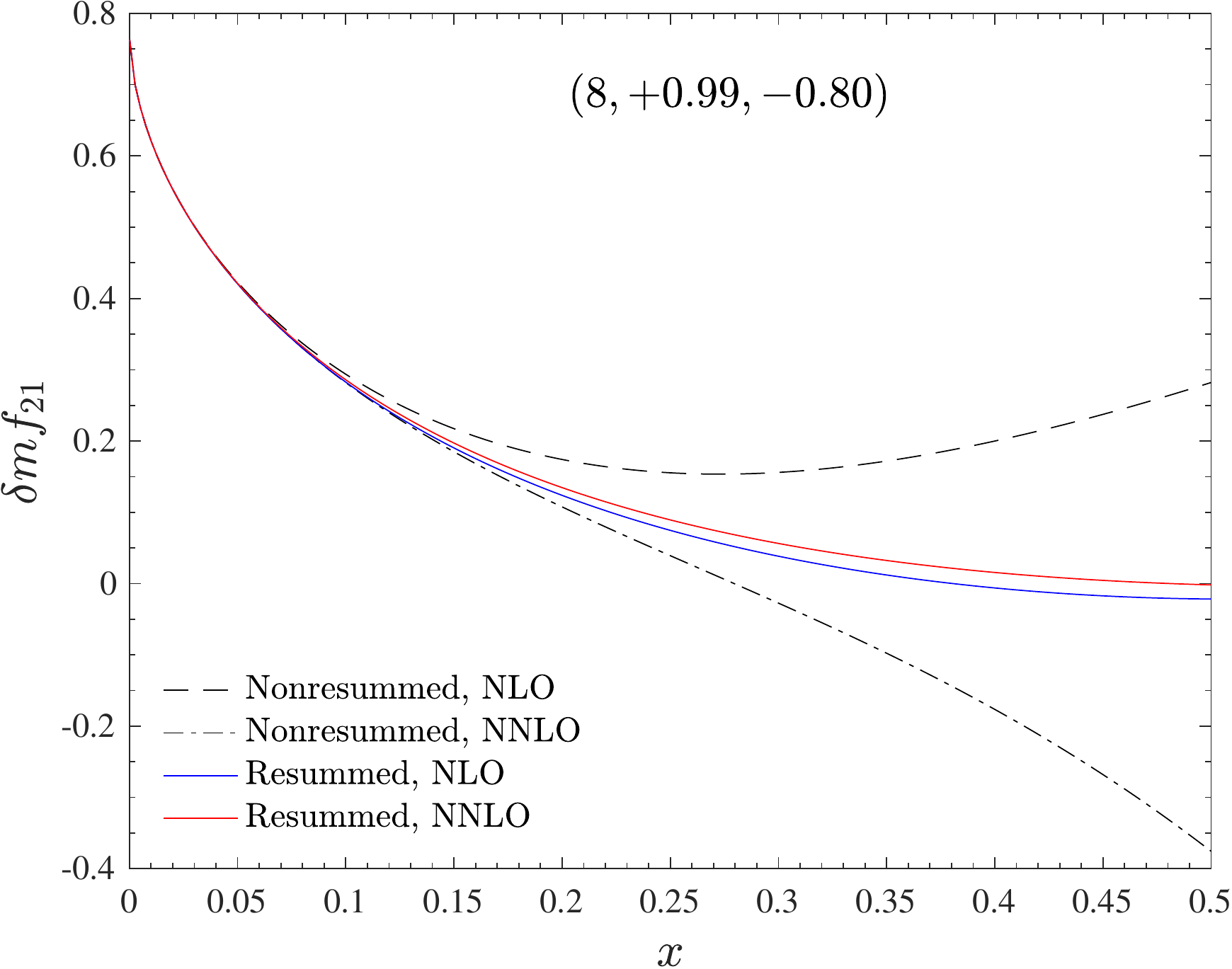}
  \caption{\label{fig:l2}Nonresummed (black) and resummed (colored) 
    waveform amplitudes $\rho_{22}$ (top panels) and $f_{21}$ (bottom panels) 
    for a few configurations. The orbital factor is taken at $3^{+3}$~PN
    {\it relative} accuracy and resummed with the Pad\'e approximants of Table~\ref{tab:rhoLSO}. 
    The consistency between NNLO and NLO truncations of the spin terms
    factor is dramatically improved when the factorization and resummation 
    procedure is applied. }
\end{figure*}
\begin{figure*}[t]
  \center
    \includegraphics[width=0.29\textwidth]{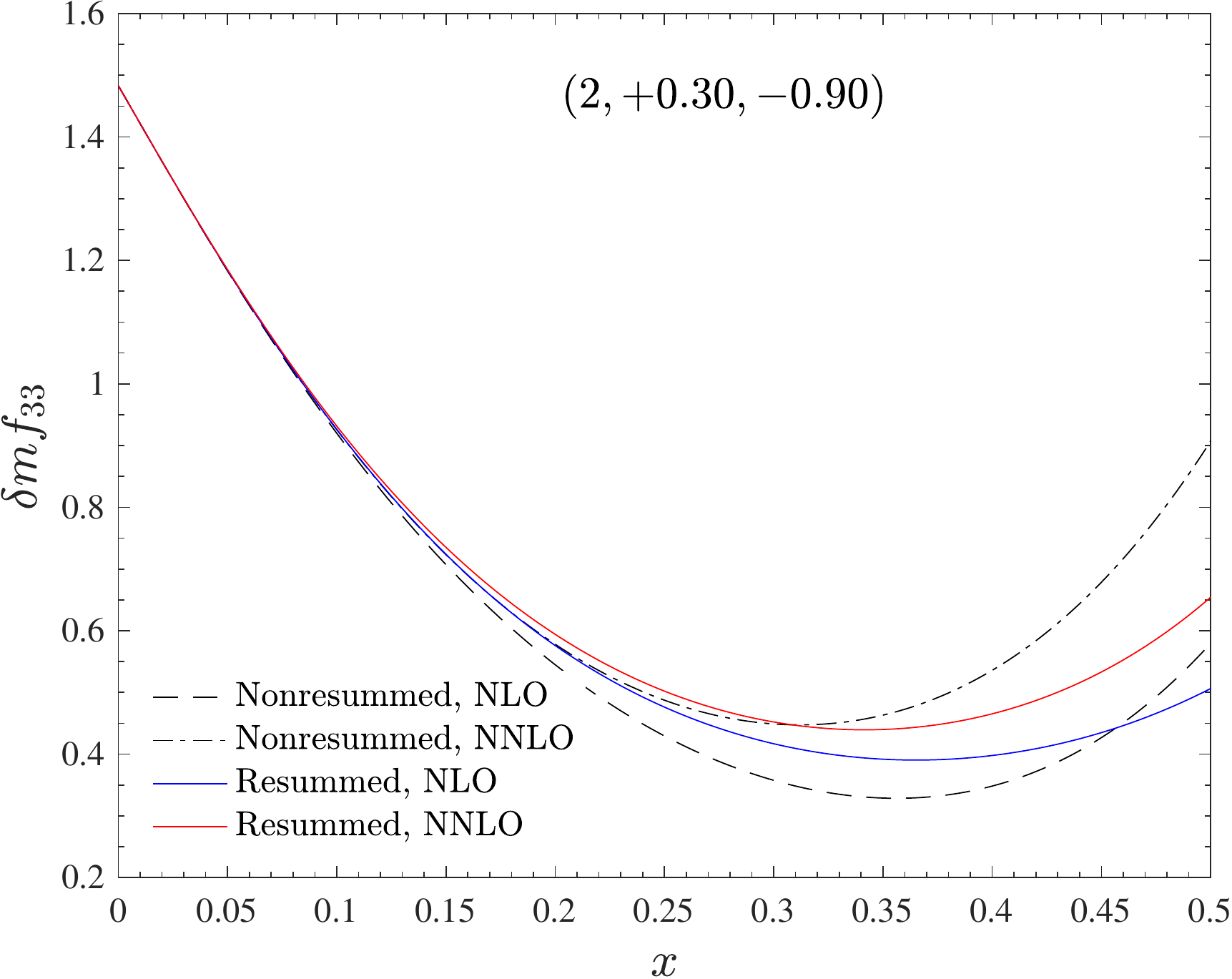}
     \hspace{1 mm}
  \includegraphics[width=0.29\textwidth]{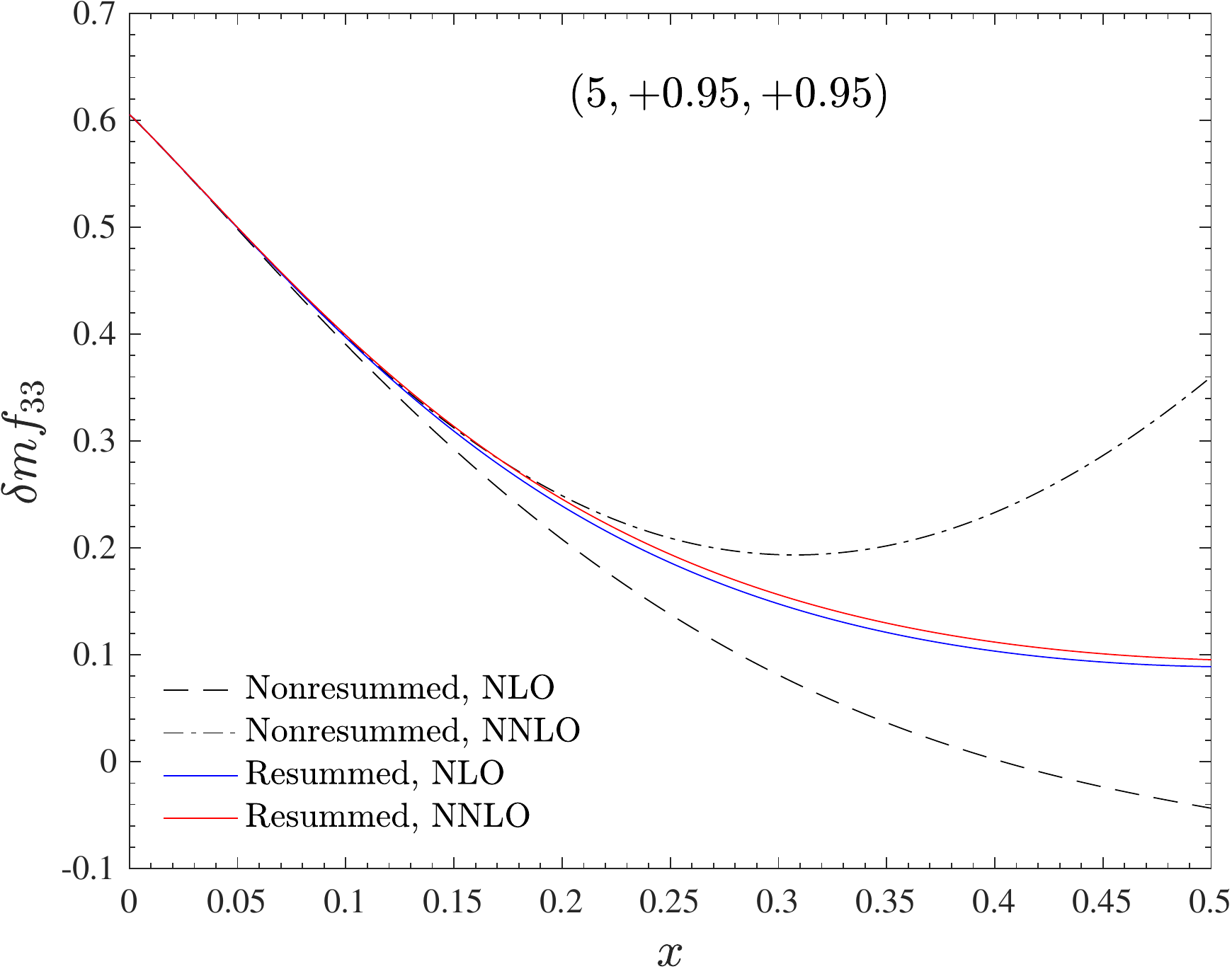}
   \hspace{1 mm}
  \includegraphics[width=0.29\textwidth]{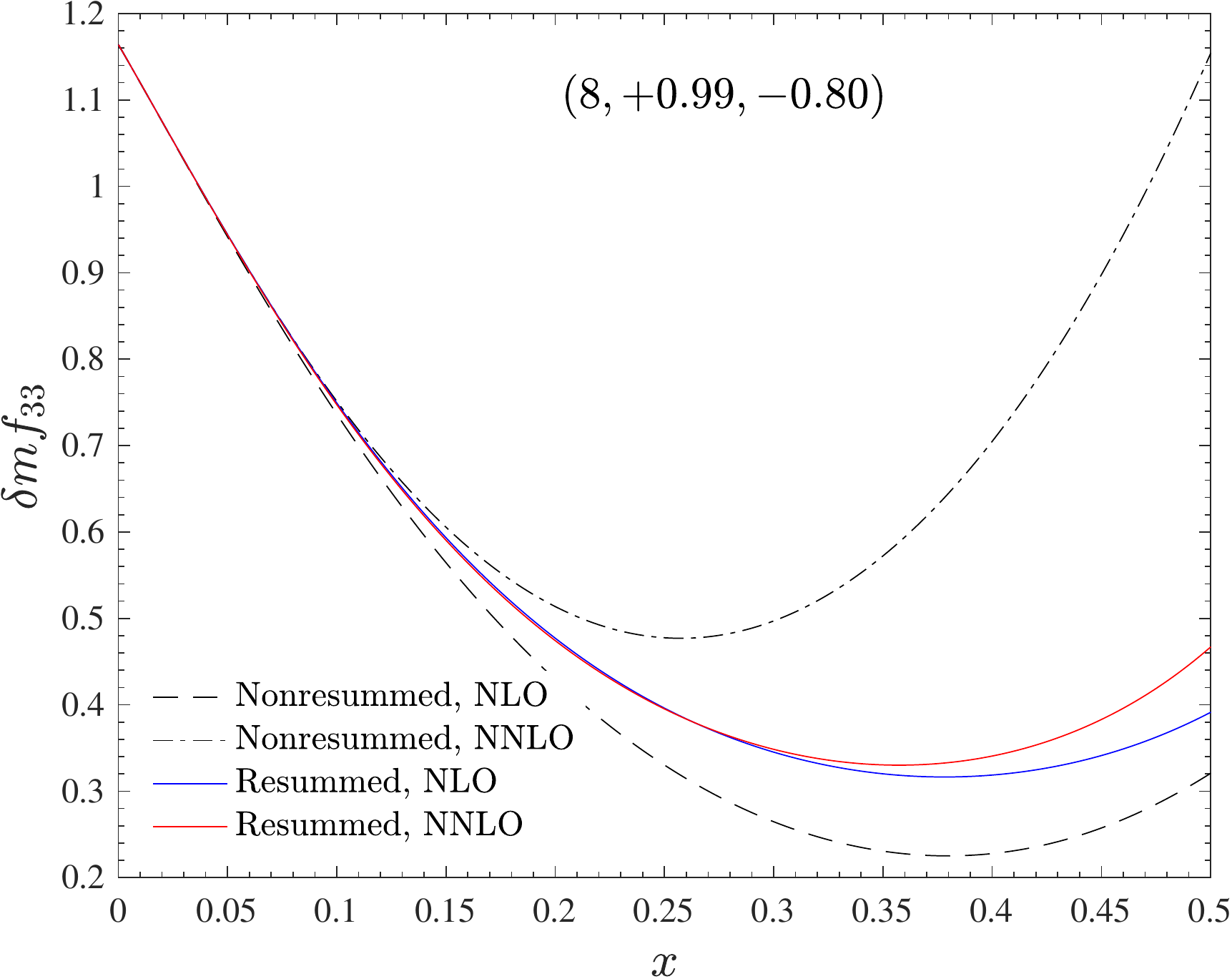}\\
     \vspace{2 mm}
      \includegraphics[width=0.29\textwidth]{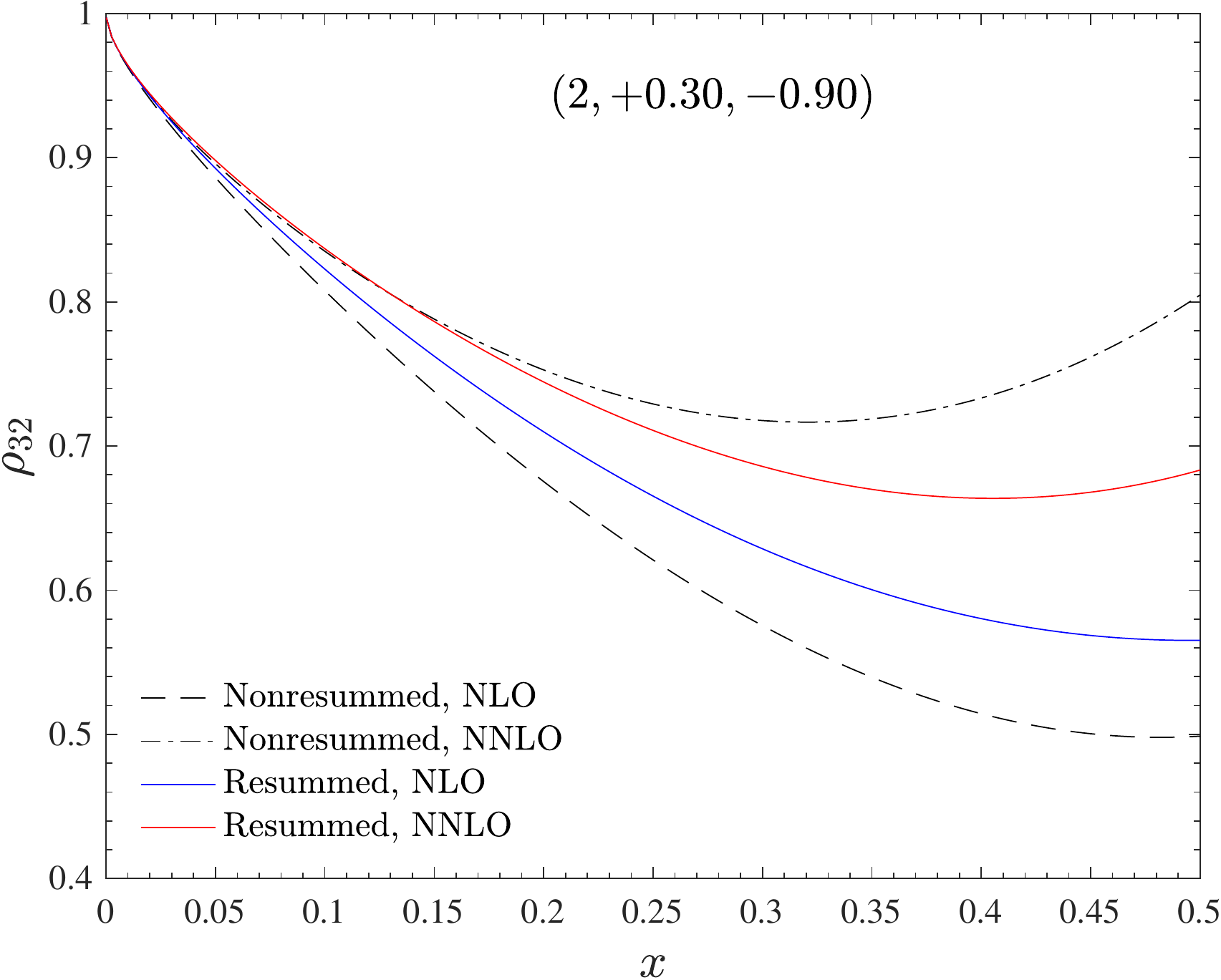}
       \hspace{1 mm}
  \includegraphics[width=0.29\textwidth]{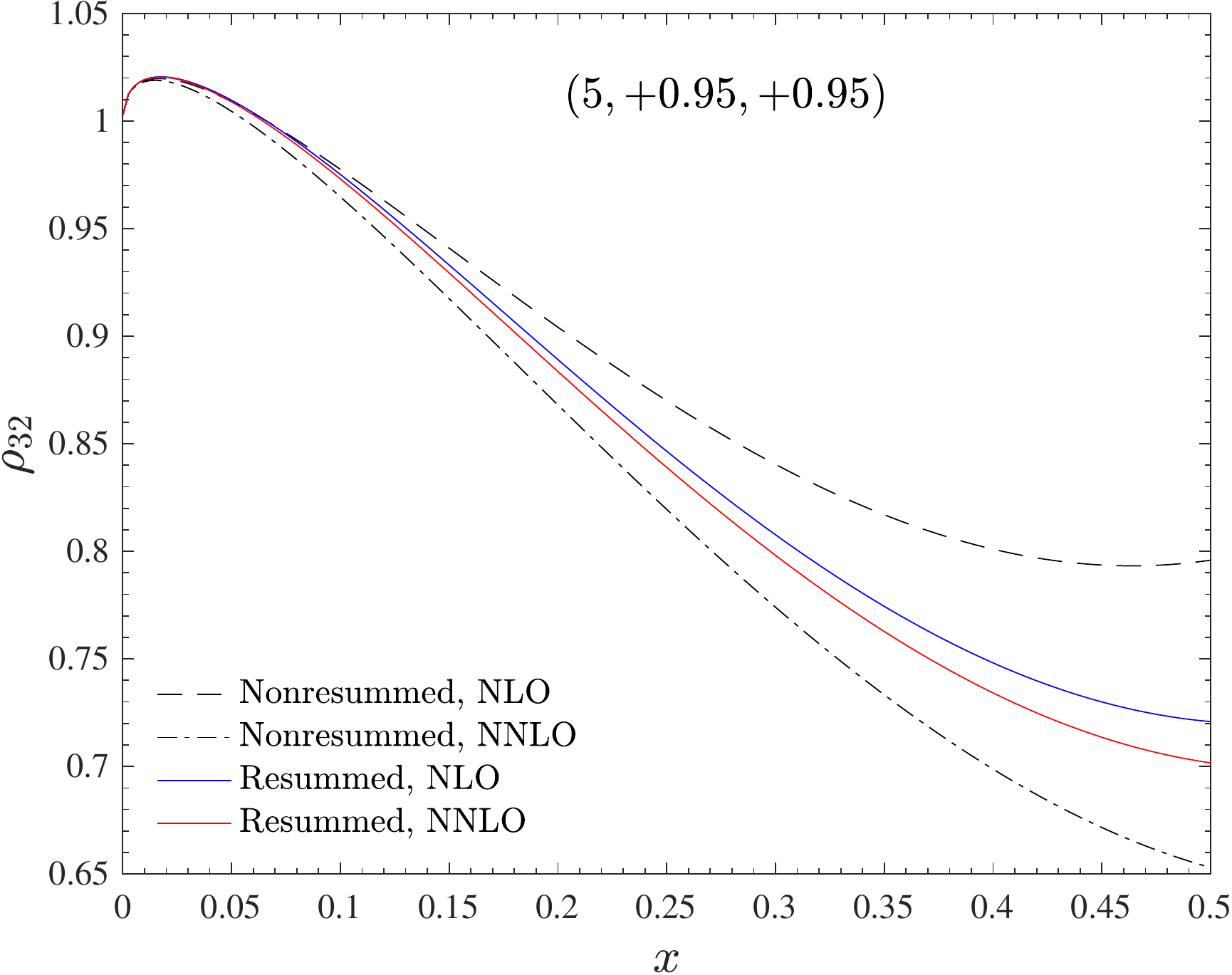}
   \hspace{1 mm}
  \includegraphics[width=0.29\textwidth]{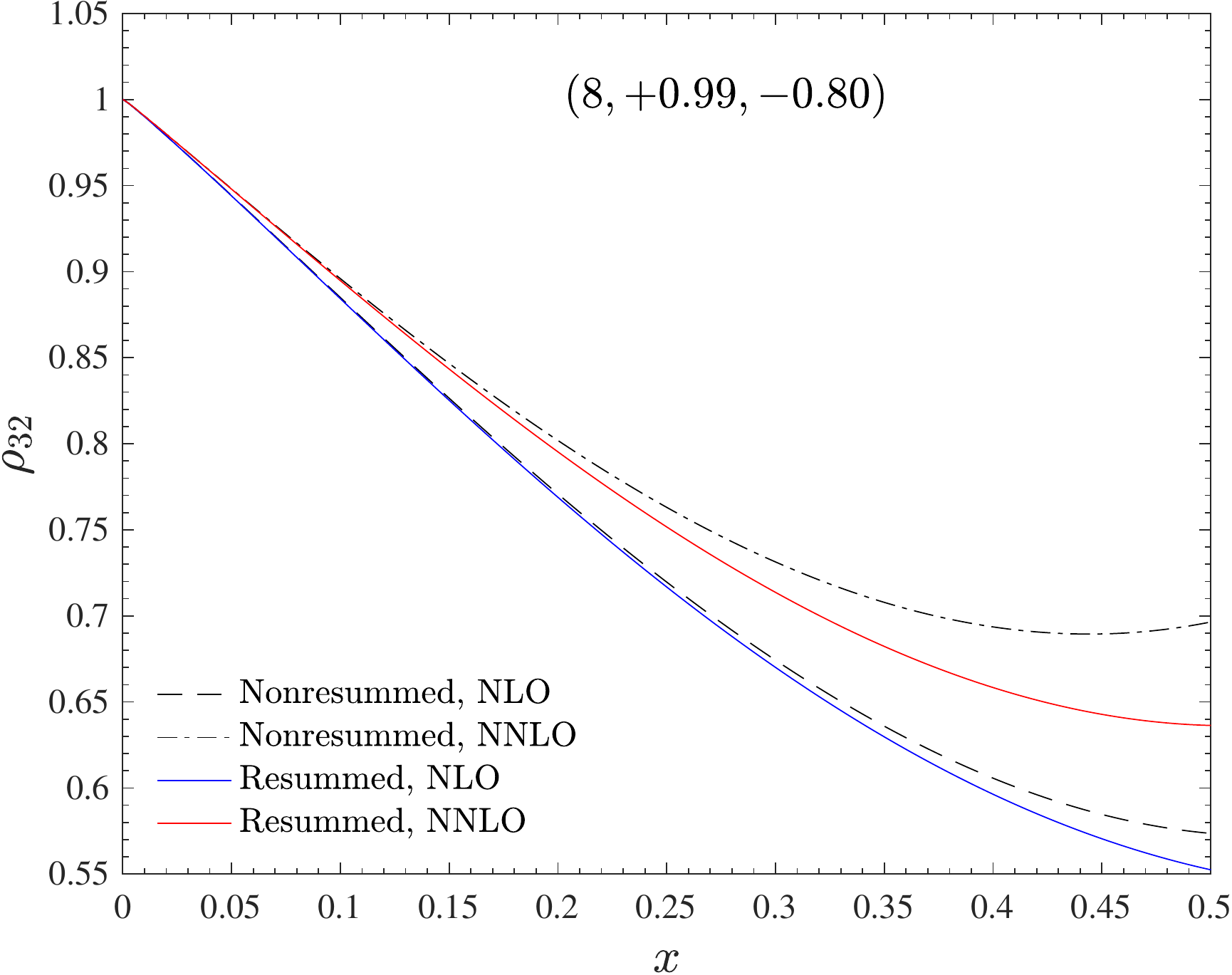}\\
     \vspace{2 mm}
        \includegraphics[width=0.29\textwidth]{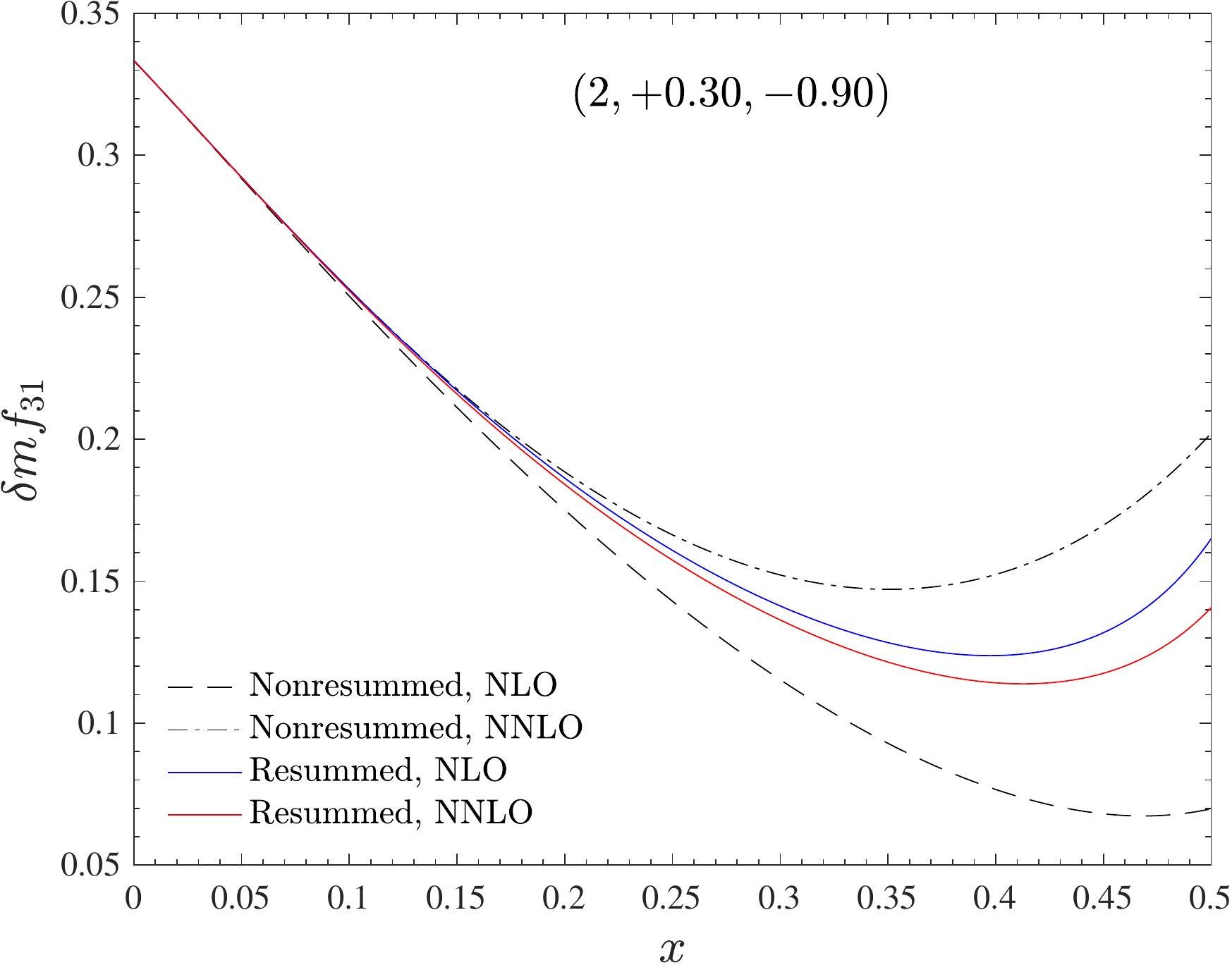}
         \hspace{1 mm}
  \includegraphics[width=0.29\textwidth]{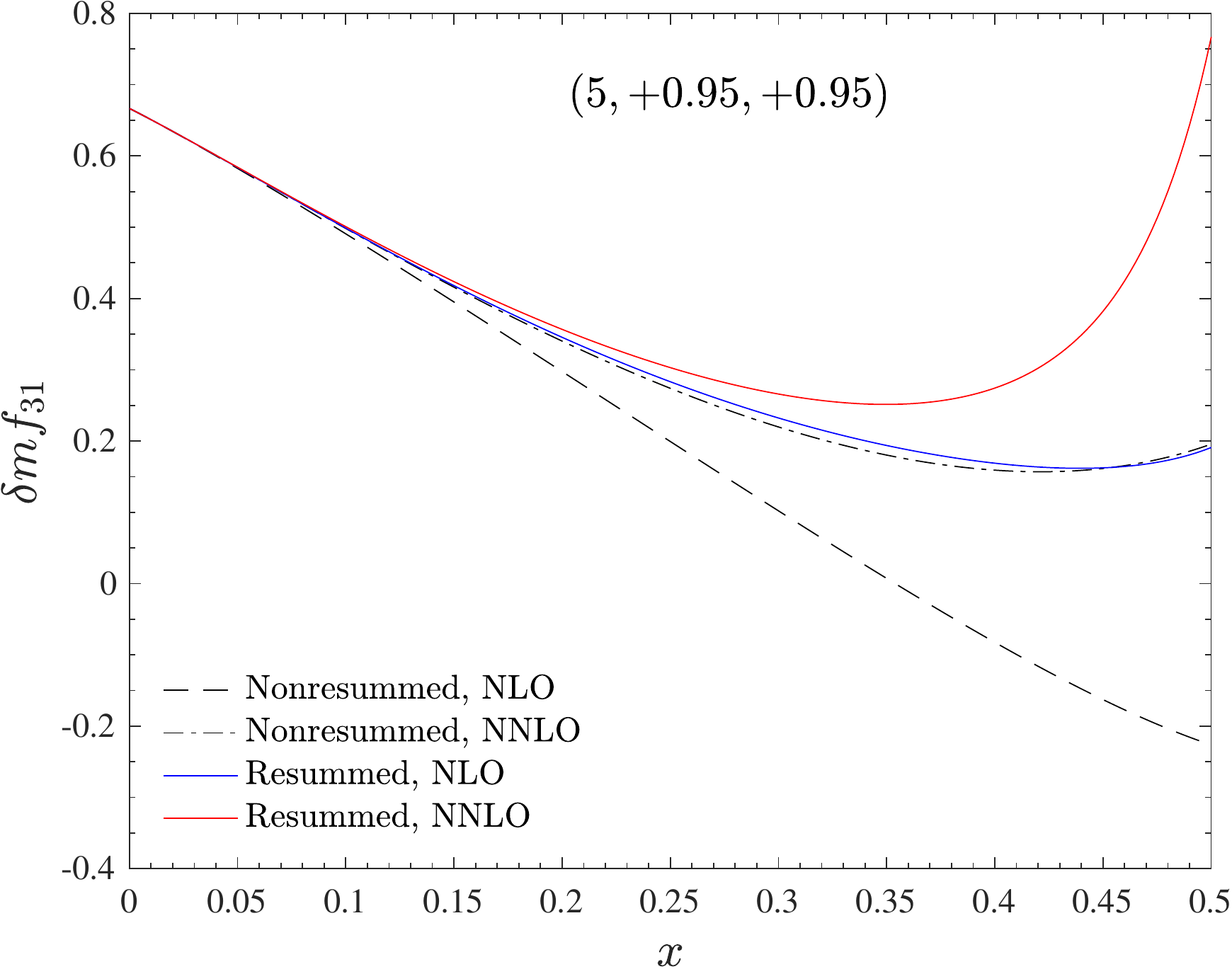}
   \hspace{1 mm}
   \includegraphics[width=0.29\textwidth]{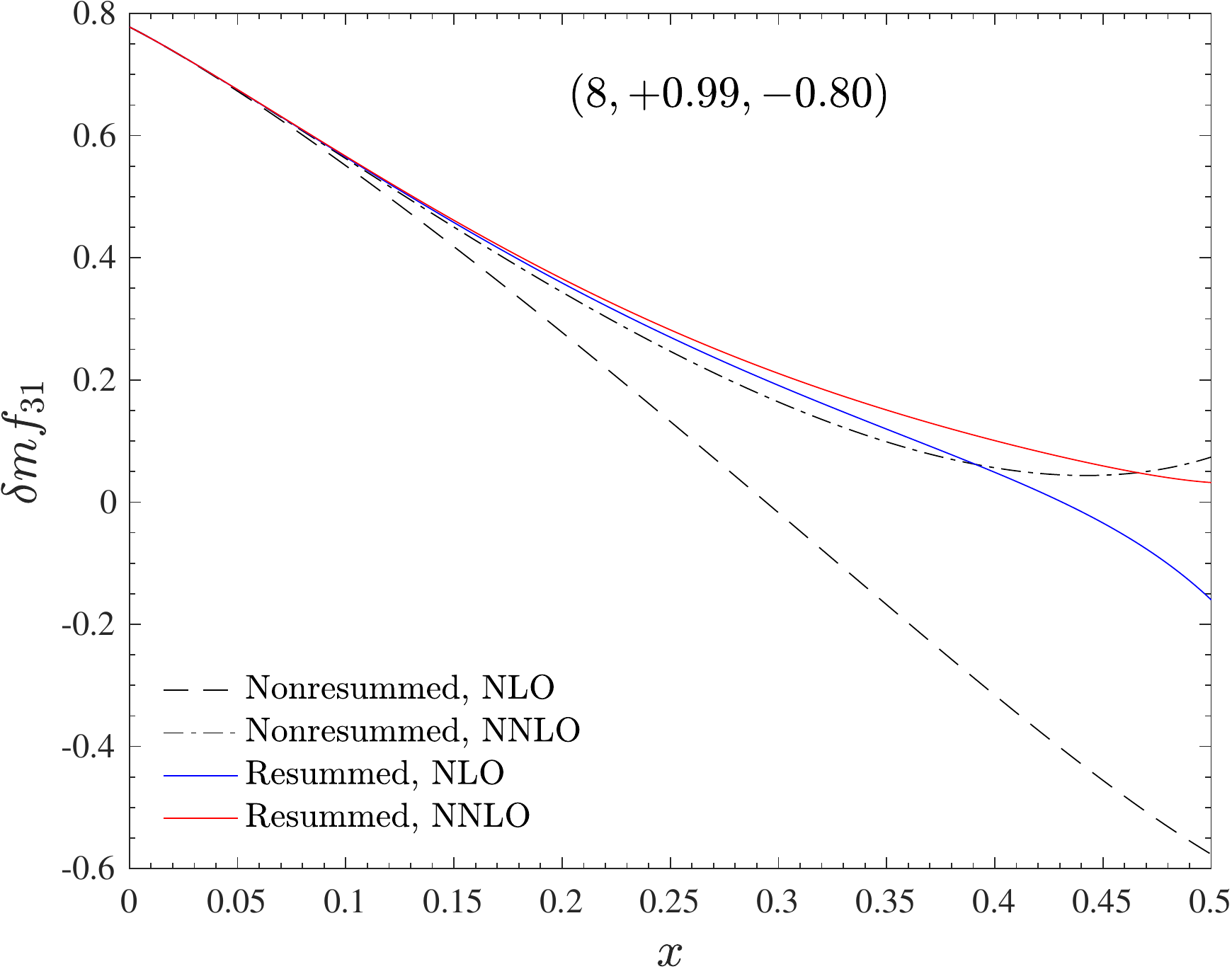}
   \caption{\label{fig:l3} Nonresummed (black) and resummed (colored) residual 
     waveform amplitudes for  $\ell=3$ multipoles. The orbital factors are taken
     at $3^{+3}$~PN relative accuracy except for the $(3,1)$ mode, that is taken
     at $3^{+2}$~PN level. Likewise the $\ell=2$ case of Fig.~\ref{fig:l2},
     the consistency between NLO and NNLO truncations of the spin terms is is
     improved by the resummation.}
\end{figure*}
We now proceed by resumming the orbital and spin factors according
to the prescriptions of Ref.~\cite{Nagar:2016ayt}, basically extending
to higher modes the treatment of the $\ell=2$ modes discussed there.
However, we want to have at least the orbital multipolar factors,
$\rho_{\ell m}^{\rm orb}$, consistent with the test-particle ones discussed
above, in order to take advantage of the high-order PN-information available
and of the robustness of its analytical representation in Pad\'e resummed form. 
To do so, we follow the, now standard, practice, originally suggested in 
Ref.~\cite{Damour:2008gu}, of {\it hybridizing} the  low-PN-order
$\nu$-dependent information available with the high-PN-order test-mass
($\nu=0$) one.
At the time of Ref.~\cite{Damour:2008gu}, the test-particle orbital
fluxes were analytically known up to 5.5PN order, which implied
that the, nonresummed, $\rho_{\ell m}$'s functions were available as 
polynomials of different order, that is $\rho_{22}^{\rm orb}=1+x+....+x^5$, 
$\rho_{21}^{\rm orb} = 1+x+ ....+ x^4$ etc., consistent with the global
5.5PN accuracy of the total flux. This prompted, at the time, the
construction of what was called the $3^{+2}$ PN approximation,
where the 3PN results were hybridized with two more test-particle
PN orders. As we saw above, the availability of PN results of high
order~\cite{Fujita:2012cm} allows us to keep more PN terms in
each $\rho_{\ell m}^{\rm orb}$'s, notably up to 6PN relative accuracy
for each $(\ell,m)$ as a good compromise between simplicity and accuracy.
Since we are working with relative PN truncations, we give here
the $3^{+3}$ PN approximation for the $\rho_{\ell m}(x;\nu)$ a
different meaning with respect to~\cite{Damour:2008gu}.
More precisely, working at $3^{+3}$PN order here means that each
$\rho_{\ell m}^{\rm orb}(x;\nu)$  carries the complete test-mass
information up to $x^6$, but whenever possible, the lower PN terms
are augmented by the corresponding $\nu$-dependent information
compatible with the $\nu$-dependent 3PN accuracy.
For example, $\rho_{22}^{\rm orb}(x;\nu)$ formally reads
\begin{align}
\rho_{22}^{\rm orb}(x;\nu)&=1+ c_1(\nu)x + c_2(\nu)x^2 + c_3(\nu;\log(x)) x^3 \nonumber\\
               &+ c^0_4(\log(x))x^4 + c^0_ 5(\log(x))x^5 \nonumber\\
               &+  c^0_ 6(\log(x),\log^2(x))x^6 ,
\end{align}
where $(c^0_4,c^0_5,c_6^0)$ are test-particle, $\nu$-independent,
coefficients with the corresponding dependence on $\log(x)$.
The function $\rho^{\rm orb}_{21}(x;\nu)$ shares the same analytical structure,
though the $\nu$-dependence of $c_3$ is currently uknown, since it is
a (global) 4PN effect. For higher modes, the $\nu$ dependence is progressively
reduced, up to only $c_1(\nu)$  for the $\ell=8$ modes~\cite{Damour:2008gu}.
Choosing the above defined $3^{+3}$~PN approximation also means that we adopt
the same Pad\'e resummation, multipole by multipole, detailed in
Table~\ref{tab:rhoLSO}. In this way we implement,
by construction, the consistency with the $\nu=0$ limit.
This choice opens the question of what would be the magnitude
of the systematic error done by neglecting such, yet-uncalculated,
$\nu$-dependent terms. Reference~\cite{Damour:2008gu} analyzed the $\nu$-dependence
of a few multipoles and concluded that, working with Taylor-expanded $\rho_{\ell m}^{\rm orb}$,
the $\nu$-dependence is mild and that the effect of the missing terms
is small enough to be considered of no importance.
We shall repeat and update that reasoning to our current choices
in the next section, though we anticipate the same conclusion
of~\cite{Damour:2008gu} remains essentially true here for all examined modes.

We turn now to discussing the resummation of the spinning factors, 
$\hat{\rho}_{\ell m}^{\rm S}$ and $\hat{f}_{\ell m}^{\rm S}$. We do so
by applying the resummation recipe of Ref.~\cite{Nagar:2016ayt},
that is: (i) for even-$m$, we  simply resum
$\hat{\rho}_{\ell m}^{\rm S}$ taking its inverse Taylor representation,
$\bar{\rho}_{\ell m}^{\rm S}(x;\,\nu)$, as in Eq.~\eqref{eq:iRrhoS};
(ii) for odd-$m$, we need to resum separately the 
two factors $\hat{f}^{\rm S_{(0)}}_{\lm}$ and $\hat{f}^{\rm S_{(1)}}_{\lm}$.
The analytical representation of the two factors we
choose depend on the multipole. More precisely: the factor
$\hat{f}_{21}^{\rm S_{(0)}}$ is  always resummed taking its inverse
Taylor representation. The same choice is also adopted to 
resum $\hat{f}_{\lm}^{\rm S_{(1)}}$ for $\ell=2$, but for $\ell\geq 3$, $m={\rm odd}$ case, the 
$\hat{f}_{\lm}^{\rm S_{(1)}}$ are kept 
in Taylor-expanded form because of the presence of spurious
poles when taking the inverse. The quality of the resummation is 
assessed in Figs.~\ref{fig:l2}  and~\ref{fig:l3} for a few 
illustrative binary configurations.
Since one does not have at hand the analogous of the  test-mass
numerical data for circularized, comparable-mass, binaries 
to compare with, our aim here is only limited to prove the 
{\it internal consistency} of the resummed  analytical expressions 
once taken at different PN orders.
To do so, by keeping the orbital part unchanged, we contrast the functions
obtained using the full NNLO information with the ones truncated at
NLO accuracy. The same figures also display the standard representation
of the $\rho_{\ell m}$'s, where no additional factorization or
resummation is adopted~\cite{Pan:2010hz}.
The plot illustrates how the spread between the NLO and NNLO
truncations in PN-expanded form is systematically much larger
than the corresponding one obtained with the factorized and
resummed functions. Interestingly, this conclusion remains
true for any configuration analyzed. This makes us conclude
that factorizing and resumming as discussed here is helpful
also in the comparable-mass case, although a precise
quantification of the improvement brought by this procedure
should be assessed through a comprehensive comparison
between an EOB model built from iResum waveforms and NR data,
in a way analogous, though more detailed,  to what briefly 
analyzed in~\cite{Nagar:2016ayt}. However, to better grasp
the meaning of this result, it is useful to remind the reader
that the merger of a binary black-hole coalescence (defined as the
peak of the $\ell=m=2$ waveform amplitude) will occur at
$x\approx 0.3$, with $x=(\omega_{22}/2)^{2/3}$ and $\omega_{22}$
the quadrupolar GW frequency\footnote{Let us recall that in
the test-particle limit this frequency approximately corresponds
to the crossing of the Schwarzschild light ring~\cite{Nagar:2006xv,Damour:2007xr}}.
As Figs.~\ref{fig:l2} and~\ref{fig:l3} illustrate, the improvement
in the consistency between PN truncations brought by the resummation
is evident precisely in a neighborhood of $0.3$. Just to pick some
random examples, this is the case for  $(1,+0.99,+0.99)$
and $(8,+0.50,0)$, configurations where the frequency parameter at
merger is $x\approx 0.38$ and $x\approx 0.32$ respectively~\cite{SXS:catalog}.

\subsection{Mild dependence of $\rho_{\ell m}^{\rm orb}$ to uncalculated $\nu$-dependent orbital terms}
\label{sec:rhoorb_nu}

As mentioned above, the idea of hybridizing test-mass orbital information with the
$\nu\neq 0$ one in the waveform amplitudes dates back to Ref.~\cite{Damour:2008gu}.
The rationale behind that choice was to show that the dependence on $\nu$ of
the coefficients in the $\rho_{\lm}^{\rm orb}$ is mild when $0\leq \nu\leq 1/4$,
so that one does not introduce a large systematical error in neglecting it.
To get to this conclusion, one was comparing the fractional variation of the
coefficients when $\nu$ is varied  between $0$ and $1/4$
(see Sec.~IVA in~\cite{Damour:2008gu}). Here we follow the same approach and
compute the fractional variation in $\nu$ for all multipoles up to $\ell=6$.
The 3PN-accurate $\nu$-dependent terms in $\rho^{\rm orb}_{31}$ and $\rho^{\rm orb}_{33}$
that were obtained only in Ref.~\cite{Faye:2014fra} are also included.
The $\log(x)$ terms are evaluated, for simplidicty at $x_{\rm LSO}^{\rm Schw}=1/6$.
The numbers listed in Table~\ref{tab:fracnu} suggest that, up to $\ell=3$, the
next missing $\nu$-dependent term might be, on average, of the order of $20\%$
larger (or smaller) then the test-mass ($\nu=0$) one (note however the larger variations
of the 2PN coefficient in $\rho_{31}^{\rm orb}$ and $\rho_{32}^{\rm orb}$.
One can then investigate the impact on $\rho_{\ell m}^{\rm orb}$ of missing $\nu$-dependent
corrections by varying the $\nu=0$ term by $\pm 30\%$. Clearly, the operation has to
be done on the $P^m_n(\rho^{\rm orb}_\lm)$ function. To be concrete on one case,
let us analyze the effect on $\rho_{21}^{\rm orb}$, whose known $\nu$-dependence stops at 2PN.
Schematically, the Taylor-expanded function reads
\begin{align}
&\rho_{21}^{\rm orb}=1 + (c^{\rm 1PN}_0 + c^{\rm 1PN}_\nu)x + (c^{\rm 2PN}_0 + c_\nu^{\rm 2PN})x^2 \nonumber\\
                 &+ \left[c^{\rm 3PN}_0(1 + \alpha)\right]x^3 + c_0^{\rm 4PN}x^4 + c_0^{\rm 5PN}x^5 + c_0^{\rm 6PN}x^6,
\end{align}
and then one takes its $P^5_1$ Pad\'e approximant. Here, $c_0^{n\rm PN}$ indicate the $\nu=0$
coefficients, while $c_\nu^{n\rm PN}$ the corresponding $\nu$-dependent terms. The effect
of the missing $\nu$-dependent information is parametrized through $\alpha$. One finds that,
even putting $\alpha=\pm 0.20$ with $\nu=1/4$, the fractional variation in
$P^5_1\left(\rho_{21}^{\rm orb}\right)$ is of the order of $0.04\%$ at the Schwarzschild
LSO $x_{\rm LSO}^{\rm Schw}=1/6$, of the order of $0.9\%$ at $x=1/3$ and of the order of $6\%$
at $x=1/2$. This value is close to the LSO location of a Kerr black hole with $\hat{a}=+0.99$
and we use here just for illustrative purposes, since a comparable mass binary, with a
nonnegligible value of $\nu$, is not expected to reach such a high frequency at merger.
Since the waveform amplitude is just $(\rho_{21})^2$, the fractional differences above get
a factor two in front, which suggests that, within the current framework, one is expecting
the 3PN correction to $\rho_{21}^{\rm orb}$ to yield an amplitude correction around merger
of just a few percents. Once the calculation of the waveform will be completed at 4PN
accuracy~\cite{Damour:2016abl,Marchand:2016vox,Damour:2017ced,Marchand:2017pir,Bernard:2017ktp}, it will be interesting
to concretely probe the reasonable
assumptions we are adopting here. In addition, inspection of the behavior of higher modes,
like $\rho_{44}^{\rm orb}$, shows that a variation of the order $20\%$ with respect to
the $\nu=0$ values has an unnaturally large effect on the global behavior of the function
in the strong-field regime ($0.3\lesssim x\lesssim 0.5$), with variations of order $8\%$
at $x=1/3$ and $\sim 30\%$ at $x=1/2$. Though we cannot make strong statements,
we are prone to think that the uncalculated $\nu$-dependent terms will provide,
on average, a correction of the order of $10\%$ to the current orbital terms,
consistently with the $\nu$-variation of the 3PN coefficients in
$\rho^{\rm orb}_{22}$ and $\rho_{33}^{\rm orb}$, as in Table~\ref{tab:fracnu}.

\begin{table}[h]
\caption{\label{tab:fracnu}Analysis of the fractional variation $\triangle c_n^{\rho_{\ell m}^{\rm orb}}(\nu)= c_n^{\rho_{\ell m}^{\rm orb}}(\nu)/ c_n^{\rho_{\ell m}^{\rm orb}}(0)-1$ of the coefficients $c_n^{\rho_{\ell m}^{\rm orb}}(\nu)$.}
\begin{ruledtabular}
\begin{tabular}{lccc}
$(\ell, m)$ & $\triangle c_1^{\rho_{\ell m}^{\rm orb}}(1/4)$ & $\triangle c_2^{\rho_{\ell m}^{\rm orb}}(1/4)$ &  $\triangle c_1^{\rho_{\ell m}^{\rm orb}}(1/4, \log(1/6))$\\
\hline
\hline
$(2,2)$ & -0.159884 & 0.185947 & -0.100421\\
$(2,1)$ & -0.649718 & 0.224005 & ...\\ 
$(3,1)$ & 0.0769231 & -18.7351 & -0.28487\\
$(3,2)$ & -0.155488 & -0.633264 & ...\\
$(3,3)$ & -0.142857 & 0.260344 & -0.0970255\\
$(4,1)$ & -0.0905316 & ... & ...\\
$(4,2)$ & -0.0181065 & 1.05237 & ...\\
$(4,3)$ & -0.141892 & ... & ...\\
$(4,4)$ & -0.230328 & 0.46265 & ...\\
$(5,1)$ & 0.219436 & ... & ...\\
$(5,2)$ & -0.117576 & ... & ...\\
$(5,3)$ & -0.0746667 & ... & ...\\
$(5,4)$ & -0.176295 & ... & ...\\
$(5,5)$ & -0.201232 & ... & ...\\
$(6,1)$ & -0.0910973 & ... & ...\\
$(6,2)$ & -0.0168919 & ... & ...\\
$(6,3)$ & -0.118343 & ... & ...\\
$(6,4)$ & -0.119186 & ... & ...\\
$(6,5)$ & -0.165766 & ... & ...\\
$(6,6)$ & -0.238208 & ... & ...\\
\end{tabular}
\end{ruledtabular}
\end{table}

\subsection{Hybridizing test-mass results: the spin information}
\label{sec:hyb-spin}
\begin{figure*}[t]
  \center
    \includegraphics[width=0.45\textwidth]{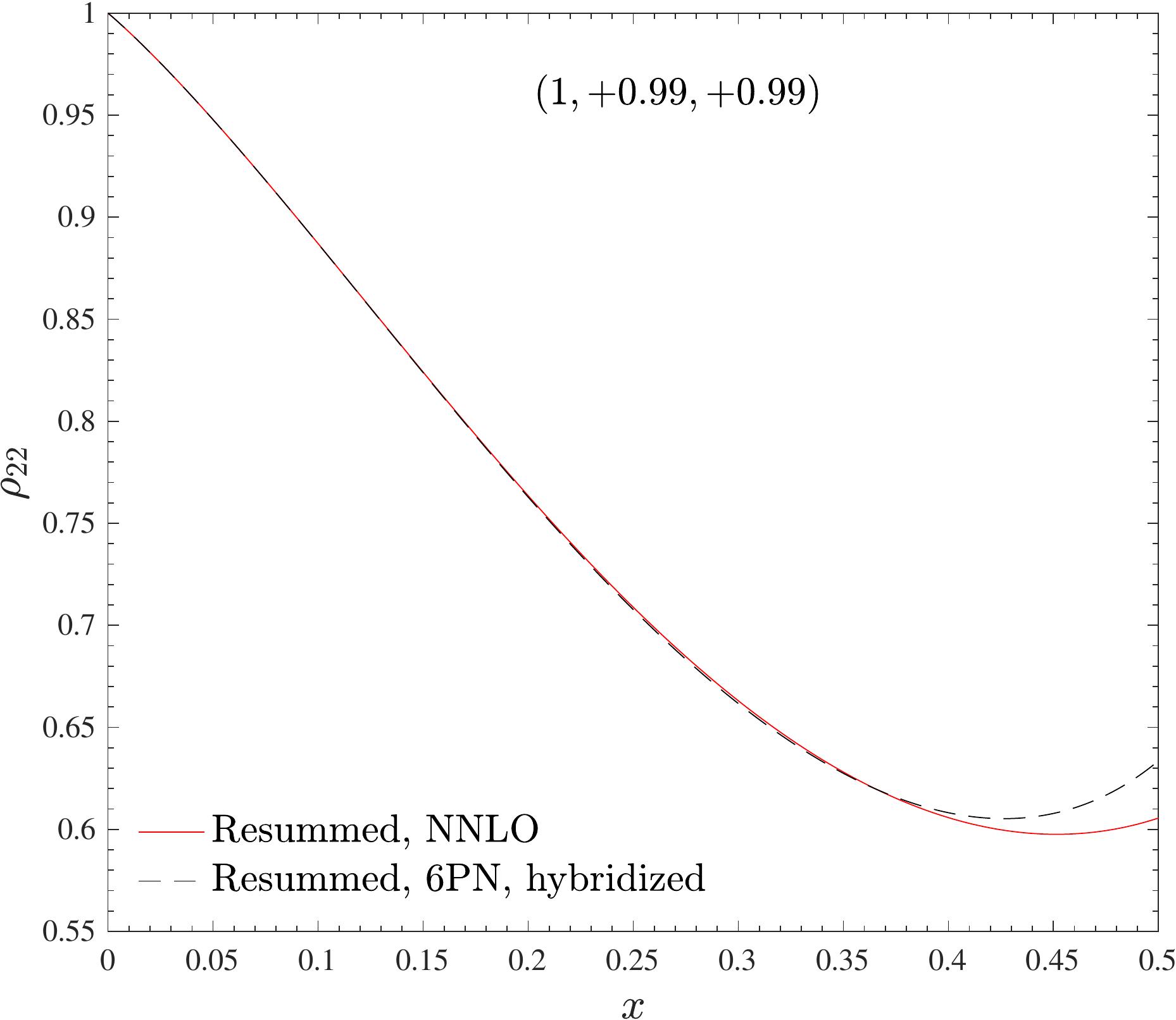}
     \hspace{1 mm}
  \includegraphics[width=0.45\textwidth]{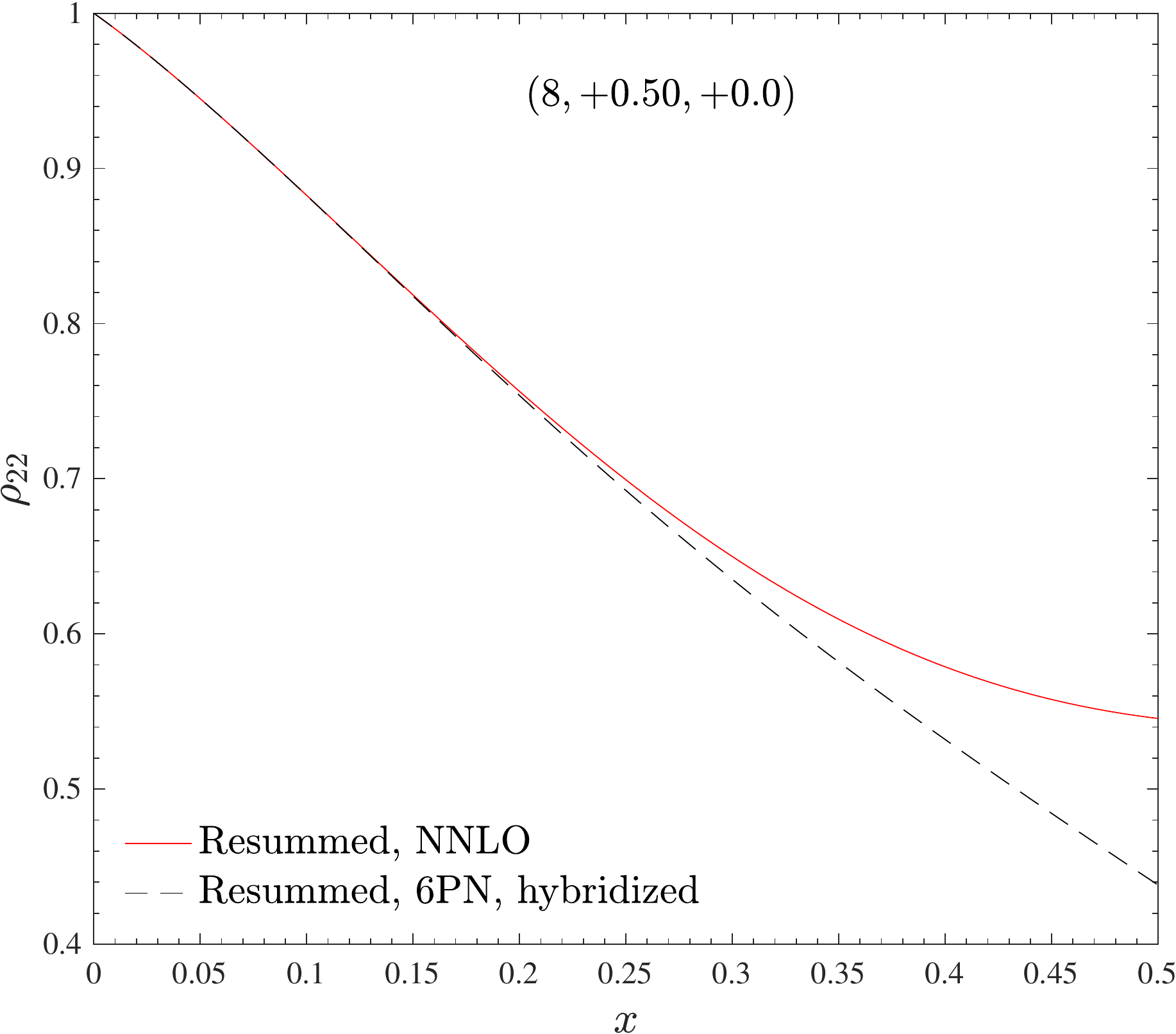}\\
  \caption{\label{fig:l2_hybrid} Modification of the results of Fig.~\ref{fig:l2}
    when (nonspinning) test-particle terms up to 6PN are hybridized with the
    NNLO $\nu$-dependent waveform (see Eqs.~\eqref{eq:c22_high}-\eqref{eq:c22SS_NNLO}
    and text). When $\nu=0$, one is using here the same spin-orbit and spin-square analytical
    information used in Fig.~\eqref{fig:rholms}. The frequency parameter approximately
    corresponding to the BBH merger is $x\approx 0.38$ for $(1,+0.99,+0.99)$ and $x\approx 0.32$
    for $(8,+0.50,0)$. In this latter case, the effect of the additional test-particle terms
    is important towards merger.}
\end{figure*}
Now that we have justified our approach of hybridizing $\nu=0$ and $\nu\neq 0$
information in the $\rho_\lm^{\rm orb}$ functions, one wonders whether an
analogous procedure exits for the $\rho^{\rm S}_{\ell m}$ (and in turn
for the factorized $\hat{\rho}_\lm^{\rm S}$).
This would allow to have EOB waveforms fully consistent and complete
all over the parameter space of nonprecessing BBHs~\footnote{Note this is
  not the case for current EOB waveform models, where the high-order test-mass
  analytical information is not incorporated~\cite{Bohe:2016gbl,Babak:2016tgq,Nagar:2017jdw}}.
Such hybridization is rather straightforward 
to do by taking advantage of the structure of the $\rho_\lm^{\rm S}$ and 
$f_\lm^{\rm S}$ functions and understanding how the
spinning test-particle limit builds up. This is especially evident using
the $\tilde{a}_i$ variables, that make the limit look apparent.
To explain the approach, let us first focus on the spin-orbit terms
entering the $\rho_{\ell m}$ functions. One sees that, at a given
$n$-PN order of the orbital part, the corresponding spin-orbit term
reads like
\begin{align}
  \label{eq:SO_npn}
&\Bigg(\hat{a}_0( c_0 + c_1 \nu + \dots+ c_{n-1} \nu^{n-1}) \nonumber\\
&+ \tilde{a}_{12}X_{12} (d_0 + d_1 \nu + \dots +d_{n-1} \nu^{n-1})\Bigg)x^{(2n+1)/2},
\end{align}
as it is clear from Eq.~\eqref{eq:rho22S} that corresponds to 3PN orbital
dynamics. Note that within our writing, the LO spin-orbit coefficients are
$\nu$-independent, the NLO are linear in $\nu$, while the NNLO are quadratic in $\nu$.
A similar, though slightly more complicated, structure is found for the
quadratic-in-spin terms, that, e.g. for $\rho_{22}^{\rm S}$, are given
as the sum of terms proportional to $\hat{a}_0^2$, $\tilde{a}_{1}\tilde{a}_2$
and $\hat{a}_0\tilde{a}_{12}X_{12}$. As for the SO case above, at LO there
is no $\nu$ dependence, while it is similarly linear-in-$\nu$ at NLO. 
The $\nu$ independent terms in Eq.~\eqref{eq:rho22S} are those 
that, combined together when $\nu\rightarrow 0$, generate the 
(spinning) test particle results, 
see e.g. Refs.~\cite{Tanaka:1996ht,Fujita:2014eta}.
Understood this, one can implement the inverse process, namely incorporate
the $\rho^{\rm S}_\lm$'s (and $f_\lm^{\rm S}$) obtained from the perturbative 
calculations of the fluxes of a spinning particle around a Kerr black hole by
imposing the structure given by Eq.~\eqref{eq:SO_npn}. This means, in particular,
replacing the dimensionelss Kerr spin  as $\hat{a}\rightarrow\tilde{a}_1$ and
the particle spin as $\sigma\rightarrow \tilde{a}_2$.
In other words, on the $\nu$-dependent side, the next-to-next-to-next-to-leading-order (NNNLO)
spin-orbit term will have the form
\be
\label{eq:C22SO_NNNLO}
c_{22}^{\rm SO^{\rm NNNLO}_\nu}=\left(c_0^{\rm NNNLO} \hat{a}_0 + X_{12}\tilde{a}_{12}d_0^{\rm NNNLO}\right)x^{9/2},
\ee
where $\left(c_0^{\rm NNNLO},d_0^{\rm NNNLO}\right)$ are the unknown $\nu$-independent, coefficients.
On the $\nu=0$ side, the corresponding spin-orbit terms reads
\be
\label{eq:c22SO}
c_{22}^{\rm SO_{0}^{\rm 4.5PN}}=\left( c_{\hat{a}}^{\rm 4.5PN} \hat{a} + c_{\sigma}^{\rm 4.5PN}\sigma\right)x^{9/2}.
\ee
By equationg the $\nu=0$ limit of Eq.~\eqref{eq:c22SO} to this equation, one finds 
\begin{align}
\label{eq:c22_high}
c_{22}^{{\rm SO}^{\rm NNNLO}_\nu }&= \Bigg(\dfrac{c_\ha^{\rm 4.5PN}+c_\sigma^{\rm 4.5PN}}{2}\hat{a}_0 \nonumber\\
                                                           &\qquad+ X_{12}\tilde{a}_{12}\dfrac{c_\ha^{\rm 4.5PN}-c_\sigma^{\rm 4.5PN}}{2}\Bigg)x^{9/2},
\end{align}
where $c_\ha^{\rm 4.5PN}$ is analytically known~\cite{Fujita:2014eta} and reads
\be
c_\ha^{\rm 4.5PN}=-\dfrac{8494939}{467775}+\dfrac{2536}{315}{\rm eulerlog}_2(x),
\ee 
while the spinning-particle term, $c_\sigma^{\rm 4.5 PN}$, is currently unknown. The same procedure can be applied
to incorporate spinning-particle spin-orbit terms at higher-PN order and can be extended to the other multipoles,
with the obvious difference that for $m$-odd multipoles the hybridization procedure applies to the $f_\lm^{\rm S}$
functions. The hybridization of the spinning-particle, spin-square terms into $\rho_{22}^{\rm S}$ is done in a similar
way. A similar calculation for the  NNLO ( relative 4PN-accuracy) spin-square term yields
\begin{align}
  \label{eq:c22SS_NNLO}
c_{22}^{\rm SS^{NNLO}_\nu}&=\Bigg[\dfrac{1}{2}\left(c_{\hat{a}^2}+c_{\sigma^2}\right)\ha_0^2+\left(c_{\ha\sigma}-c_{\ha^2}-c_{\sigma^2}\right)\tilde{a}_1\tilde{a}_2\nonumber\\
                                               &+\dfrac{1}{2}\left(c_{\ha^2}-c_{{\sigma}^2} \right)\hat{a}_0\tilde{a}_{12}X_{12}\Bigg]x^4,
\end{align}                            
where $(c_{\sigma^2},c_{\ha\sigma},c_{\sigma^2})$ are the coefficients entering the spin-square spinning-particle
term at 4PN, that will have the structure
\be
c_{22}^{\rm SS_0}=\left(c_{\ha^2}\ha^2 + c_{\ha\sigma}\ha\sigma + c_{\sigma^2}\sigma^2\right)x^4,
\ee
where only $c_{\ha^2}$ is currently analytically known and reads~\cite{Fujita:2014eta}
\be
c_{\ha^2} = \dfrac{18353}{21168}.
\ee
This approach gives us a consistent way of hybridizing the test-mass result above with the low-PN
$\nu$-dependent information. Even if the spinning-particle terms are not currently published
starting from 4PN order, it could be instructive to investigate the robustness of the results
of Fig.~\ref{fig:l2} under the incorporation of the {\it nonspinning} test-particle terms.
To do so, we replicate the procedure done above to incorporate the 4PN and 4.5PN test-particle
terms for the spin-square 5PN and 6PN terms as well as for the 5.5PN spin-orbit term,
that will have the same relation given in Eqs.~\eqref{eq:C22SO_NNNLO}-\eqref{eq:c22SS_NNLO}
with the test-particle coefficients. After this is done, we factorize and resum the
hybrid $\rho_{22}$ as before. Such,
test-particle-improved, $\bar{\rho}^{\rm S}_{22}$ is consistent with the $\nu=0$ function
discussed in Sec.~\ref{sec:iRlm} except for the obvious absence of the spin-cube terms
coming beyond the NLO as well as of the terms involving higher powers of the spins up to
the sixth-power, that enters at LO in the 6PN term. The effect of the additional
$\nu=0$ terms is illustrated in Fig.~\ref{fig:l2_hybrid} for $(1,+0.99,+0.99)$
and $(8,+0.50,0)$, where the hybridized function is contrasted with the NNLO one
of Fig.~\ref{fig:l2}. The figure shows that the effect is quantitatively important
for the case $(8,+0.50,0)$, notably towards the merger freuquency $x\approx 0.32$. 
We postpone to future work a deeper study of the effect of the $\nu=0$ terms
all over the parameter space and of their importance in comparisons with NR waveform
data. Such study will be performed by including {\it also} some of the higher-order
spin-orbit terms for a circularized spinning particle that are currently not available
in the literature and that are being calculated~\cite{Kavanagh:2018}.

\section{Conclusions}
\label{sec:conclusions}
In this paper we have improved and generalized the factorization and resummation procedure
of waveform amplitudes introduced in Ref.~\cite{Nagar:2016ayt}. The key conceptual step
of the approach relies on factorizing the orbital and the spin-dependence into two separate
factors that can then be resummed separately in various ways.
Our results can be summarized as follows:
\begin{itemize}
\item[(i)] Concerning a circularized, (nonspinning) particle orbiting a Kerr black hole,
  we have shown that the (relative) 6PN-accurate $\rho_\lm$ functions can be factorized
  and resummed in a form that yields a more than satisfactory agreement (of the order
  of a few $\%$) with the corresponding numerical (exact) function up to the last-stable-orbit.
  This is notably true for the case of a quasi-extremal black hole with dimensionless
  spin parameter $\hat{a}=+0.99$. One of the novelties with respect to previous
  work~\cite{Nagar:2016ayt} is that the 6PN-accurate orbital function $\rho_{22}^{\rm org}$
  is resummed with a Pad\'e approximant (typically $P^4_2$).
  The same recipe proved to work essentially the same way for all subdominant modes
  up to $\ell=6$, modulo a few exceptions where working at either higher or lower
  PN-information proved a better choice~\ref{tab:rhoLSO}). More concretely, the
  factorization-resummation procedure allows us to obtain an analytic flux, summed over
  all multipoles up to $\ell=6$, that is consistent at the $5\%$ fractional level
  up to the LSO, for the most demanding case of $\hat{a}=+0.99$. This result is
  accomplished relying only on purely analytical information, without any
  additional fit to the numerical fluxes. We recall that this route was followed instead
  in Ref.~\cite{Taracchini:2013wfa} , where several higher-PN terms, unknown at the time,
  were calibrated to the same Teukolsky data we are using here. The fits were able to
  improve the standard $\rho_\lm$-nonresummed flux so to have a fractional difference
  at the LSO of the order (or below) $1\%$. We have briefly illustrated
  (for the $\ell=m=2$ mode only, see end of Sec.~\ref{sec:iRlm})
  that an analogous route can be followed also in our framework and that it is easy
  to reach analytical/numerical fractional differences in $\rho_{22}$ of the order of
  $0.1\%$ at the LSO for $\hat{a}=+0.99$ by just choosing one effective parameter
  entering at 6.5PN order in the resummed spin factor $\bar{\rho}_{22}^{\rm S}$.
  Alternatively, we want to remind that we have explored only a few of
  the many possibile choices. Focusing on the $(2,2)$ mode only for
  definitess, the logic driving our approach is: first (i), to simplify
  things we choose to keep the same PN-order for both the orbital and
  PN factors; then (ii), as a similarly simple choice to reduce the growth
  of the  spin factor in the strong-field regime (see discussion in
  Ref.~\cite{Nagar:2016ayt}), we resummed it with its inverse-Taylor approximant;
  (iii) finally, we found that a good match with the exact numerical
  data was found by taking the $P^4_2$ Pad\'e approximant of the orbital factor. 
  Once the factorization paradigm is accepted, any of the points (i)-(iii) above could
  be, in principle, changed. For example, for $(3,1)$ we found that
  the numerical/analytical agreement gets improved by keeping the spin factor at
  8PN accuracy and the orbital factor just at 4PN accuracy resummed
  with a $(3,1)$ Pad\'e approximant.
  Similarly, for some multipoles like $(3,3)$ and $(4,4)$ things are such that
  the straight, Taylor-expanded, form of the orbital factor yields a better
  agreement with the numerical results. These facts suggest that it
  might be possible that there are some special combinations of Pad\'e approximant
  for the orbital part and PN-truncation of the spin factor that could further reduce
  the analytical/numerical disagreement in the near-LSO regime. Seen the large amount 
  of PN-knowledge that is available (up to 20PN~\cite{Shah:2014tka}),
  this would require a specific, dedicated study
  depending on the spin regime where one would like to use the resummed flux.
  For example, the resummed analytical fluxes we present here could be use to
  improve the radiation reaction force used to drive the quasi-circular transition
  from inspiral to plunge in the test-particle limit~\cite{Harms:2014dqa}.
  Several studies showed the limits of the standard, nonresummed, analytical
  approach at 5PN~\cite{Harms:2014dqa} and proposed to improve it with
  effective fits~\cite{Taracchini:2013wfa}. The approach we present here
  offers an alternative, though slightly less accurate, to the effective
  fits of~\cite{Taracchini:2013wfa}. Whether such difference is relevent
  or not will depend on the specific configuration and/or the problem
  under consideration. For example, it will be interesting to investigate
  to which extent the perturbative calculation of the recoil velocity
  of Ref.~\cite{Harms:2014dqa} can be improved by the use of the new
  radiation reaction presented here, in particular when the central
  Kerr black hole is quasi-extremal with spin aligned with the orbital
  angular momentum.
  
\item[(ii)]. We have extended the factorization and resummation procedure
  to all the existing $\nu$-dependent spin terms. This means that we go
  up to NNLO in the spin-orbit coupling, up to NLO in the spin-spin coupling
  and up to LO in the spin-cube coupling. This is done consistently for
  all multipoles currently known above the LO contributions ($\ell=4$).
  In doing so, we propose to use the orbital part of the waveform in hybridized
  form, where currently known, $\nu$-dependent
  orbital terms are hybridized with the test-mass term, as proposed long ago
  in~\cite{Damour:2008gu}. The novelty here is that, to maintain the consistency
  with the choices done in the test-particle limit, each $\rho_{\lm}$ is kept up to relative
  6PN order (but a few exceptions), with the Pad\'e-approximant that was chosen
  in the test-particle limit, however maintaining the full $\nu$-dependence that is currently
  available in the low-order terms. By contrast, as a first choice, the spin-dependent factor
  is not hybridized with high-order test-particle results, but it is inverse resummed
  at the currently available $\nu$-dependent PN order. We have explored the robustness of
  this choice on an indicative sample of binary configurations,
  contrasting the resummed amplitude with the plain, Taylor-expanded, $\rho_\lm$.
  Since we do not have circularized comparable-mass BH data to compare with,
  the only effect that we could investigate for is the {\it consistency}
  between NNLO and NLO truncations of the waveform, as an indication of the
  analytical robustness of the resummed expressions.
  Our Figs.~\ref{fig:l2} (for $\ell=2$) and~\ref{fig:l3} (for $\ell=3$)
  show that, for the same binary, differences that are large for the
  Taylor-expanded $\rho_\lm$ or $\delta m f_\lm$ are either very much
  reduced, or practically negligible, in the resummed representation
  of the same functions. This effect is very striking on $\delta m f_{21}$,
  where not only one can see this effect, but the function is also qualitatively
  close to the numerical one (see for example Fig.~5 of Ref.~\cite{Nagar:2016ayt}. 
 
  These findings suggest that the resummed waveform amplitudes should be incorporated
  within the EOB approach as a new, state-of-the-art, analytical waveform paradigm.
  This was pointed out  already in Ref.~\cite{Nagar:2016ayt}, but here we reinforce
  that statement after a deeper and more detailed analysis. In particular, we expect
  that next-to-quasi-circular corrections~\cite{Damour:2007xr} to the waveforms will
  generically have a smaller impact than in current EOB models~\cite{Nagar:2017jdw},
  because they will hopefully have to bring just small corrections to the already good
  strong-field behavior of the analytical waveform. This was briefly pointed out
  already in Ref.~\cite{Nagar:2017jdw}, but we are planning to investigate this
  extensively in future work.

 (iii) Following Ref.~\cite{Nagar:2016ayt}, we wrote all spin-dependent expressions using as spin
  variables the Kerr parameters of the individual black holes divided by the total mass of
  the system, $\tilde{a_i}\equiv a_i/M=S_i/(m_i M)$. The use of this quantities to parametrize
  a spin-dependent function was already suggested in Ref.~\cite{Nagar:2015xqa} in the context
 of informing a next-to-next-to-next-to-leading order spin-orbit effective parameter using NR simulations;
 similarly, the same spin variables allow for a simple recasting of the NLO correction to the centrifugal radius,
 that has rather complicated coefficients when written using the dimensionless spin variables $\chi_i = S_i/m_i^2$,
 see Eq.~(58)-(65) of~\cite{Nagar:2015xqa}. When using $\tilde{a}_i$, Eq.~(58) of~\cite{Nagar:2015xqa} reduces
 to the following very compact form
 \begin{align}
   \delta a^2&=\dfrac{M}{r^3}\bigg\{\dfrac{5}{4}\tilde{a}_{12}\hat{a}_0X_{12}-\left(\dfrac{5}{4}+\dfrac{\nu}{2}\right)\hat{a}_0^2\nonumber\\
             &\qquad\quad+\left(\dfrac{1}{2}+2\nu\right)\tilde{a}_1\tilde{a}_2\bigg\}.
 \end{align}
 The use of the $\tilde{a}_i$'s in our context, on top of providing similar simplifications
 in writing the formulas, is extremely convenient since these variables are natuarally
 connected to the (spinning) test-particle limit, that can be obtained straightforwardly
 by just putting $\nu=0$ in the equations. On top of this, since our analytical writing of
 the fluxes makes absolutely transparent
 which terms combine to generate the (spinning) test-particle limit, it is technically
 clear how to hybridize the $\nu=0$ information with $\nu\neq 0$ one {\it also} in the
 presence of spin, in order to have a waveform model that is fully consistent with the
 test-mass results discussed at point (i) above. This analysis suggests then that, since
 the structure of the expansion of the functions $\rho_{22}^{\rm S}$ (or $\hat{\rho}_{22}^{\rm S}$)
 is clear, one can have access to the leading-order, $\nu$-independent terms, by
 using perturbative, spinning, test-particle analytical calculations~\cite{Tanaka:1996ht,Fujita:2014eta}
 and then promoting the BH dimensionless spin $\hat{a}=S_1/m_1^2$ to $\tilde{a}_1$
 and the spin of the particle $\sigma=S_2/(m_1 m_2)$ to $\tilde{a}_2$, though keeping
 the additional constraint that a special structure with $X_{12}\tilde{a}_{12}$ exists
 in the $\nu$-dependent case. This constraint implies that the $\nu$-independent terms
 entering $\rho_{\ell m}(x;\nu)$ are  obtained as linear combinations of the {\it spinning and
 nonspinning} test-particle perturbative results. Unfortunately, since the current accuracy of
 the fluxes of a spinning-particle obtained using perturbative calculations is still at
 2.5~PN order~\cite{Tanaka:1996ht}, currently we can only rely on nonspinning test-particle
 perturbative to explore the effect of the hybridization. As a preliminary analysis,
 focusing only on the $(2,2)$ mode,  we hybridized the (nonspinning) particle ($\nu=0$)
 spin-orbit and spin-spin analytical information on a Kerr black hole up to 6PN with the
 $\nu$-dependent analytical pieces up to NNLO in the spin-orbit coupling (i.e., 3.5PN).
 The modifications we found for large values of $x$ are quantitatively more important
 when the mass-ratio is large than for the equal-mass case, see Fig.~\eqref{fig:l2_hybrid}.
 However, since the spinning-particle information is not incorporated,
 since $c_\sigma=c_{\sigma^2}=c_{\hat{a}\sigma}=0$ for all higher PN orders considered, this result should
 be taken only as illustrative of the effect and of the general strategy that might be
 used to take into account spinning $\nu=0$ information. Deeper analytical explorations
 are necessary to understand whether the test-mass information should be incorporated
 in this form in current EOB waveform models and whether it is of any help/relevance
 for LIGO/Virgo sources. In this respect it will be extremely useful to have perturbative
 analytical calculation of the energy fluxes emitted by a spinning particle on circular
 orbits around a Schwarzschild (or even a Kerr) black hole, a work that is currently
 in progress~\cite{Kavanagh:2018}. In the former case, for instance restricted to the
 simpler case of working at linear order in the particle's spin, we could already
 complement the current $\nu=0$ spin-orbit knowledge, and possibly improve, in a more
 consistent way, the analysis sketched in Fig.~\ref{fig:l2_hybrid}. In addition, the
 fluxes of a spinning particle on Kerr would give us access to some leading-order
 $\tilde{a}_1\tilde{a}_2$ terms. Though stating whether such test-particle information
 will have any important impact on LIGO/Virgo targeted waveform models requires deeper
 investigations, it will certainly allow us to improve their self-consistency all
 over the binary parameter space.
\end{itemize}

\acknowledgements
We are grateful to A.~Boh\'e and S.~Marsat for sharing with us the result of their
PN calculation of the multipolar waveform and fluxes before publication. We especially
acknowledge discussions with S.~Marsat and help in providing cross checks of our
analytical expressions. We also thank S.~Hughes for providing us with the numerical
fluxes used to compute $(\rho_\lm,f_\lm)$ in the test-particle limit. We are also
indebted to T.~Damour for discussions and constructive criticisms on the manuscript
and to C.~Kavanagh for sharing with us preliminary calculation of the fluxes of
a spinning particle on Schwarzschild. F.~M. thanks IHES for hospitality during
the final stages of development of this work.

\appendix
  \section{Multipolar fluxes}
  In this Appendix we explicitly report, for completeness, the PN-expanded, complete, Newton-normalized multipoles of the energy flux up to 
  NNLO in the spin-orbit coupling, NLO in the spin-spin coupling and LO in the spin-cube couplings. Though these expressions are obtained 
  as the square of the Newton-normalized waveform multipoles of Eqs.~\eqref{eq:h22}-\eqref{eq:h41}, it is convenient to have them written
  down explicitly.  Each multipolar contribution to the flux is written as the product of the Newtonian prefactor $F_{\ell m}^{N}$ and the PN 
  correction  $\hat{F}_{\ell m}$ as
   \be
  F_{\ell m}\equiv F_{\ell m}^{(N,\epsilon)} \hat{F}_{\ell m}^{(\epsilon)},
  \ee
  where the PN correction factors explicitly read
  \begin{widetext}
\begin{align}
  \hat{F}_{22}^{(0)}&=1+\left(-2\ha_0+4\pi-\frac{2}{3}\ta_{12} X_{12}\right)x^{3/2}+\left(-\frac{107}{21}+\frac{55}{21}\nu\right)x
  +\left(\frac{4784}{1323}+2\ha_0^2 -\frac{87691}{5292}\nu+\frac{5851}{1323}\nu^2\right)x^2\nonumber\\
&+\biggl[\left(-\frac{428}{21}+\frac{178}{21}\nu\right)\pi +\ha_0\left(\frac{158}{63}-\frac{257}{63}\nu\right)-\ta_{12} X_{12}\left(\frac{50}{63}+\frac{187}{63}\nu\right)\biggr]x^{5/2}\nonumber\\
&+\biggl[\frac{99210071}{1091475}+\frac{1650941}{349272}\nu-\frac{669017}{19404}\nu^2+\frac{255110}{43659}\nu^3+\left(\frac{16}{3}+\frac{41}{48}\nu\right)\pi^2\nonumber\\
&-\left(8\ha_0 +\frac{8}{3}\ta_{12} X_{12}\right)\pi+\frac{257}{63} \nu  \left(\to^2+\frac{962 \to \tt}{257}+\to^2\right)+\frac{284}{63}X_{12}
   \left(\to^2-\tt^2\right)\nonumber\\
   &-\frac{53}{7} \left(\to^2+\frac{1072 \to
   \tt}{477}+\tt^2\right)-\frac{1712}{105}{\rm eulerlog}_2(x)\biggr]x^3\nonumber\\
&+\biggl[\left(\frac{19136}{1323}-\frac{144449}{2646}\nu+\frac{33389}{2646}\nu^2\right)\pi+\ha_0\left(\frac{29234}{1323}+\frac{77212}{1323}\nu+\frac{1747}{1323}\nu^2\right)\nonumber\\
&+\ta_{12} X_{12}\left(\frac{37858}{3969}+\frac{96575}{7938}\nu-\frac{23182}{3969}\nu^2\right)-\frac{2}{3}(\ha_0^3 +3\ha_0^2\ta_{12} X_{12})\biggr]x^{7/2},
\end{align}

\begin{align}
\hat{F}_{21}^{(1)}&=1-4\nu-3\ta_{12} X_{12}x^{1/2}+\left(-\frac{17}{14}+\frac{44}{7}\nu-\frac{40}{7}\nu^2+\frac{9}{4}\ta_{12}^2\right)x+\biggl[2\pi(1-4\nu)+\ha_0\left(-\frac{43}{7}+\frac{172}{7}\nu\right)\nonumber\\
&+\ta_{12} X_{12}\left(\frac{195}{28}+\frac{18}{7}\nu\right)\biggr]x^{3/2}+\biggl\lbrace-\frac{2215}{7056}-\frac{13567}{1764}\nu+\frac{65687}{1764}\nu^2-\frac{853}{147}\nu^3\nonumber\\
&-\frac{26}{7}\left(\to^2-\frac{68}{13}\to\tt+\tt^2\right)-\frac{323}{14}\left(\to^2+\frac{26}{323}\to\tt+\tt^2\right)\nu+ X_{12}\biggl[\frac{157}{14}(\to^2-\tt^2)-6\pi\ta_{12}\biggr]\biggr\rbrace x^2\nonumber\\
&+\biggl\lbrace\left(\frac{9}{2}\ta_{12}^2-\frac{17}{7}+\frac{81}{7}\nu-\frac{52}{7}\nu^2\right)\pi+\ha_0\left(\frac{2131}{882}-\frac{1165}{147}\nu-\frac{3068}{441}\nu^2\right)\nonumber\\
&-\ta_{12} X_{12}\left(\frac{10121}{1764}-\frac{30595}{882}\nu-\frac{16927}{3528}\nu^2\right)+3\ha_0\ta_{12}^2+\frac{3}{2}X_{12}[5(\to^3-\tt^3)+\to(\to\tt-\tt^2)]\biggr\rbrace x^{5/2},
\end{align}

\begin{align}
\hat{F}_{31}^{(0)}&=1-4\nu+\left(-\frac{16}{3}+20\nu+\frac{16}{3}\nu^2\right)x+\biggl[2\pi(1-4\nu)+\ha_0\left(\frac{1}{2}-2\nu\right)+\ta_{12} X_{12}\left(-\frac{9}{2}+13\nu\right)\biggr]x^{3/2}\nonumber\\
&+\biggl[\frac{437}{33}-\frac{5164}{99}\nu-\frac{523}{99}\nu^2+\frac{812}{99}\nu^3+3\ha_{0}^2(1-4\nu)-8X_{12}(\to^2-\tt^2)\biggr]x^2\nonumber\\
&+\biggl[\left(-\frac{32}{3}+39\nu+\frac{44}{3}\nu^2\right)\pi+\ha_0\left(-\frac{103}{18}+\frac{283}{6}\nu-\frac{874}{9}\nu^2\right)+\ta_{12} X_{12}\left(\frac{145}{6}-\frac{1211}{18}\nu-\frac{41}{3}\nu^2\right)\biggr]x^{5/2},
\end{align}

\begin{align}
\hat{F}_{32}^{(1)}&=1+\frac{1}{1-3\nu}\biggl\lbrace 2(\ha_0 -\ta_{12} X_{12})x^{1/2}+\left(-\frac{193}{45}+\frac{145}{9}\nu-\frac{73}{9}\nu^2\right)x\nonumber\\
&+\frac{1}{(1-3\nu)}\biggl[4\pi(1-3\nu)^2+\ha_0\left(-\frac{778}{45}+\frac{715}{9}\nu-\frac{730}{9}\nu^2\right)+\ta_{12} X_{12}\left(\frac{538}{45}-\frac{313}{9}\nu-\frac{44}{9}\nu^2\right)\biggr]x^{3/2}\biggr\rbrace,
\end{align}

\begin{align}
\hat{F}_{33}^{(0)}&=1-4\nu+(-8+36\nu-16\nu^2)x+\biggl[6\pi(1-4\nu)+\ha_0\left(-\frac{7}{2}+14\nu\right)+\ta_{12} X_{12}\left(-\frac{1}{2}+5\nu\right)\biggr]x^{3/2}\nonumber\\
&+\biggl[\frac{1003}{55}-\frac{18352}{165}\nu+\frac{8937}{55}\nu^2-\frac{6188}{165}\nu^3+3\ha_0^2(1-4\nu)\biggr]x^2\nonumber\\
&+\biggl[(-48+213\nu-84\nu^2)\pi+\ha_0\left(\frac{559}{30}-\frac{843}{10}\nu+\frac{586}{15}\nu^2\right)+\ta_{12} X_{12}\left(-\frac{59}{30}-\frac{183}{10}\nu+\frac{391}{15}\nu^2\right)\biggr]x^{5/2}\ ,
\end{align}

\begin{align}
\hat{F}_{41}^{(1)}&=1-4\nu+\frac{1}{-1+2\nu}\biggl\lbrace\biggl[\ha_0\left(-\frac{5}{2}+10\nu\right)+\ta_{12} X_{12}\left(\frac{5}{2}-5\nu\right)\biggr]x^{1/2}\nonumber\\
&+\biggl[\frac{202}{33}-\frac{2627}{66}\nu+\frac{2188}{33}\nu^2-\frac{664}{33}\nu^3\biggr]x\biggr\rbrace\ ,\\
\nonumber\\
\hat{F}_{42}^{(0)}&=1+\frac{1}{1-3\nu}\biggl\lbrace\left(-\frac{1311}{165}+\frac{805}{33}\nu-\frac{19}{11}\nu^2\right)x\nonumber\\
&+\biggl[4\pi(1-3\nu)-\frac{4}{5}\ha_0\left(\frac{1}{3}-\nu\right)-\ta_{12} X_{12}\left(\frac{76}{15}-\frac{52}{5}\nu\right)\biggr]x^{3/2}\nonumber\\
&+\frac{1}{(1-3\nu)}\left(\frac{7199152}{275275}-\frac{140762423}{825825}\nu+\frac{37048126}{117975}\nu^2-\frac{3504901}{31460}\nu^3-\frac{65037}{7865}\nu^4\right)x^2\biggr\rbrace,\\
\nonumber\\
\hat{F}_{43}^{(1)}&=1-4\nu+\frac{1}{-1+2\nu}\biggl\lbrace\biggl[\ha_0\left(-\frac{5}{2}+10\nu\right)+\ta_{12} X_{12}\left(\frac{5}{2}-5\nu\right)\biggr]x^{1/2}\nonumber\\
&+\left(\frac{78}{11}-\frac{3139}{66}\nu+\frac{932}{11}\nu^2-\frac{1048}{33}\nu^3\right)x\biggr\rbrace, \\
\nonumber\\
\hat{F}_{44}^{(0)}&=1+\frac{1}{1-3\nu}\biggl\lbrace\left(-\frac{1779}{165}+\frac{1273}{33}\nu-\frac{175}{11}\nu^2\right)x\nonumber\\
&+\biggl[8\pi(1-3\nu)-\ha_0\left(\frac{76}{15}-\frac{76}{5}\nu\right)-\ta_{12} X_{12}\left(\frac{4}{15}-\frac{28}{5}\nu\right)\biggr]x^{3/2}\nonumber\\
&+\frac{1}{(1-3\nu)}\biggl(\frac{2187772}{55055}-\frac{261025727}{825825}\nu+\frac{95748634}{117975}\nu^2-\frac{4466169}{6292}\nu^3+\frac{1119423}{7865}\nu^4\biggr)x^2\biggr\rbrace.
\end{align}
The Newtonian prefactor can be written in closed form as
\be
\label{eq:FlmNewt}
F^{(N,\epsilon)}_{\ell m} = \dfrac{1}{8\pi} x^3 m^2(-)^{\ell +\epsilon}\left\vert {\cal R} h_{\ell m}^{(N,\epsilon)}\right\vert^2\ ,
\ee
where the Newtonian waveform multipole $h_\lm^{(N,\epsilon)}$ is explicitly given as~\cite{Damour:2008gu}
\be
\label{eq:hlmNexpl}
{\cal R} h_\lm^{(N,\epsilon)} = M\nu \,n_\lm^{(\epsilon)}c_{\ell+\epsilon}(\nu)x^{(\ell + \epsilon)/2}Y_{\ell - \epsilon,-m}(\pi/2,\phi)\ ,
\ee
with
\be
c_{\ell+\epsilon}(\nu)=X_2^{\ell+\epsilon-1}+(-)^{\ell+\epsilon}X_1^{\ell + \epsilon-1}
\ee
and
\begin{align}
n_\lm^{(0)}&=({\rm i} m)^\ell\dfrac{8\pi}{(2\ell + 1)!!}\sqrt{\dfrac{(\ell+1)(\ell+2)}{\ell(\ell-1)}}\ ,\\
n_\lm^{(1)}&=-({\rm i}m)^\ell\dfrac{16\pi{\rm i}}{(2\ell+1)!!}\sqrt{\dfrac{(2\ell+1)(\ell+2)(\ell^2-m^2)}{(2\ell-1)(\ell+1)\ell(\ell-1)}}\ .
\end{align}
The explicit evaluation of Eq.~\eqref{eq:FlmNewt} for the multipoles of interest here gives
\begin{align}
\hat{F}_{22}^{\rm Newt}&=\frac{32}{5}\nu^2 x^5\ ,\\
\hat{F}_{21}^{\rm Newt}&=\frac{8}{45} (1-4 \nu ) \nu ^2 x^6 \ ,\\
\hat{F}_{31}^{\rm Newt}&=\frac{(1-4 \nu ) \nu ^2}{1260}x^6 \ ,\\
\hat{F}_{32}^{\rm Newt}&=\frac{32}{63} (1-3 \nu )^2 \nu ^2 x^7 \ ,\\
\hat{F}_{33}^{\rm Newt}&=\frac{243}{28} (1-4 \nu ) \nu ^2 x^6 \ ,\\
\hat{F}_{41}^{\rm Newt}&=\frac{(1-4 \nu ) (1-2 \nu )^2 \nu ^2}{44100}x^8 \ ,\\
\hat{F}_{42}^{\rm Newt}&=\frac{32 (1-3 \nu )^2 \nu ^2}{3969}x^7 \ ,\\
\hat{F}_{43}^{\rm Newt}&=\frac{729}{700} (1-4 \nu ) (1-2 \nu )^2 \nu ^2 x^8 \ ,\\
\hat{F}_{44}^{\rm Newt}&=\frac{8192}{567} (1-3 \nu )^2 \nu ^2 x^7 \ .
\end{align}

\end{widetext}

\bibliography{refs20180108}

\end{document}